\documentclass[nofootinbib,superscriptaddress,a4paper, floatfix]{revtex4-2}
\usepackage[english]{babel}
\usepackage[utf8]{inputenc}
\usepackage[colorinlistoftodos, color=green!40, prependcaption]{todonotes}
\usepackage{amsthm}
\usepackage{mathtools}
\usepackage{physics}
\usepackage{braket}
\usepackage{listings}
\usepackage{xcolor}
\usepackage{graphicx}
\usepackage[left=23mm,right=13mm,top=35mm,columnsep=15pt]{geometry} 
\usepackage{adjustbox}
\usepackage{placeins}
\usepackage[scaled=1.09,proportional,lightcondensed]{zlmtt}
\usepackage[T1]{fontenc}

\usepackage{lipsum}
\usepackage{csquotes}
\usepackage[pdftex, pdftitle={Article}, pdfauthor={Author}]{hyperref} 
\usepackage{booktabs}
\usepackage{amssymb}
\usepackage{bm}
\usepackage{caption}
\usepackage{subcaption}
\usepackage{textcomp}
\usepackage{multirow}
\usepackage{nicefrac}
\usepackage{multirow}
\usepackage{dsfont}

\usepackage{tikz}
\usetikzlibrary{quantikz}

\definecolor{codegreen}{rgb}{0,0.6,0}
\definecolor{codegray}{rgb}{0.5,0.5,0.5}
\definecolor{codepurple}{rgb}{0.58,0,0.82}
\definecolor{tqblue}{HTML}{08293d}
\definecolor{backcolour}{HTML}{fefdf5}

\lstdefinestyle{mystyle}{
    backgroundcolor=\color{backcolour},   
    commentstyle=\color{codegreen},
    keywordstyle=\color{magenta},
    numberstyle=\tiny\color{codegray},
    stringstyle=\color{codepurple},
    basicstyle=\ttfamily\footnotesize\color{tqblue},
    breakatwhitespace=false,         
    breaklines=true,
    postbreak=\mbox{\textcolor{magenta}{$\hookrightarrow$}\space},                 
    captionpos=b,                    
    keepspaces=true,                 
    numbers=left,                    
    numbersep=5pt,                  
    showspaces=false,                
    showstringspaces=false,
    showtabs=false,                  
    tabsize=2
}

\newcommand{\orb}[1]{$\chi_{#1}$}
\newcommand{\amp}[2]{t_{#1}^{#2}}
\newcommand{\opwithouthat}[1]{\mathrm{#1}}
\newcommand{\operator}[1]{\hat{\opwithouthat{#1}}}
\newcommand{\quoperator}[1]{\mathsf{#1}}
\newcommand{\uunitary}{\quoperator{U}}
\newcommand{\hada}{\quoperator{H}}
\newcommand{\ham}{\operator{H}}
\newcommand{\exgen}{\operator{A}}
\newcommand{\imag}{\ensuremath{i}}
\newcommand{\identity}{\mathds{1}}
\newcommand{\pauli}[1]{\operator{\sigma}_{#1}}
\newcommand{\cre}[1]{\operator{a}^{\dagger}_{#1}}
\newcommand{\an}[1]{\operator{a}_{#1}}
\newcommand{\half}[1]{\frac{#1}{2}}
\newcommand{\nicehalf}[1]{\nicefrac{#1}{2}}
\newcommand{\thetahalf}{\half{\theta}}

\newcommand{\qpx}{\quoperator{X}}
\newcommand{\qpy}{\quoperator{Y}}
\newcommand{\qpz}{\quoperator{Z}}
\newcommand{\phiref}{\ensuremath{\Phi_\text{ref}}}
\newcommand{\ketphiref}{\ensuremath{\ket{\phiref}}}

\definecolor{codegreen}{rgb}{0,0.6,0}
\definecolor{codegray}{rgb}{0.5,0.5,0.5}
\definecolor{codepurple}{rgb}{0.58,0,0.82}
\definecolor{tqblue}{HTML}{08293d}
\definecolor{backcolour}{HTML}{fefdf5}

\lstdefinestyle{mystyle}{
    backgroundcolor=\color{backcolour},   
    commentstyle=\color{codegreen},
    keywordstyle=\color{magenta},
    numberstyle=\tiny\color{codegray},
    stringstyle=\color{codepurple},
    basicstyle=\ttfamily\footnotesize\color{tqblue},
    breakatwhitespace=false,         
    breaklines=true,
    postbreak=\mbox{\textcolor{magenta}{$\hookrightarrow$}\space},                 
    captionpos=b,                    
    keepspaces=true,                 
    numbers=left,                    
    numbersep=5pt,                  
    showspaces=false,                
    showstringspaces=false,
    showtabs=false,                  
    tabsize=2
}

\usepackage[normalem]{ulem}
\newcommand{\reviewed}[2]{#2}
\newcommand{\rereviewed}[2]{#2}

\newcommand*{\citen}{}
\DeclareRobustCommand*{\citen}[1]{%
  \begingroup
    \romannumeral-`\x 
    \setcitestyle{numbers}%
    \cite{#1}%
  \endgroup
}

\usepackage{epstopdf}

\usepackage{listings}
\definecolor{cream}{RGB}{222,217,201}
\definecolor{codegreen}{rgb}{0,0.6,0}
\definecolor{codegray}{rgb}{0.5,0.5,0.5}
\definecolor{codepurple}{rgb}{0.58,0,0.82}
\definecolor{tqblue}{HTML}{08293d}
\definecolor{backcolour}{HTML}{fefdf5}

\begin{document}

\title{A Quantum Computing View on Unitary Coupled Cluster Theory}

\author{Abhinav Anand~$^\&$}
\email[E-mail:~]{abhinav.anand@mail.utoronto.ca}
\affiliation{Chemical Physics Theory Group, Department of Chemistry, University of Toronto, Canada.}

\author{Philipp Schleich~$^\&$}
\email[E-mail:~]{philipps@cs.toronto.edu}
\affiliation{Department of Computer Science, University of Toronto, Canada.}
\affiliation{Applied and Computational Mathematics, Department of Mathematics, RWTH Aachen University, Aachen, Germany.}
\affiliation{Vector Institute for Artificial Intelligence, Toronto, Canada.}

\author{Sumner Alperin-Lea~$^\&$}
\affiliation{Chemical Physics Theory Group, Department of Chemistry, University of Toronto, Canada.}

\author{Phillip W. K. Jensen~$^\&$}
\affiliation{Chemical Physics Theory Group, Department of Chemistry, University of Toronto, Canada.}

\author{Sukin Sim}
\affiliation{Department of Chemistry and Chemical Biology, Harvard University, USA.}

\author{Manuel D\'{i}az-Tinoco}
\affiliation{Chemical Physics Theory Group, Department of Chemistry, University of Toronto, Canada.}
\affiliation{Department of Physical and Environmental Sciences, University of Toronto Scarborough, Canada.}

\author{Jakob S. Kottmann}
\affiliation{Chemical Physics Theory Group, Department of Chemistry, University of Toronto, Canada.}
\affiliation{Department of Computer Science, University of Toronto, Canada.}

\author{Matthias Degroote}
\affiliation{Chemical Physics Theory Group, Department of Chemistry, University of Toronto, Canada.}
\affiliation{Department of Computer Science, University of Toronto, Canada.}

\author{Artur F. Izmaylov}
\email[E-mail:~]{artur.izmaylov@utoronto.ca}
\affiliation{Chemical Physics Theory Group, Department of Chemistry, University of Toronto, Canada.}
\affiliation{Department of Physical and Environmental Sciences, University of Toronto Scarborough, Canada.}

\author{Al\'{a}n Aspuru-Guzik}
\email[E-mail:~]{aspuru@utoronto.ca}
\affiliation{Chemical Physics Theory Group, Department of Chemistry, University of Toronto, Canada.}
\affiliation{Department of Computer Science, University of Toronto, Canada.}
\affiliation{Vector Institute for Artificial Intelligence, Toronto, Canada.}
\affiliation{Canadian  Institute  for  Advanced  Research  (CIFAR)  Lebovic  Fellow,  Toronto,  Canada}
\date{\today}

\def\thefootnote{\&}\footnotetext{These authors contributed equally to this work}\def\thefootnote{\arabic{footnote}}

\begin{abstract}
    We present a review of the Unitary Coupled Cluster (UCC) ansatz and related ans\"atze which are used to variationally solve the electronic structure problem on quantum computers. A brief history of coupled cluster (CC) methods is provided, followed by a broad discussion of the formulation of CC theory.
    This includes touching on the merits and difficulties of the method and several variants, UCC among them, in the classical context, to motivate their applications on quantum computers.
    In the core of the text, the UCC ansatz and its implementation on a quantum computer are discussed at length, in addition to a discussion on several derived and related ans\"atze specific to quantum computing. The review concludes with a unified perspective on the discussed ans\"atze, attempting to bring them under a common framework, as well as with a reflection upon open problems within the field.
\end{abstract}

\maketitle

\tableofcontents

\section{Introduction}\label{sec:introduction}

The simulation of physical ensembles like molecules at the quantum mechanical level has been a motivating target of quantum computational research since Feynman and Manin first independently introduced the notion of a quantum computer~\cite{manin1980computable,feynman1982simulating}. 
In addition, great strides have been made in classical electronic structure computation, facilitating a great flourishing of computational chemistry as a whole. 
The flourishing of both technologies has advanced their overlap, with considerable development toward performing electronic structure calculations on quantum computers~\cite{cao2019quantum,mcardle2020quantum,bauer2020quantum}. 
In particular, variational quantum algorithms have emerged at the forefront of both quantum computing generally and quantum chemistry in particular as a focus for development in the current and near-term era of quantum computing~\cite{bharti2021noisy,cerezo2020variational}. 
Within the regime of variational quantum computational electronic structure, one ansatz, the Unitary Coupled Cluster (UCC) ansatz, has emerged as a focal point for the development of new algorithms and even for experiments~\cite{romero2018strategies}.
In this review, we explore the coupled cluster algorithmic framework, how energies can be measured on quantum computers, and a broad selection of UCC-derived ans\"atze, in order to give the reader a sense of the state and trajectory of a centrepiece method for quantum computational chemistry at present.

\subsection{A brief history of Coupled Cluster theory}
First introduced in 1960 by Coester and Kummel \cite{coester1960} for calculating nuclear binding energies, coupled cluster (CC) theory soon found itself at the forefront of \textit{ab initio} quantum chemistry. It remains enthroned as a \textit{gold standard}, striking a balance between accuracy and efficiency. Work by Sinano\u{g}lu and coworkers in the early sixties~\cite{sinanoglu1962many} brought great focus to the problem of many-electron correlations, and first introduced an approximate cluster-pair treatment thereof. In 1966, {\v C}{\'i}{\v z}ek \cite{cizek1966}  first introduced the explicit formulation of the theory for the wavefunctions of electrons; further work by {\v C}{\'i}{\v z}ek and Paldus \cite{paldus1972correlation,paldus1977correlation,paldus1978correlation} applied the formalism to problems of electronic structure. Beginning in the late 1970's, CC theory began to generalise and proliferate throughout quantum chemistry, in no small part due to the work of Bartlett, Purvis, Pople, and others~\cite{Bartlett1978manybody,pople1978electron,bartlett1980molecular,purvis1982full,yoon1984coupled}. Throughout the 1980's and 1990's, CC techniques were applied to more diverse systems, such as quasi-degenerate states or relativistic fields in heavy atoms, and to a broad range of calculations, such as the calculation of excitation energies, vibronic spectra, raman spectral intensities, valence bond solutions, and more ~\cite{meissner1988quasidegenerate,abe2014relativistic,hideo1984linear,christiansen2004vibrational,perera1999raman,cullen1996generalized,bishop1991overview}. As computers became ever-more powerful, CC continued to gain ascendancy through the general rise of computational methods in chemistry~\cite{bartlett2007review}, demonstrating great promise in the accurate calculation of quantum energies of small and weakly-correlated molecular systems and materials. It should come as little surprise, then, that the forefront method of quantum chemistry on near-term quantum computers should have its roots in such a powerful and popular formalism. We shall explore it now.

\subsection{Assumed background and notation conventions}
Throughout this review, we assume the reader is familiar with second quantisation~\cite{jorgensen2012second, surjan2012second, shavitt2009many, helgaker2014molecular}, the Hartree-Fock (HF) method~\cite{szabo1982modern, helgaker2014molecular}, and passing knowledge of the Full Configuration Interaction (FCI)~\cite{szabo1982modern, helgaker2014molecular} method for single-reference states in quantum chemistry. Additionally, we assume the reader is at least at the level of the well-versed beginner in their understanding of quantum computation, perhaps at the level of a beginning graduate student or an advanced undergraduate~\cite{nielsen2000quantum}.

We formulate our notation in the following manner: General operators are denoted in upright font accentuated by a hat, e.g. $\ham$ for the Hamiltonian or $\cre{p}, \an{q}$ for creation and annihilation operators at sites $p,q$. Additionally, we distinguish quantum operations in form of unitary circuits from general operators by using sans-serif font, e.g. $\quoperator{U}$. Our formalism generally follows that of Helgaker et. al,~\cite{helgaker2014molecular} as well as that of Szabo and Ostlund~\cite{szabo1982modern}  with few exceptions:
\orb{} indicates a molecular spin-orbital, generally composed of linear combinations of atomic orbitals taken from a pre-defined basis set. Indices \textit{p},\textit{q}, and so forth will indicate arbitrary spin orbitals; indices \textit{i},\textit{j}, etc. will indicate spin-orbitals occupied in the HF state, while indices \textit{a},\textit{b}, etc. indicate unoccupied (virtual) orbitals in the HF state.
We will denote all amplitudes -- the coefficients which multiply excitation terms -- as $\amp{p,q,r,s\ldots}{t,u,v,w\ldots}$ where lower indices are annihilated and upper indices are created. In the case of pure-excitation ans\"atze, the only non-vanishing amplitudes are of the form $\amp{i,j,k\ldots}{a,b,c\ldots}$, where appropriate we will define sums over cluster operators along these indices, implicitly removing trivial terms, reserving the more general form for ans\"atze for more generalised formulations. Later (from Section~\ref{subsec:ucc-ansatz}) in the review, we use $\theta$ instead of $\amp{}{}$ to denote amplitudes. \reviewed{}{However, in cases when dealing with quantum circuits or parametrizations in a generator formalism (i.e., when talking about some unitary operation $\quoperator{U}= \exp{\thetahalf \operator{A}}$ with some antihermitian generator $\operator{A}$) acting as multi-qubit rotations, we use $\thetahalf$ since this typically corresponds to a rotation by an angle $\theta$. We further point out that this scaling by a constant factor is merely cosmetic and does not change the final results of the procedure outlined in this review, assuming consistency within the procedure for each expression.}

We employ a ``big endian'' description of quantum states on quantum computers, that is, a string of 4 qubits $q_1,q_2,q_3,q_4$ is written as a ket-state $\ket{q_4 q_3 q_2 q_1}$.

\subsection{Coupled Cluster theory}\label{subssec:cctheory}

\subsubsection{Formulation}
Within a second-quantised framework, the coupled cluster wavefunction is described as an exponentiated excitation operator acting upon a reference determinant, usually the Hartree-Fock state. In terms of second quantisation, the Hartree-Fock state is the single-occupation vector with the lowest energy; generally, it also is a close approximation to the ground state, though only the minimality of its energy, among single-occupation vectors in the chosen basis, is guaranteed. We note for that multi-reference variants of CC -- in which the wavefunction is defined by the action of operators upon a linear combination of reference states, irreducible to each other -- have long been an open area of exploration within quantum chemistry~\cite{monkhorst1981coupled,piecuch2002multiref}, including but not limited to equation-of-motion, Hilbert-space, and Fock-space based methodologies.\cite{mukherjee1975correlation,kutzelnigg1982fock1,mukhopadhyay1989size,jeziorski1989valence, meissner1989multireference,meissner1991transformation,rittby1991multireference,kaldor1991fock,meissner2012various}. Such methods, however, lay well afield of the scope of this review.  In a fixed spin-orbital basis \{$\chi_p$\}, the coupled cluster wavefunction is given as
\begin{align}
\ket{\text{CC}}=e^{\operator{S}} \ketphiref \label{eq_exponen_ansatz}
\end{align}
where the operator $\operator{S}$ is called a cluster operator, and transforms state $\ketphiref$ into a state composed of a linear combination of determinants. The exponential map of the operator is defined through its Taylor expansion. The choice of cluster operator in conjunction with a strategy for determining the optimal amplitudes (see below) defines a given method, and the general form of the state created by the application of the cluster operator is called an ansatz; the terms method and ansatz are, in common parlance, often used interchangeably. If the operator $\hat{S}$ consists only of fermionic excitations from occupied to virtual orbitals, the generated ansatz is referred to as traditional coupled-cluster (TCC). In this case, the generating cluster operator is:
\begin{align}\label{eq:cluster-op}
\operator{T} = \sum_{k=1}^{n} \operator{T}_{k}, \quad \operator{T}_k &= \frac{1}{(k!)^2} \sum_{ij\hdots}^{\text{occ}} \sum_{ab\hdots}^{\text{vir}} t^{ab\hdots}_{ij\hdots} \hat{\tau}^{ab\hdots}_{ij \hdots} \rereviewed{}{\equiv   \sum_{i<j<\hdots}^{\text{occ}} \sum_{a<b<\hdots}^{\text{vir}} t^{ab\hdots}_{ij\hdots} \hat{\tau}^{ab\hdots}_{ij \hdots}}  , \\ \nonumber  \operator{\tau}^{ab\hdots}_{ij \hdots} &= \cre{a}\cre{b} \hdots  \an{j}\an{i},
\end{align}
where each $\operator{T}_k$ includes a product of $2k$ annihilation and creation operators; \emph{k} annihilation operators acting on orbitals occupied in the reference state, and \emph{k} creation operators which act on unoccupied orbitals of the reference state.  The choice of different values of \emph{n} yields different truncated ans\"atze. If \emph{n} equals the number of electrons, the FCI wave function lies within the manifold of wavefunctions represented by the TCC wave function and no approximation is made. Upon finding the correct amplitudes, the TCC energy is able to reach the FCI energy and hence, the lowest energy state possible for the system under analysis in a fixed basis with a single reference determinant. If the cluster operator is antihermitian, $\operator{S}^\dagger = - \operator{S}$, then $e^{\operator{S}}$ is unitary. In general, the term Unitary Coupled Cluster (UCC), introduced in the late 1980's by Bartlett, Noga, Simons, Hoffmann and Kutzelnigg~\cite{bartlett1988,hoffmann1988unitary, bartlett1989, kutzelnigg1991}, is reserved for cases where $\operator{S} =  \operator{T} - \operator{T}^\dagger$, though any antihermitian cluster operator generates a unitary operator when exponentiated. 
In addition to TCC and UCC, several other CC-ans\"atze have been explored. We note, for completeness, such ans\"atze as the Extended Coupled Cluster (ECC) \cite{Arponen1983Variational,arponen_extended_1987}, Quadratic Coupled Cluster \cite{voorhis2000quadratic}, and Generalized Coupled Cluster (GCC)\cite{Nooijen2000Brueckner}. ECC was designed to be a variational method capable of exploring time-dependent properties and processes such as bond-breaking; Quadratic CC is itself an approximation to ECC, meant to eliminate the calculation of certain computationally costly quantities. GCC, which includes a generalised excitation operator (no distinction between occupied and unoccupied orbitals), partially annihilates the reference state but couples certain excited determinants to each other, is more accurate than TCC at lower truncations, though the inclusion of more amplitudes is, of course, computationally more expensive all the same. 

Coupled cluster theory derives its power and ubiquity from a number of sources. Chiefly, CC is powerful because of the properties of truncated CC methods; in the limit of the inclusion of all degrees of excitation operators, CC will return the same energies as FCI. However, unlike truncated CI, truncated coupled cluster is \reviewed{size extensive}{size-consistent} -- that is to say, it is suitable for treating systems like crystals, quantum gases, polymers, liquids, and so forth, which truncated CI is inappropriate for at any degree of truncation. A computational method is said to be \reviewed{size-extensive}{size-consistent} if the wavefunction of a non-interacting system is given as a product state of the non-interacting fragments by said method, and the exact energy can be written in an additive form as the energies of the fragment. For example, consider the subsystems \emph{A} and \emph{B} placed infinitely far apart such that they do not interact. In that case, the actual -- not merely the approximate -- wavefunction can be factorised into two wavefunctions, $\ket{\Psi_{AB}} = \ket{\Psi_A}\ket{\Psi_B}$, localised of the individual subsystems leading to the energy being the sum of the individual subsystems energies, $\big(\ham_A + \ham_B\big)\ket{\Psi_{AB}} = \big(E_A + E_B \big)  \ket{\Psi_{AB}}$. If these requirements are fulfilled we say the computational method is \reviewed{size-extensive}{size-consistent}. The exponential ansatz (\ref{eq_exponen_ansatz}) has such separability built into it. This is in contrast to truncated CI which lack \reviewed{size-extensivity}{size-consistency}.  For example, the product state of the subsystems \emph{A} and \emph{B} using CC is
\begin{align}
\exp\big( \operator{T}_A \big)\ket{\Phi_A} \otimes  \exp\big( \operator{T}_B \big)  \ket{\Phi_B} &= \exp(\operator{T}_A +  \operator{T}_B) \ket{\Phi_A}\ket{\Phi_B},
\end{align}
with $\operator{T}_A$ and $\operator{T}_B$ describing electronic excitations on the individual subsystems. The resultant state under truncated CI however is given as
\begin{align}
 \big(1&+ \sum_i \operator{C}_{A,i} \big) \ket{\Phi_A} \otimes \big(1+ \sum_i  \operator{C}_{B,i} \big) \ket{\Phi_B}  =\\ \notag \big( 1 &+ \sum_i \operator{C}_{A,i} + \sum_i \operator{C}_{B,i} \big) \ket{\Phi_A}\ket{\Phi_B} + \sum_{ij} \operator{C}_{A,i}\operator{C}_{B,j}\ket{\Phi_A}\ket{\Phi_B}, 
\end{align}
which does not result in a separable expression on the right hand side, due to the last term in the expression. The term $\operator{C}_{A,i}\operator{C}_{B,j}$ contains excitations not included in the wavefunction (as best approximated in a fixed finite single-reference basis), and, consequently, truncated CI lacks \reviewed{size-extensivity}{size-consistency}. This assurance of \reviewed{size-extensivity}{size-consistency} can be thought of as originating in the non-linearity of the coupled cluster ansatz, even though \reviewed{size-extensivity}{size-consistency} is itself linear seperability of non-interacting systems under an ansatz. 

If the FCI ansatz is in intermediate normalisation (that is to say, $\braket{\text{FCI} | \text{HF}} = 1$ where
$\ket{\text{FCI}} = \sum_{i=0}^{N_{occ}} \operator{C}_i \ket{\text{HF}}$ with each $\operator{C}_i$ being the operator which generates all \textit{i}-fold configuration terms), then an interesting relationship among the $\operator{C}_i$  and $\operator{T}_i $  emerges. We note the first several here:
 \begin{align}
     \operator{C}_0 &=1 \\
     \operator{C}_1 &= \operator{{T}}_1 \\
     \operator{C}_2 &= \operator{{T}}_2 + \frac{1}{2} \operator{{T}}_{1}^{2} \\
     \operator{{C}}_3 &= \operator{{T}}_3 + \operator{{T}}_1\operator{{T}}_2 + \frac{1}{6} \operator{{T}}_{1}^{3} \\
     \operator{{C}}_4 &= \operator{T}_4 +  \operator{T}_1\operator{T}_3 + \frac{1}{2} \operator{T}_{2}^{2} + \frac{1}{2} \operator{T}_{1}^{2}\operator{T}_{2} + \frac{1}{24} \operator{T}_{1}^{4} 
 \end{align}
 whence it can be seen that for a given $\operator{C}_i$, all the $\operator{T}_j$ for $j\leq i$ are contributing thereto. Accordingly, even truncating at singles and doubles only, truncated CC contains portions of every configuration term. Contributions to the energy (which are necessarily indirect, since only doubly-excited determinants can overlap with the Hartree-Fock state in the equation for the cluster energy, eq. \eqref{energy}, by Brillouin's theorem~\cite{szabo1982modern}) from higher excitations will compete with the direct contributions from the lower excitations in determining the optimal CC amplitudes. This implicit connection to all higher order configurations grants truncated CC a considerable degree of accuracy, outperforming essentially all other standard single-reference methods with comparable computational cost. Because a FCI calculation is intractably hard for all but the smallest of problems, this leaves truncated CC one of the most powerful methods whose scaling is admissible. The quartic truncation of the \reviewed{Baker-Campbell-Hausdorff expansion}{Hadamard lemma} (see[ \eqref{eq:bch} below) renders the scaling of projective TCC rather modest; typically, CCSD is considered to having a scaling that is $~n^2 N^4$; for some truncation up to order $k$ the scaling is in general $~n^k N^{k+2}$, with n the number of orbitals and N the number of electrons. This modest scaling, combined with \reviewed{size-extensivity}{size-consistency}, allow truncated CC to greatly improve on Hartree-Fock (or, indeed, any single-reference method) without committing correspondingly vast additional computational resources.

In its most general form, coupled cluster can include both excitation and de-excitation operators, where (de)-excitation is understood as an action that annihilates electrons in the (higher) lower-energy orbitals to the reference states and correspondingly creates them in the (lower) higher-energy orbitals. In the case of single reference coupled cluster with the Hartree-Fock determinant as the reference state, however, the excitation-only ans\"atze, as well as the mixed excitation and de-excitation ans\"atze are both exact,\cite{van_voorhis2001two-body} and will generate the same determinants; excitation-only and mixed methods differ in the wavefunction just in the amplitudes that multiply individual Slater determinants. As UCC on quantum computers is a single-reference ansatz with e.g a Hartree-Fock reference determinant, we will explore herein only the strictly occupied-to-virtual-excitation-operator formulation of the coupled cluster ansatz. 

\subsubsection{Projective approaches}\label{subsec:projectiveclassical}
Traditional CC, unlike FCI and its truncations, is not typically solved by variational methods; the non-Hermiticity of the exponentiated excitation operator makes variational minimization difficult\reviewed{}{as in this case the energy is not bounded from below}. Instead, projective methods are employed. By insertion into the Schr{\"o}dinger equation under the system Hamiltonian $\ham$, we obtain the following equation for an electronic eigenstate:
\begin{equation}
    \ham e^{\operator{S}} \ketphiref = E e^{\operator{S}} \ketphiref \label{energy}
\end{equation}
for a cluster operator $\operator{S}$ whose parameters are optimal. Additionally, the optimal excitation amplitudes induce the constraint of the orthogonality of any excited determinant with the reference state. Letting $\ket{\mu}$ be any member of the set of excited determinants, the optimal amplitudes of $\hat{T}$ are found by solving one of two sets of coupled equations. One projects onto the manifold of excited determinants, given by:
\begin{equation}
    \label{mu} \bra{\mu} = \bra{\phiref} \operator{\tau}_{\mu}^{\dagger}
\end{equation}
where each $\ket{\mu} \in \{\ket{\mu}:\hat{\tau}_{\mu}\ket{\Phi} \}$ with $\tau_{\mu}$ any of the individual strings of annihilation-creation operators contained in  the $\hat{T}_k$.
Projection onto this manifold generates a set of coupled equations:
\begin{align}
    \bra{\phiref}\ham e^{\operator{S}} \ketphiref &= E \label{unlinked_e} \\
    \bra{\mu}\ham e^{\operator{S}} \ketphiref &= E \bra{\mu} e^{\operator{S}} \ketphiref \label{unlinked}
\end{align}
This set of equations is most generally solved iteratively. Equations \eqref{unlinked_e} and \eqref{unlinked} are referred to as the ``unlinked'' coupled cluster equations, in reference to diagrammatic coupled cluster theory~\cite{shavitt2009many}, because therein both ``unlinked'' and ``linked'' diagrams contribute to the energy. Another form of projected equations can also be constructed, and is in fact the more commonly used form for TCC and its truncations. By multiplying the Schr\"odinger equation \eqref{energy} by the operator $e^{-\operator{S}}$ on the left, one generates
\begin{align}
    \bra{\phiref}e^{-\operator{S}}\ham e^{\operator{S}}\ket{\phiref} &= E \label{linked_E} \\
    \bra{\mu} e^{-\operator{S}}\operator{{H}}e^{\operator{S}}\ket{\phiref} &= 0 \label{linked}
\end{align}
for all $\ket{\mu}$. Equations \eqref{linked_E} and \eqref{linked} are called the ``linked'' equations, as they do not give rise to ``unlinked'' diagrams in the diagrammatic treatment. The action of the exponential operators in \eqref{linked_E} and \eqref{linked} on the Hamiltonian is a similarity transformation, yielding an \reviewed{effective Hamiltonian}{similarity transformed Hamiltonian}, $\Tilde{\opwithouthat{H}} = e^{-\operator{S}}\ham e^{\operator{S}}$. It must be noted that this transformed Hamiltonian is no longer strictly Hermitian, because the operator by which it is transformed is not unitary, an issue addressed in the following subsection. 
The use of \eqref{linked_E} and \eqref{linked} requires some expression for $\Tilde{\opwithouthat{H}}$. The most straightforward approximation is to use the \reviewed{Baker-Campbell-Hausdorff approximation (BCH)}{Hadamard lemma}. It is $\textit{prima facie}$ possible that this approximation to $\Tilde{\opwithouthat{H}}$ never truncates, giving:
\begin{align}
    \label{eq:bch}
    \Tilde{\opwithouthat{H}} = e^{-\operator{S}} \ham e^{\operator{S}} = \ham + [\ham,\operator{S}] + \frac{1}{2}[[\ham,\operator{S}],\operator{S}] + \hdots 
\end{align}
However, when $\operator{S}=\operator{T}$, then the series truncates at the fourth order, and is given as below: 
\begin{align}
     e^{-\operator{T}} \ham e^{\operator{T}} =  &\ham + [\ham, \operator{T}] + \frac{1}{2} [[\ham,\operator{T}],\operator{T}] \\ \notag &+ \frac{1}{6} [[[\ham,\operator{T}],\operator{T}],\operator{T}] + \frac{1}{24}[[[[\ham,\operator{T}],\operator{T}],\operator{T}],\operator{T}]. 
\end{align}
The reason the expansion truncates at the quartic order is that the molecular Hamiltonian
 \reviewed{only has one- and two-body interactions}{contains at most two-body operators}. \reviewed{}{Since the molecular Hamiltonian contains at most two creation and two destruction operators it can be contracted with at most four $\operator{T}$ operators -  detailed derivation in second-quantized and diagrammatic language can be found in~\cite{shavitt2009many}}. Once the coupled projective equations are obtained, any of several classical methods may be employed in order to solve said equations. Among the most standard of these is the quasi-Newton direct inversion in the subspace (DIIS) method. 

\subsubsection{Variational approaches}
\label{sec:variational_approaches}
It is possible to optimise a coupled-cluster expectation value variationally. In such a formalism, the goal is to minimise the Rayleigh-Ritz energy functional: 
\begin{align}
E(\operator{S}) &= \frac{\braket{\Phi|e^{\operator{S}^\dagger} \ham e^{\operator{S}}| \Phi}}{\braket{\Phi|e^{\operator{S}^\dagger}  e^{\operator{S}}| \Phi}}. \label{rayleigh-ritz}
\end{align}
One might attempt to apply the \reviewed{BCH formula}{Hadamard lemma} to the transformed Hamiltonian and to the product term $e^{\operator{S}^{\dagger}}  e^{\operator{S}}$, in order to render the variational formulation more tractable. Note, however, that the transformation of the Hamiltonian is no longer the \reviewed{effective Hamiltonian}{similarity transformed Hamiltonian} $\Tilde{\opwithouthat{H}}$, cf. \eqref{eq:bch}; furthermore, if $\operator{S}$ is not antihermitian, the evaluation of the denominator of \eqref{rayleigh-ritz} is not trivial. In general -- including for both the pure excitation operator and for the usual UCC cluster operator $\operator{S}=\operator{T}-\operator{T}^\dagger$, no convergent truncation for the Rayleigh-Ritz functional is known. \reviewed{}{The reason there is no truncation is the operator cannot be separated into only virtual creation and occupied destruction operators.} An advantage with variational over projected methods are they furnish an upper bound to the FCI energy, and of that reason we expect better agreement with FCI results. The deviation from the variational and projected energy, however, may be small and of little practical consequence --- often, this has been observed in the weakly correlated regime~\cite{jorgensen2012second}. If the single-reference state recovers mostly of the ground state then the CC methods would differ only slightly. The variational coupled-cluster (VCC) method, where $\operator{S} = \operator{T}$ in \eqref{rayleigh-ritz}, shows within numerical studies in \cite{harsha_difference_2018, cooper_benchmark_2010, evangelista_alternative_2011} better agreement with FCI results over variational unitary coupled-cluster (VUCC), where $\operator{S} = \operator{T}-\operator{T}^\dagger$. Within these studies this was in particular observed in the strongly correlated regime. For instance the Lipkin model \cite{harsha_difference_2018}, which is a chain of \emph{N} sites with each two levels and \emph{N} particles that are allowed to hop between levels but between different sites. They analysed the model in both weakly and strongly correlated limit, and found that VCC performs slighty better than VUCC, and the variational generalized CC method outperformed all the CC methods. This is consistent with observations for smaller molecules in \cite{cooper_benchmark_2010} and \cite{evangelista_alternative_2011}: At equilibrium bond distances, the differences between the various CC methods were found to be small~\cite{cooper_benchmark_2010,evangelista_alternative_2011} due to weak correlation. Going to the strong correlation limit by stretching bonds, the variational methods performed better than the projected methods, and, in general, VCC was found to be better than VUCC. This comparison, however, required truncating the Rayleigh-Ritz formula in addition to the truncation of the excitation operator.

Examination of both the variational and projective approaches in TCC, as well as between VTCC and VUCC on classical computers, reveals that each method, though powerful, suffers certain qualitative and quantitative limitations, which prevent any single method from being pronounced ``superior'' to the others in general. In addition to the projective and variational methods mentioned above, we note the existence of methods for representing CC energies through the eigenvalues of small-dimensitonality active-space effective Hamiltonians. \cite{kowalski2018properties,kowalski2021dimensionality}
 
 \subsubsection{Moving beyond the classical}
 While the non-linearity of the CC ansatz grants \reviewed{size-extensivity}{size-consistency} in truncation, it also denies easy access to variationality. This lack of variationality means that two energies from projective coupled cluster methods cannot be straightaway compared, as can HF, FCI, or truncated CI; without reference to experiment (or to FCI directly) the quality of an energy returned by projective coupled cluster cannot be verified ascertained. In practice, the increased accuracy overall of projective coupled cluster relative to methods like truncated CI or DFT outweighs the uncertainty induced by a lack of variationality, in part because physical properties like rate constants depend on differences in energies, rather than their magnitudes. Complicating matters, however, coupled cluster, at all levels of truncations, may suffer from an ``explosive'' problem: convergence to finite energies is not guaranteed; even for only modestly correlated systems, coupled cluster may return infinitely negative energies, whereas the other standard electronic structure methods can, in general, guarantee convergence, even when accuracy is far out of reach.
There are some recent advances in the field which can counteract this exploding behaviour, while achieving accurate ground state energies as shown in Ref.~\citen{eriksen2020ground}
 Though VCC and VUCC, performed on classical computers, guarantee a finite energy and maintain \reviewed{size-extensivity}{size-consistency} with respect to truncation with respect to the operator $\operator{T}$, approximations to \eqref{rayleigh-ritz} introduce new errors.
 Furthermore, convergence of an optimisation of such variational methods may be slow or difficult, leading to prohibitive cost; the advent of quantum computation might lead to this becoming a tractable problem. 
 
 \subsubsection{Moving beyond the classical}
 While the non-linearity of the CC ansatz grants size-consistency in truncation, it also denies easy access to variationality. This lack of variationality means that two energies from projective coupled cluster methods cannot be straightaway compared, as can HF, FCI, or truncated CI; without reference to experiment (or to FCI directly) the quality of an energy returned by projective coupled cluster cannot be verified ascertained. In practice, the increased accuracy overall of projective coupled cluster relative to methods like truncated CI or DFT outweighs the uncertainty induced by a lack of variationality, in part because physical properties like rate constants depend on differences in energies, rather than their magnitudes. Complicating matters, however, coupled cluster, at all levels of truncations, may suffer from an ``explosive'' problem: convergence to finite energies is not guaranteed; even for only modestly correlated systems, coupled cluster may return infinitely negative energies, whereas the other standard electronic structure methods can, in general, guarantee convergence, even when accuracy is far out of reach.
 Though VCC and VUCC, performed on classical computers, guarantee a finite energy and maintain size-consistency with respect to truncation with respect to the operator $\operator{T}$, approximations to \eqref{rayleigh-ritz} introduce new errors. Furthermore, convergence of an optimisation of such variational methods may be slow or difficult, leading to prohibitive cost; the advent of quantum computation might lead to this becoming a tractable problem. 

\section{Estimating energies on quantum computers}
A fundamental goal of quantum chemistry is to solve the electronic structure problem, i.e., given a molecular Hamiltonian $\operator{H}$, finding the energy eigenstates $\ket{E_i}$ and the corresponding energy eigenvalues $E_i$.
The first proposal for calculating energy eigenvalues of molecular Hamiltonians on a quantum computer was put forward by Aspuru-Guzik et~al. in Ref.~\citen{aspuru2005simulated} and further developed by Whitfield et~al. in Ref.~\citen{whitfield2011simulation}.
These proposals involved the use of the Trotter-Suzuki relations~\cite{suzuki1976generalized} and the Quantum Phase Estimation (QPE) algorithm~\cite{kitaev_quantum_1997, abrams_simulation_1997, abrams_quantum_1999} for computing the ground state energies of electronic Hamiltonians. 
However, the methods based on QPE require significant quantum resources ($\mathcal{O}(\log\frac{1}{\epsilon})$ qubits and controlled-$U$ operations within an additive error $\epsilon$) and are unlikely to be used in the near future. 
An alternative method for calculating molecular energies is the Variational Quantum Eigensolver (VQE), also proposed by Aspuru-Guzik and co-workers~\cite{peruzzo2014variational}, which involves using a parameterized circuit to prepare a quantum state, and measure the expectation value with respect to a Hamiltonian. The expectation value is then variationally minimised iteratively to obtain the approximate ground state energy. We describe the two methods briefly in the following sub-sections.

\subsection{Quantum Phase Estimation}
The quantum phase estimation algorithm is a quantum algorithm to estimate the eigenvalues of a unitary operator $\quoperator{U}$. 
For any unitary operator $\quoperator{U}$ with eigenstate $\ket{\Psi}$ and eigenvalue $\lambda$, one can write:
\begin{align}
    \quoperator{U} \ket{\Psi} = \lambda \ket{\Psi} = e^{2\pi i \theta}\ket{\Psi},
\end{align}
where the eigenvalue can be expressed as a phase $\lambda = e^{2\pi i \theta}$, which is the origin of the name ``phase estimation''.
QPE uses a unitary operator $\quoperator{O}$ to prepare an approximation to the eigenstate $\ket{\psi} = \quoperator{O}\ket{0}^{\otimes n}$ on a quantum computer and perform the following transformation:
\begin{align}
    \ket{\psi}\ket{0}^{\otimes m} \rightarrow \ket{\psi}\ket{\Tilde{\theta}},
\end{align}
where $\ket{\Tilde{\theta}}$ is a state that contains a binary approximation to the phase. 
The set of qubits with the approximate eigenstate $\ket{\psi}$ is referred to as the state register, and the qubits with the state $\ket{\Tilde{\theta}}$ as energy registers.
This transformation is realised by the application of multiple unitaries $\quoperator{U}^{2^n}, n \in \{0, .., m-1\}$ on the state register controlled by the qubits in the energy register, followed by an inverse quantum Fourier transformation (QFT) on the energy register. 
The energy register with state $\ket{\Tilde{\theta}}$ can then be measured to extract the phase information. Intuitively, we probe the system with $\quoperator{U}^{2^n}$ $m$ times, where each application adds an additional bit of precision to the phase of the state, and then extract the information by QFT into the energy register.

For a given molecular Hamiltonian $\ham$, one uses the unitary operator $\quoperator{U} = e^{-i\ham t}$ since its phase corresponds to the eigenvalues of $\ham$; this unitary can be implemented using Trotter-Suzuki relations~\cite{suzuki1976generalized}. 
The approximate eigenstate $\ket{\psi}$ can be prepared using an ansatz $\uunitary(\theta)$ in the VQE framework which is described briefly in the next section. 
In summary, quantum phase estimation for extracting the spectrum of a Hamiltonian can be represented as performing the following operation,
\begin{align}
    \text{QPE}(\ham) \quoperator{O} \ket{0}^{\otimes (m+n)} =  \text{QPE}(\ham) \ket{\psi} \ket{0}^{\otimes m} \approx \ket{\psi}\ket{\Tilde{\theta}}. 
\end{align}
Measuring $\ket{\Tilde{\theta}}$ results in $m$~bits which can used to recover the corresponding eigenvalue $\lambda$.

Let us now consider a single-qubit example with the Hamiltonian $\ham = \frac{1}{2}\left(1 - \pauli{x} \right)$. This Hamiltonian has eigenvalues $0,1$ and the eigenstates $\ket{\pm}=\frac{1}{\sqrt{2}}\left(\ket{0}\pm\ket{1}\right)$.  Since in this special case, the eigenvalues are already binary, one can simply represent them as $\ket{0}, \ket{1}$ on the energy register without worrying about numerical accuracy in the binary representation. This is also the reason why a single appliction of $e^{-i\ham t}$ suffices. The full QPE circuit for this example then is:
\begin{equation}
     \uunitary_{\text{QPE}} = 
\begin{tikzcd}
 & \gate{\hada} & \ctrl{1} & \gate{\hada}  &\meter{0/1}   \\
 & \gate{\quoperator{O}} & \gate{e^{-\imag \ham t}} & \qw 
\end{tikzcd} \nonumber
\end{equation}
where the circuit $\quoperator{O}$ creates the trial state and $\hada$ are Hadamard gates. 
Taking $\ket{0}=\frac{1}{\sqrt{2}}\left(\ket{+}+\ket{-}\right)$ for the trial state (i.e., $\quoperator{O} = \identity$), the QPE circuit creates the total wavefunction
\begin{align}
    \uunitary_\text{QPE}\ket{00} &= \uunitary_\text{QPE}\frac{1}{\sqrt{2}}\left(\ket{+} + \ket{-}\right)\otimes\ket{0}\nonumber\\
    &= \frac{1}{\sqrt{2}} \left( \ket{+0} + \ket{-1} \right),
\end{align}
which couples the eigenvalues (0,1) in state register with the eigenstates $(\ket{+},\ket{-})$ in the energy register. With a more general initial state $a\ket{+}+b\ket{-}$, the final state of the QPE circuit would be $\frac{1}{\sqrt{2}} \left( a\ket{+0} + b\ket{-1} \right)$. The probabilities of measuring the eigenenergy $0$ or $1$ in the energy register and collapsing into the corresponding eigenstate in the state register are $\left| a \right|^2$ and $\left|b\right|^2$ respectively; in this initial example, both 50\%. In general, the Hamiltonians of interest have more than two eigenvalues which are not 0 or 1 in their natural units. This is circumvented by shifting and re-scaling the Hamiltonian of interest, in order to move the eigenvalues into the interval $\left[0,1\right]$ and the use of an $m$-bit binary approximation. The Hamiltonian above is, in fact, a shifted and re-scaled version of the $\pauli{x}$ operator.

\subsection{Variational Quantum Eigensolver}
The Variational Quantum Eigensolver was proposed in Ref.~\citen{peruzzo2014variational}, as a method to find approximate ground states of molecular Hamiltonians, with near-term quantum hardware in mind. It is based on the Hybrid Quantum-Classical (HQC) approach, wherein one uses quantum resources in tandem with classical computers in order to exploit the available quantum resources to the utmost extent while outsourcing the optimisation task to a classical machine. VQE uses the quantum computer for state preparation using a parameterized unitary operation $\uunitary\left(\theta\right)$, commonly referred to as \textit{ansatz}, and the estimation of the expectation value of an observable $\ham$. The expectation value is then minimised as per the variational principle using a classical optimisation algorithm. The parameters of the unitary operator are optimised iteratively until convergence in the energy. The whole process can be summarised as:
\begin{align}
    E_\mathrm{min} &= \min_\theta \langle \ham \rangle_{\uunitary(\theta)}, \\
    \langle \ham \rangle_{\uunitary(\theta)} &\equiv  \langle 0 \rvert \uunitary^\dagger\left(\theta\right) \ham  \uunitary\left(\theta\right) \lvert 0\rangle.
\end{align}
Using the optimal parameter values $\theta^* = \text{argmin}_\theta\{\langle \ham \rangle_{\uunitary(\theta)}\}$, one can write the ground state of the system as $|\Psi \rangle = \uunitary\left(\theta^* \right) \lvert 0\rangle$. Note that here, if we choose a UCC-type ansatz, $\uunitary(\theta)$ needs to contain the reference preparation for the HF state.

\subsection{Tackling challenges within QPE and VQE by UCC}
The success of both quantum phase estimation and the variational quantum eigensolver depends on having a good approximation to the eigenstate of the Hamiltonian of which the energy is to be estimated, since the performance of both algorithms is sensitive to the initial guess provided. 
Thus, it is of utmost importance to use an efficient state preparation procedure, and that the ansatz with the right parametrisation is capable of reaching the final state with a good overlap.
One such ansatz inspired from classical quantum chemistry methods is the unitary coupled cluster approach, which offers a physically motivated framework that can approximate the exact eigenstates by increasing the rank of the excitation operators of UCC.
The UCC approach can be readily implemented on a quantum computer and has been used for VQE based methods, which can be further used as initial state for more advanced algorithms like QPE or other time-evolutions. 

The reader is directed to these reviews~\cite{cao2019quantum, mcardle2020quantum, bharti2021noisy, cerezo2020variational} to get a detailed understanding of the VQE algorithm, Refs.~\citen{cao2019quantum} and ~\cite{mcardle2020quantum} provide a detailed overview of general quantum chemistry on quantum computer, Ref.~\citen{bharti2021noisy} provide extended details of variational and NISQ algorithms and Ref.~\citen{cerezo2020variational} a more compact introduction to variational algorithms.
In what follows we describe in detail the implementation of the UCC method on a quantum computer.

\section{The Unitary Coupled Cluster ansatz for quantum computing}\label{subsec:ucc-ansatz}
On a quantum computer, the UCC ansatz is implemented by constructing a unitary operator $U$ as described in sec~\ref{subssec:cctheory}, which can be written as follows:
\begin{align}
    \uunitary = e^{\operator{T} - \operator{T}^{\dagger}}
\end{align}
where $\operator{T} = \sum_i \operator{T}_i$ is the fermionic excitation operator and consists of single, double, triple, \ldots, up to $n$-fold excitations. 
The unitary operator $\uunitary$ then has to be decomposed into operations that can be implemented on a quantum device.
Since individual excitations do not commute, one needs to invoke approximations, which is often done using the Trotter decomposition, which relates the exponential of a sum to a ordered product of individual exponentials. This provides a reasonable  approximation to the full unitary in terms of local unitary operations. 
The first-order decomposition formula is given by:
\begin{align}
    e^{\operator{T} - \operator{T}^{\dagger}} &= e^{\sum_i \theta_i (\operator{T}_i - \operator{T}_i^{\dagger})} \\
    &\approx \left(\prod_i e^{\frac{\theta_i}{t} (\operator{T}_i - \operator{T}_i^{\dagger})}\right)^t + \mathcal{O}\left(\frac{1}{t}\right),\label{eq:firstordertrotter}
\end{align}
where $t$ is the Trotter number (order of decomposition) and $\operator{T_i}$ are the excitation operators. To reduce the error throughout this approximation, higher-order decomposition formulas can be employed~\cite{childs2021trotter}.
However, they require additional quantum resources and are not used in general when constructing the circuit corresponding to the UCC ansatz~\cite{romero2018strategies}. This is also due to the fact, as found out in Refs.~\citen{Barkoutsos2018,romero2018strategies}, that the optimisation procedure as part of VQE accounts \rereviewed{}{partially} for the error due to Trotterisation; in this sense, the Trotterised ansatz is to be understood as an ansatz on its own. However this only holds, if the order of individual operators in the ordered product is suitable, since the choice of order in the product in Eq.~\eqref{eq:firstordertrotter} is not unique. This will be further discussed in section~\ref{subsubsec:order}. 

Using the first-order approximation, we can decompose our parametrised unitary in terms of elementary quantum gates efficiently with respect to system size. However, this leads in general to relatively large circuits.
The resulting wavefunction that can be prepared on a quantum computer can be expressed as
\begin{align}
    \ket{\psi_{\mathrm{UCC}}(\theta)} &= \uunitary \ket{\Phi_{\mathrm{ref}}} = 
    e^{\sum_{(\mathbf{p}, \mathbf{q}) \in \mathcal{E}}
    (\nicehalf{\theta_{\mathbf{p}\mathbf{q}}}) \operator{A}_{\mathbf{p}\mathbf{q}}      }  \ket{\Phi_{\mathrm{ref}}}
    \\ \notag
    &\approx   \prod_{(\mathbf{p},\mathbf{q}) \in \mathcal{E}} e^{  (\nicehalf{\theta_{\mathbf{p}\mathbf{q}}}) \operator{A}_{\mathbf{p}\mathbf{q}}      }  \ket{\Phi_{\mathrm{ref}}},
\end{align}
where $\mathcal{E}$ is the set of all possible electronic excitations, with parameters $\theta$ and  antihermitian excitation generators $\exgen$ as defined in~\eqref{eq:fermionic-ex-gen}. E.g. for UCCSD,
\begin{align}\label{eq:excitation-set-sd}
\mathcal{E}_{\mathrm{SD}} = \left\{  \{(p, q), 1\le p < q \le M \} \cup \{ (pq  ,rs), \color{white}\}\right\}\normalcolor
\\ \color{white}\left\{\{\normalcolor 1\le p<q \le M, 1 \le r<s \le M  \}     \right\} \notag
\end{align}
or generally up to $n$-fold excitations,
\begin{equation}
\mathcal{E}_n = \left\{ (\mathbf{p}, \mathbf{q}):  1\le \abs{\mathbf{p}},\abs{\mathbf{q}}  \le n , 1\le p_i < q_i \le M  ~\forall i=1,\ldots,n   \right\}
\end{equation}
where $M$ is the number of spin-orbitals. To allow a   more compact notation later on, we introduce multi-indices $\mu$ replacing the $(\mathbf{p}, \mathbf{q})$-tuples above. This results in general $n$-fold excitation generators \begin{equation}
    \operator{A}_\mu = \cre{\mu_1} \an{\mu_{1+n}} \dots \cre{\mu_n} \an{\mu_{2n}} -\mathrm{h.c.}
\end{equation}
where $\cre{}$ and $\an{}$ are the fermionic creation and annihilation operators respectively. The generators are ordered with respect to particle affinity and sets of excitations
\begin{equation}
    \mathcal{E}_n=\{ \mu: 1\le \nicefrac{\abs{\mu}}{2} \le n, 1 \le \mu_{i} < \mu_{ i+n } \le n ~\forall i=1,\ldots,n    \} 
\end{equation}

We next describe the fermionic excitation generators, their exponential maps and corresponding circuit elements in detail.

\subsection{Fermionic excitations: The basic building blocks}\label{subsec:ferm-exc}

A general $n$-fold fermionic excitation operator that excites electrons between spin-orbitals $\mathbf{p}=\left\{p_0\dots p_n\right\}$ and $\mathbf{q}=\left\{q_0\dots q_n\right\}$ can be written as:
\begin{align}
    \exgen_{\mathbf{p}\mathbf{q}} = \prod_{k=1}^{n}\cre{p_k}\an{q_k} - \prod_{k=1}^{n}\cre{q_k}\an{a_k}  \label{eq:fermionic-ex-gen} 
\end{align}
To ease readability, we call $\exgen_{\mathbf{p},\mathbf{q}}$ excitations, even though they contain both excitations and de-excitations. Given that for $n=1$, $\exgen_{{p},{q}}$ corresponds to orbital rotations, the term rotation could be employed as well. Throughout this work, we consider antihermitian excitation generators $\exgen_{\mathbf{p}\mathbf{q}}$, such that their exponential yields a unitary operation. Alternatively, one may choose Hermitian excitation generators by inclusion of the imaginary unit into the generators and an additional factor $-\imag$ in the exponent. Further, within this section, we keep the division of the amplitude by 2 as a convention since this facilitates the compilation into gate sequences in section~\ref{sec:qubit_perspective}. From section~\ref{sec:ucc-flavors}, we will not follow this convention strictly to allow for easier readability, yet keeping in mind that a mapping $\theta\to\thetahalf$ is necessary in case one uses the described implementation techniques.

\rereviewed{
}{
It was shown by the authors in Ref.~\citen{kottmann2021feasible} that if we decompose this unitary $\uunitary_{\mathbf{p}\mathbf{q}}\left(\theta\right)$ as
\begin{align}\label{eq:generator_decomp_param_excit}
    \uunitary_{\mathbf{p}\mathbf{q}}\left(\theta\right) = e^{\frac{\theta}{4}\left( \exgen_{\mathbf{p}\mathbf{q}}+ \imag\operator{P}_{\mathbf{p}\mathbf{q}}\right)} e^{\frac{\theta}{4}\left( \exgen_{\mathbf{p}\mathbf{q}} - \imag\operator{P}_{\mathbf{p}\mathbf{q}}\right)},
\end{align}
we can utilise the \textit{shift-rule}~\cite{schuld2019evaluating} to evaluate analytical gradients efficiently for optimisation, as the operators $\exgen_{\mathbf{p}\mathbf{q}} \pm \imag \operator{P}_{\mathbf{p}\mathbf{q}}$ are involutions, i.e.,
\begin{align}
    \left(\exgen_{\mathbf{p}\mathbf{q}} \pm \imag \operator{P}_{\mathbf{p}\mathbf{q}}\right)^2 = \identity.
\end{align}

One can then use the above expressions to write the exponential map of a parametrised $n$-fold fermionic excitation operator, $e^{\theta \operator{A}_{\mathbf{p}\mathbf{q}}}$, as
\begin{align}
    \uunitary_{\mathbf{p}\mathbf{q}}\left(\theta\right) &= e^{\thetahalf \operator{A}_{\mathbf{p}\mathbf{q}}} \label{eq:basic_fermionic_unitary}\\
    &= \cos\left({\thetahalf}\right)\identity + \sin\left(\thetahalf\right)\operator{A}_{\mathbf{p}\mathbf{q}} + \left(1-\cos\left(\thetahalf\right)\right)\operator{P}_{\mathbf{p}\mathbf{q}},\nonumber
\end{align}
where $\operator{P}_{\mathbf{p}\mathbf{q}} = \identity -\prod_{k=1}^{n}      \cre{p_k}\an{p_k} \an{q_k}\cre{q_k} -\prod_{k=1}^{n} \cre{q_k}\an{q_k} \an{p_k}\cre{p_k}$ is the projector onto the nullspace of the generator $\operator{A}_{\mathbf{p}\mathbf{q}}$ \cite{Evangelista2019,xu2020test,chen2021quantum,kottmann2021feasible}.
The unitary $\uunitary_{\mathbf{p}\mathbf{q}}\left(\theta\right)$ can be regarded as a rotation gate, rotating within all configurations made up of \textit{all} $\mathbf{p}$ spin-orbitals occupied and \textit{all} $\mathbf{q}$ spin-orbitals, as well as their complements where the reverse conditions are fulfilled.}

This formulation allows for a feasible automatically differentiable framework for quantum circuits built up from primitive excitations as in Eq.~\eqref{eq:basic_fermionic_unitary}, described in detail in Ref.~\citen{kottmann2021feasible} and applied in a similar way in Ref.~\citen{anselmetti2021local}. This was further generalized in Refs.~\citen{izmaylov2021analytic,wierichs2021general}.

\subsection{Qubit perspective on fermionic excitations}\label{sec:qubit_perspective}
In order to implement a unitary operation $ \uunitary_{\mathbf{p}\mathbf{q}}\left(\theta\right)$ on a quantum computer, one needs to map the fermionic operation to an operation that is native to a quantum computer, i.e., what is typically called a qubit operator. To do so, there exist several encodings, such as Jordan-Wigner(JW)~\cite{jordan1928pauli}, Bravyi-Kitaev(BK)~\cite{bravyi2002fermionic}, among others~\cite{setia2018bravyikitaevsuperfast,cao2019quantum,mcardle2020quantum}. 
In the Jordan-Wigner transformation, creation and annihilation operators can be expressed using qubit operators $\pauli{\pm} = \frac{1}{2}(\pauli{x} \pm \imag \pauli{y}) $ and $\pauli{z}$:
\begin{align}
    \cre{k} =& \identity^{\otimes k-1}\otimes \pauli{+}^k \otimes \pauli{z}^{\otimes n-k} \\
    \an{k} =& \identity^{\otimes k-1} \otimes \pauli{-}^k  \otimes \pauli{z}^{\otimes n-k},
\end{align}
where $\identity$ is the identity operator. The ladder operators $\pauli{\pm}^k$ create/annihilate particles at site $k$, while the $\pauli{z}$ operations encode parity information in the phase. 
The various mapping strategies allow to write  unitary operators corresponding to fermionic (de-)excitations  in terms of exponentials of Pauli operators, which can then be transformed to circuit elements. 

We next illustrate the construction of quantum circuits corresponding to the exponentiation of different Pauli strings; we say an operator $\operator{P}$ is a Pauli string  over $N$ qubits, if
\begin{equation}
    \operator{P} = \bigotimes_{j=1}^N \pauli{j}^{(j)}, \quad \text{where~} \pauli{j}\in\{ \identity, \pauli{x}, \pauli{y}, \pauli{z}\}. 
\end{equation}
To construct the quantum circuit corresponding to some exponential map $e^{- \imag \frac{\theta}{2} \operator{P} }$, we use an entangling CNOT-ladder between all Pauli operators $\pauli{j}^{(j)}\in \operator{P}$, where we choose the convention to start at the smallest $j$ and go until the largest. In the middle of this ladder, there is a  $\quoperator{R}_z(\theta)$ gate which performs a $z$-rotation by the parameter $\theta$, followed by a disentangling CNOT ladder, in reverse order as before. To implement operators other than $\pauli{z}$, one needs to add respective basis transformations before the circuit and their inverses after. The rotatation between $\pauli{x}$ and $\pauli{z}$ is achieved by the Hadamard transformation $\hada = \quoperator{R}_y(\half{\pi})$ and $\quoperator{R}_x(\frac{\pi}{2})$ transforms from the $\pauli{y}$ to $\pauli{z}$ basis.   

As a brief example, we show the quantum circuit for $e^{- \imag \frac{\theta}{2} \pauli{z}^{(0)} \pauli{z}^{(1)} }$:
\begin{align}
e^{- \imag \frac{\theta}{2} \pauli{z}^{(0)} \pauli{z}^{(1)} } = \raisebox{-17.5pt}{\includegraphics[width=0.35\columnwidth]{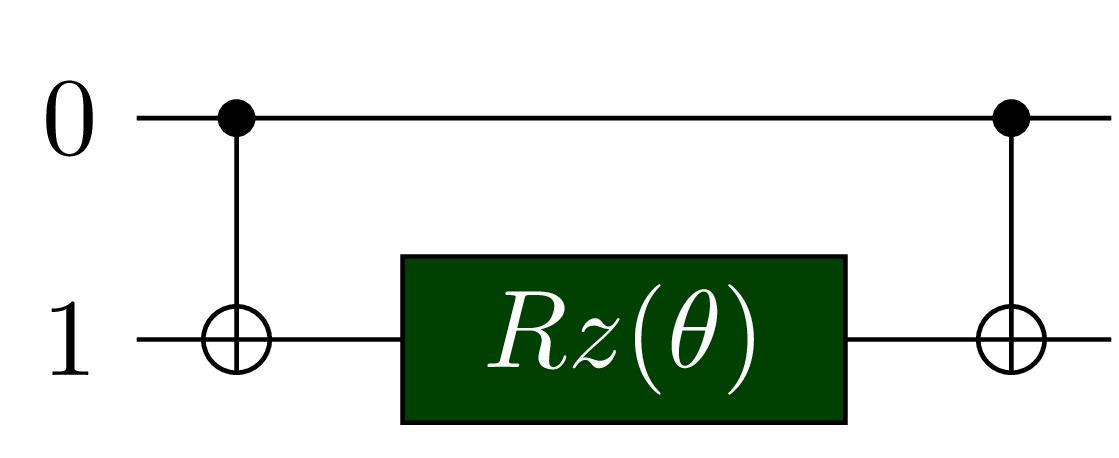}} 
\end{align}

\begin{figure}[htbp]
\centering
    \begin{tabular}{c c}
    \toprule
    Exponential maps & Circuit\\
    \midrule
    $e^{- \imag \frac{\theta}{2} \pauli{z}^{(0)} \pauli{z}^{(1)} \pauli{z}^{(2)} \pauli{z}^{(3)}}$ &
    \includegraphics[height=50pt]{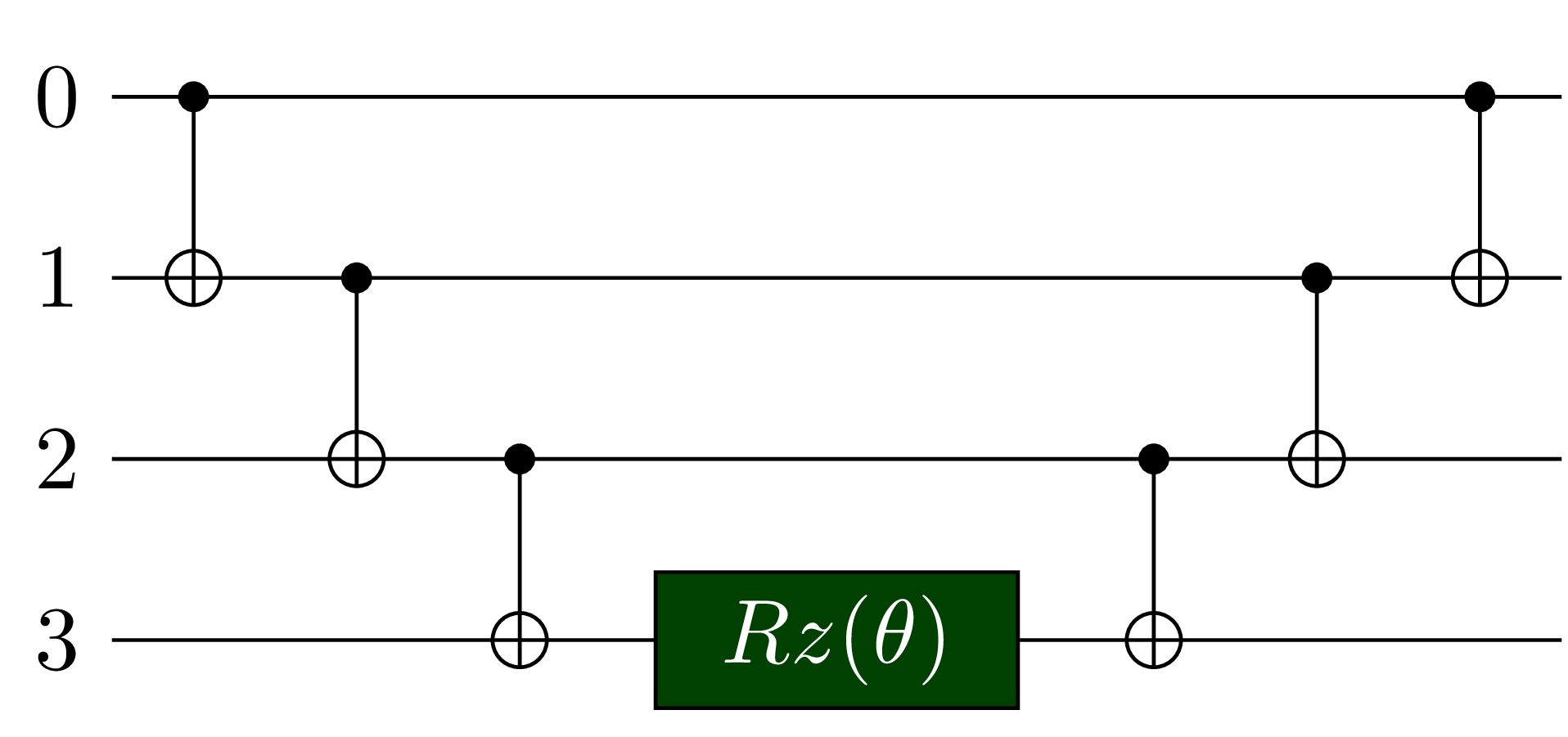} \\
    \midrule
    $e^{- \imag \frac{\theta}{2} \pauli{x}^{(0)} \pauli{z}^{(1)} \pauli{z}^{(2)} \pauli{z}^{(3)}}$ &
    \includegraphics[height=50pt]{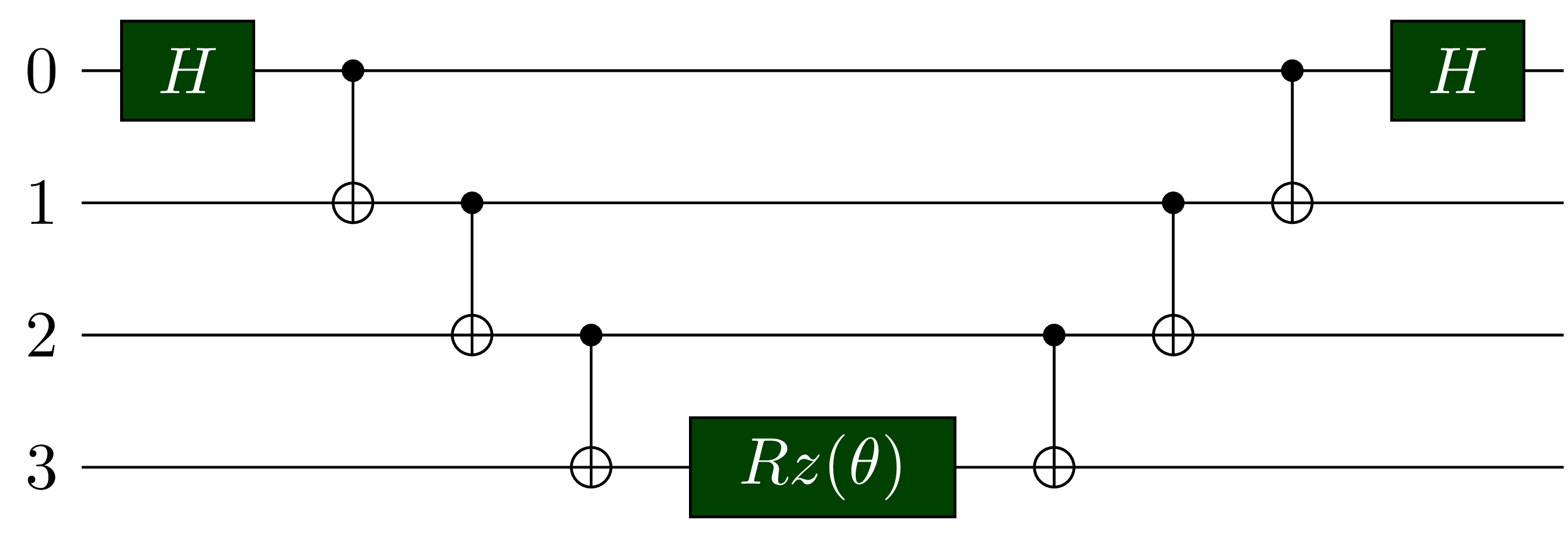} \\
    \midrule
    $e^{- \imag \frac{\theta}{2} \pauli{y}^{(0)} \pauli{z}^{(1)} \pauli{z}^{(2)} \pauli{z}^{(3)}}$ &
    \includegraphics[height=50pt]{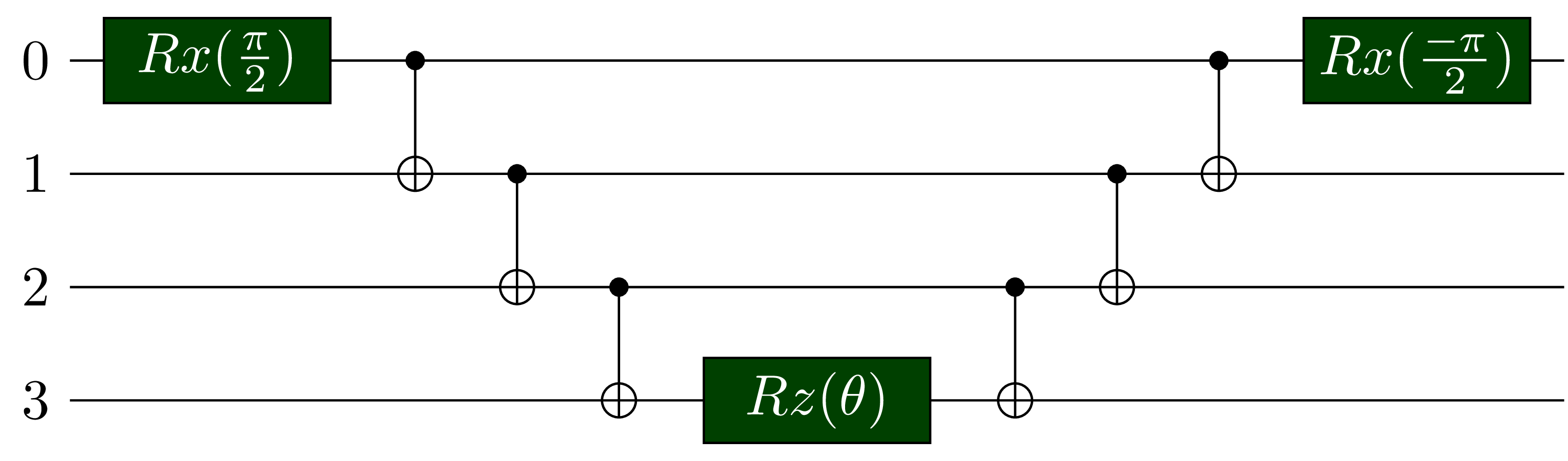} \\
    \midrule
    $e^{- \imag \frac{\theta}{2} \pauli{x}^{(0)} \pauli{x}^{(1)} \pauli{y}^{(2)} \pauli{x}^{(3)}}$ &
    \includegraphics[height=50pt]{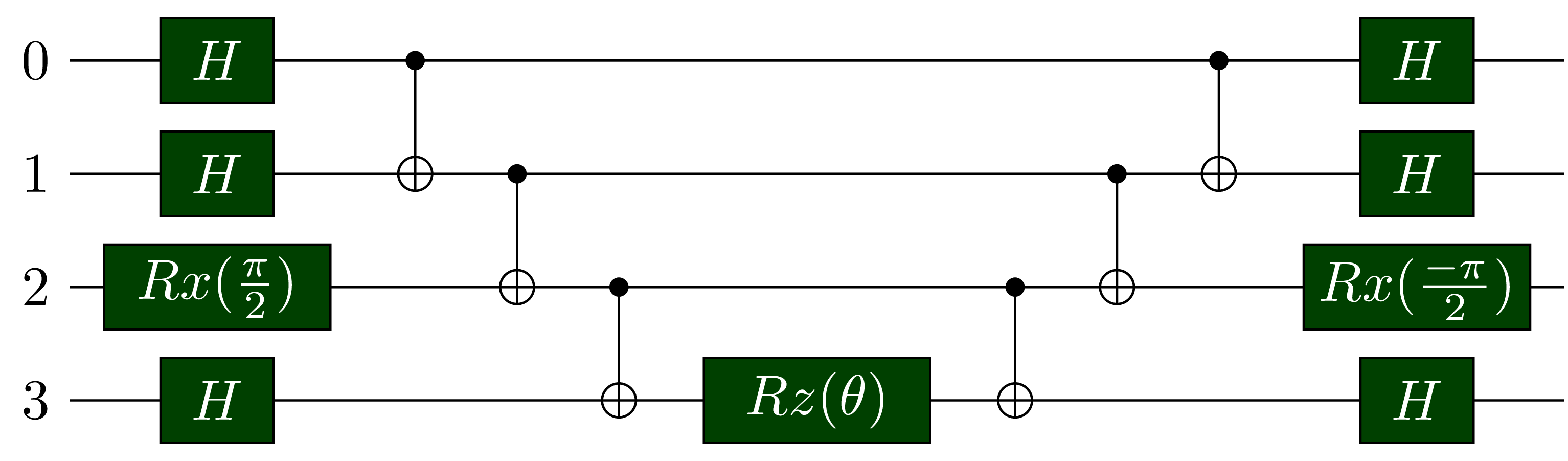} \\
    \midrule
    \end{tabular}
    \caption{The quantum circuits corresponding to the exponential maps of different Pauli strings}
    \label{fig:exp_map}
\end{figure}

For a detailed understanding of the exponential map of second-quantized operators, the reader is redirected to~\cite{whitfield2011simulation,yordanov2020efficient}. Some more exemplary circuits are illustrated in Figure~\ref{fig:exp_map}.

\subsection{Commutativity of subterms in fermionic excitation operators}

As we have seen earlier, applying a qubit mapping such as Jordan-Wigner~(JW) or Bravyi-Kitaev~(BK), a single fermionic excitation operator decomposes into different qubit operators. If the qubit operators comprising a fermionic excitation commute with each other then the exponential function of the fermionic excitation is a product of exponential functions of the qubit operators (individual Pauli strings).
We follow Ref.~\citen{romero2018strategies} to look at the commutativity of the subterms in the excitation operators.

Consider the commutator of two arbitrary Pauli strings of $\pauli{x}$ and $\pauli{y}$ operators, $\operator{P}_A = \otimes_{j=1}^N \pauli{a_j}^{(j)}$ and $\operator{P}_B = \otimes_{j=1}^N \pauli{b_j}^{(j)}$, each of acting on the same set of $N$ qubits:
\begin{align}
    [\operator{P}_A, \operator{P}_B] = \otimes_{j=1}^N ( \pauli{a_j}^{(j)}\pauli{b_j}^{(j)} ) - \otimes_{j=1}^N ( \pauli{b_j}^{(j)} \pauli{a_j}^{(j)} ),
\end{align}
where $\pauli{a_j}, \pauli{b_j} \in \{\pauli{x}, \pauli{y} \}$, and the superscript $j$ represent the qubit on which the Pauli operators act. The products $\pauli{a_j}^{(j)}\pauli{b_j}^{(j)}$ can be simplified as
\begin{equation}\label{eq:product_paulis}
    \pauli{a_j}^{(j)}\pauli{b_j}^{(j)} = 
        \begin{cases}
         1 & \text{if } a_j = b_j \\
         \imag \pauli{z} & \text{if } a_j = x \text{ and } b_j = y \\
         -\imag \pauli{z} & \text{if } a_j = y \text{ and }  b_j = x
        \end{cases}.
\end{equation}
Note that we do not need to consider $\pauli{z}$ operations here as they do not appear in this context as they don't contribute to the commutation relation.
The commutator can then be re-written as
\begin{align}
    [\operator{P}_A, \operator{P}_B] = (-1)^{n_y^A - c_y}\{ 1 - (-1)^{n_y^B -n_y^A} \} (-\imag)^{N - c_x - c_y} \operator{P},
\end{align}
where $n_y^A$ and $n_y^B$ are the number of $\pauli{y}$ operators in strings $A$ and $B$, respectively, $c_x$ is the number of instances when $a_i = b_i = x$,  $c_y$ respectively the number of instances when $a_i = b_i = y$ and $\operator{P}$ is the Pauli string generated by the multiplication. 

For single and double fermionic excitation operators with real amplitudes, it was shown in Ref.~\citen{romero2018strategies}, that the qubit operators in the Jordan-Wigner transformation have both $n_y^A$ and $n_y^B$ odd. Thus, the value of $n_y^B -n_y^A$ is always even, and consequently $[\operator{P}_A, \operator{P}_B] = 0$. This suggests that all the subterms of a fermionic excitation operator commute.

Next, note that if two Pauli strings $\operator{P}_A, \operator{P}_B$ commute, this property is conserved under unitary transformation of the Pauli strings, i.e., 
 $ [\operator{P}_A, \operator{P}_B] = 0 \Rightarrow  [\operator{P}_A^{'}, \operator{P}_B^{'}] = 0$, where $\operator{P}_A^{'} = \uunitary^\dagger \operator{P}_A \uunitary$ and $\operator{P}_B^{'} = \uunitary^\dagger \operator{P}_B \uunitary$. More generally, this property would be preserved for any similarity transformation of the Pauli strings.
Thus, the terms arising from the single and double fermionic excitation operators commute in any qubit-fermion mapping, since there exist unitary Clifford transformations between Pauli strings in various qubit-fermion mappings. 

\subsection{Impact of Trotterisation and operator ordering}\label{subsubsec:order}
The UCC ansatz is typically implemented by decomposition into a set of easier-to-implement modules, carried out by a Trotter decomposition. 
However, individual excitation generators  $\exgen_\mu$, where $\mu$ denotes an arbitrary excitation, generally do not commute. 
Consequently, a \rereviewed{}{truncated} Trotter decomposition of $e^{\sum_\mu \exgen_\mu}$ is not unique. 
The same problem arises, whenever an exponential of non-commuting operators is to be Trotterised, such as in Hamiltonian simulation $e^{-\imag \ham t}$, whenever individual terms of the Hamiltonian do not commute \cite{Tranter2019}.

Reference~\cite{Grimsley2020ordering} investigated the influence of operator ordering in a Trotterised UCCSD ansatz.
This paper shows that especially in the case of static correlation, particular orderings lead to chemically meaningful improvements in energies. 
A simple example of an unfavourable or inefficient ordering is starting with an excitation that does not include occupied reference orbitals.
They numerically demonstrated the importance of ordering by computing the variance in the resulting energy values by randomly shuffling orders of operators then optimising the energies. The authors noted that it is generally favourable to first apply doubles excitations to the reference state before applying the singles excitations to minimise errors and variances in energy under shuffling operators.
They additionally showed that the $k$-UpCCGSD ansatz exhibits lower variance in energy under shuffling of operators with growing $k$, or the number of repeating ansatz layers.
From this observation, the authors suggested the role of excess parameters, introduced by $k$-fold repetition of the circuit with independent parameters (see also Section~\ref{subsubsec:k-up}), in helping lower the energy variance.

Reference~\cite{Evangelista2019} examined the capability of UCC to exactly parametrise an arbitrary fermionic wavefunction, and compare it to classical CC and a disentangled form of UCC. To this end, they derive a condition based on a differential equation for the path from a reference state $\ket{\Phi_\mathrm{ref}}$ to some arbitrary state $\ket{\psi}$, which states that the gradient of the parametrisation along the path must not be orthogonal to the analytical gradient of the path, hence being able to span the full path. Grounded on numerical studies, they argue, that UCC is indeed capable of exactly parametrising an arbitrary state, as long as said condition is not violated. 

The situation is different with disentangled or factorised UCC that appears after the Trotterisation 
\begin{equation}\label{eq:disucc}
    \ket{\psi_\text{dUCC}} = \Big(\prod_{\mu\in \mathcal{A}} e^{(\reviewed{\nicehalf{\theta_\mu}}{\theta_\mu}) \exgen_\mu}\Big) \ket{\phiref}.
\end{equation}
Reference~\cite{Evangelista2019} shows, that disentangled forms of UCC generally do not provide exact parametrisation even if all excitations up to the number of electrons are taken. Also, even if one uses all excitations up to the number of electrons, the disentangled form is still order dependent. Yet, for some model systems there were found certain orders of $\exgen_\mu$ which generated exact results. 
\rereviewed{}{Apart from that, we point out that the wavefunction generated by the procedure outlined in Ref.~\citen{smart2021quantum} in order to solve the anti-Hermitian contracted Schr\"odinger equation forms a similar factorised structure as in~\eqref{eq:disucc}.}

These problems of disentangled form were further investigated in Ref.~\citen{izmaylov2020order}, it was found that the reason for the order-dependence of factorised UCC can be attributed to the absence of commutation closure for a set of used excitations. In other words, commutators of second and higher order excitations do not form a closed Lie algebras. To construct Lie algebras from them, one needs to add other operators that can be generated by commutators $[\exgen_\mu,\exgen_\nu]$. Once these extra operators are included in the 
generator set, the disentangled UCC expression becomes
order independent and can provide the exact parametrisation.~\cite{izmaylov2020order}   
To avoid augmenting the unitary transformation by too many  additional elements, Ref.~\citen{izmaylov2020order} suggests to 
use only symmetry adapted linear combinations of generators that commute with the system symmetries (e.g. electron number and spin operators).

\rereviewed{}{
Beyond the order problem, Trotterisation might also impact the symmetries of the wavefunction associated to the ansatz. For symmetries of the initial wavefunction to be conserved, unitary fragments or their generators need to commute with the respective symmetry operators, i.e., particle number or spin symmetries.
}

\subsection{Compilation strategies}
The exponential map of Pauli strings is implemented on a quantum computer by decomposition into single- and two-qubit gates as shown above in Section~\ref{sec:qubit_perspective}. 
A standard implementation~\cite{whitfield2011simulation} of the exponential map using the Jordan-Wigner transformation leads to an additional factor in the gate count, scaling linearly with the number of qubits $N$. The overhead can be reduced to a logarithmic scaling~\cite{bravyi2002fermionic} using the Bravyi-Kitaev transformation. 
Ref.~\citen{hastings2014improving} presented a new technique, where they use a modified circuit with an ancilla qubit that controls the application of a term to achieve improvements in the gate overhead.
Their technique is based on the fact that the Pauli strings generated from the Jordan-Wigner transformation compute the parity of a given set of qubits. Thus one can reduce the computational cost by removing the redundant parity calculations. 
It was implemented for the UCC ansatz in Ref.~\citen{Kuhn.JCTC.2019} and was shown to achieve significant reduction in the number of two-qubit-gates.

Recently, a technique based on partitioning the Pauli strings into commuting sets and then performing simultaneous diagonalisation of each set was proposed in Ref.~\citen{van_den_Berg_2020} for reducing the gate count in context of Hamiltonian simulation. 
A similar approach to that of Ref.~\citen{van_den_Berg_2020} is proposed in the Appendix which reduces the CNOT gate count by more than 50\% and removes the compiling bottlenecks of partition into commuting Pauli strings or complex compilation algorithms.

More recently, in Ref.~\citen{cowtan2020generic} the authors proposed a compilation strategy for the Trotterised UCC ansatz into one- and two-qubit gates using techniques from  ZX-calculus~\cite{vandewetering2020zxcalculus}. They use approximate solutions to the graph colouring problem to partition the Pauli strings into commuting sets. Then, they fix the commuting terms in a greedy approach and proceed using techniques from ZX-calculus to simplify the unitaries obtained by the greedy approach to  lower the computational cost. 

Other approaches~\cite{yordanov2020efficient, Zhang_2021, wang2021resourceoptimized, motta2021low} have been proposed as well to reduce the resource requirements for implementing a UCC ansatz on near-term quantum devices. 
These include the construction of efficient fermionic excitation operators~\cite{yordanov2020efficient}, graph clustering algorithm~\cite{Zhang_2021} and generalised fermionic to qubit transformations~\cite{wang2021resourceoptimized}.
The above mentioned compilation techniques represent a continued effort 
to reduce the quantum resources necessary to implement certain quantum circuit 
-- and here, we are in particular interested in UCC ans\"atze -- such that eventually,  
implementations on physical quantum devices allow to access classically intractable regimes.
\rereviewed{}{Even though a considerable amount of work has been done in making the compilation of single and double excitation operators more efficient, there is still need for improved compilation techniques when higher-order excitations are included in the ansatz as their respective circuits tend to grow significantly larger.}

\subsection{Example}

We now show an illustrative example of how to construct a circuit corresponding to the single and double fermionic excitation operator $\uunitary_{\mathbf{p}\mathbf{q}}\left(\theta\right) = e^{\thetahalf \exgen_{\mathbf{p}\mathbf{q}}}$ with $(\mathbf{p}, \mathbf{q})  = ([0], [1])$ and $([0,2], [1,3])$. Additionally, we look at the full UCC operator for the Hydrogen molecule in a minimal basis. The examples are realised in the software package \textsc{Tequila}~\cite{kottmann2021tequila}. 

\begin{figure}[htbp]
\centering
     \begin{subfigure}[b]{\columnwidth}
         \centering
         \includegraphics[width=0.225\columnwidth]{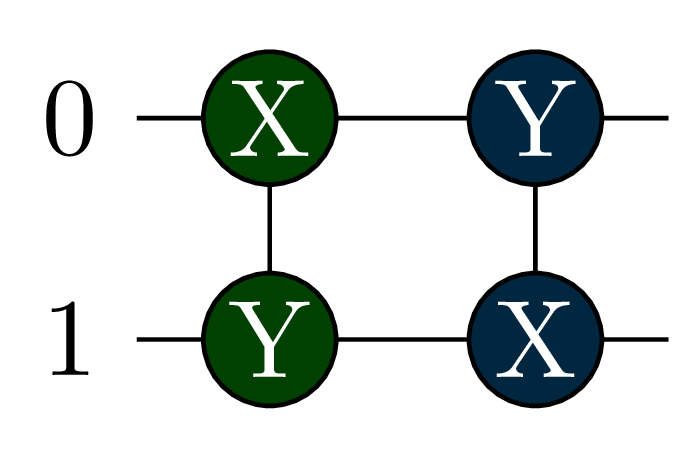}
         \caption{Compact representation}
     \end{subfigure}
     \hfill
     \begin{subfigure}[b]{\columnwidth}
         \centering
         \includegraphics[width=0.95\columnwidth]{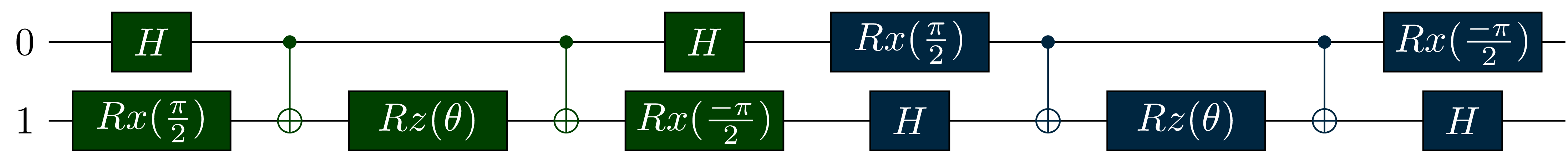}
         \caption{Full representation}
     \end{subfigure}
\caption{Circuit for the unitary operator $\uunitary_{\mathbf{p}\mathbf{q}}\left(\theta\right) = e^{\thetahalf \exgen_{\mathbf{p}\mathbf{q}}}$, where $\mathbf{p} = [0]$ and $\mathbf{q} = [1]$.}
\label{fig:ucc_ex_gate}
\end{figure}

We first show how to initialize a hydrogen molecule of a given geometry and then construct the circuit corresponding to an arbitrary single excitation operator below. The resulting circuit is shown in Figure~\ref{fig:ucc_ex_gate}. 
\begin{lstlisting}[language=Python]
import tequila as tq

# initialize molecular data
geom = "H 0.0 0.0 0.0\nH 0.0 0.0 0.7"
mol = tq.chemistry.Molecule(geometry=geom, basis_set="sto-3g", transformation="jordan_wigner")

# construct the unitary for the single excitation fermionic operator
idx = [(0),(1)]
excitation_gate = mol.make_excitation_gate(idx, angle="s")
tq.draw(excitation_gate)



\end{lstlisting}

\begin{figure}[]
\centering
    \begin{subfigure}[b]{\columnwidth}
    \centering
    \includegraphics[width=0.65\columnwidth]{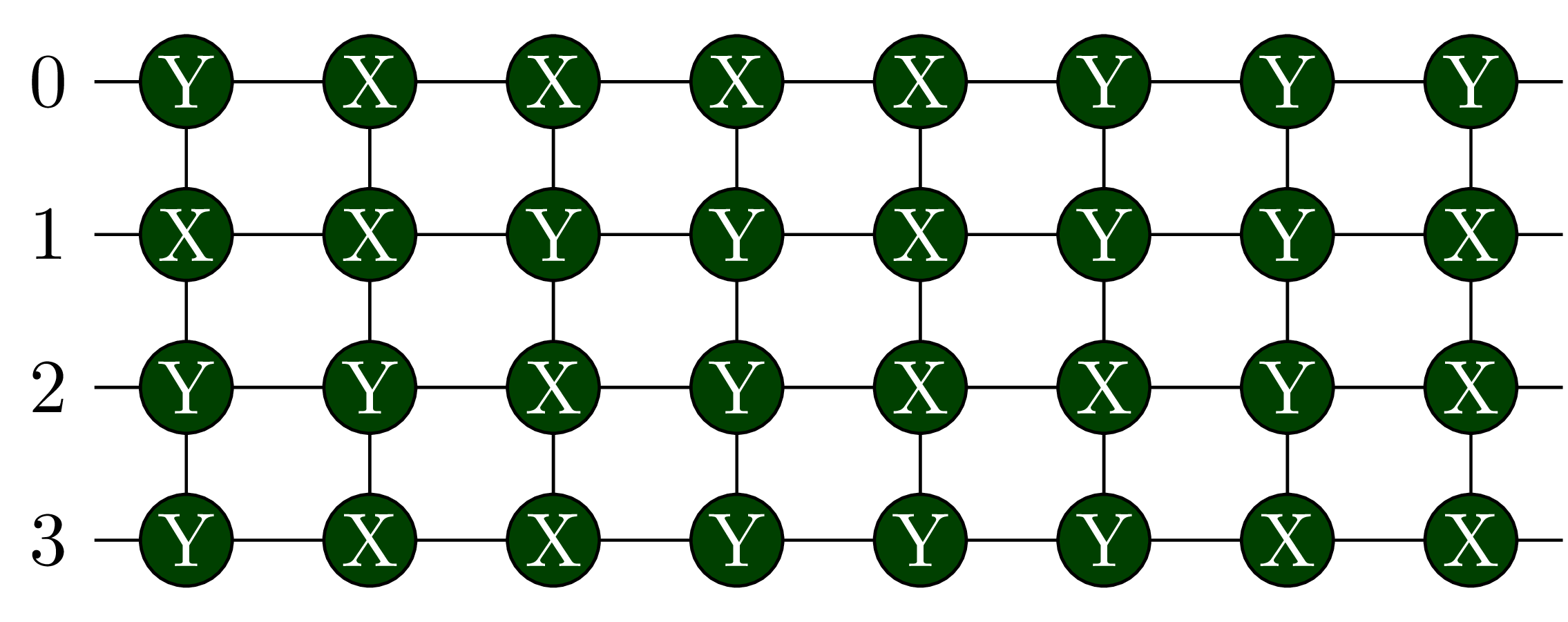}
    \caption{Jordan-Wigner encoded quantum circuit for the unitary operator $\uunitary_{\mathbf{p}\mathbf{q}}\left(\theta\right) = e^{\thetahalf \exgen_{\mathbf{p}\mathbf{q}}}$, where $\mathbf{p} = [0,2]$ and $\mathbf{q} = [1,3]$. }
    \label{fig:ucc_d_exc}
    \end{subfigure}
    \begin{subfigure}[b]{\columnwidth}
    \centering
    \includegraphics[width=0.7\columnwidth]{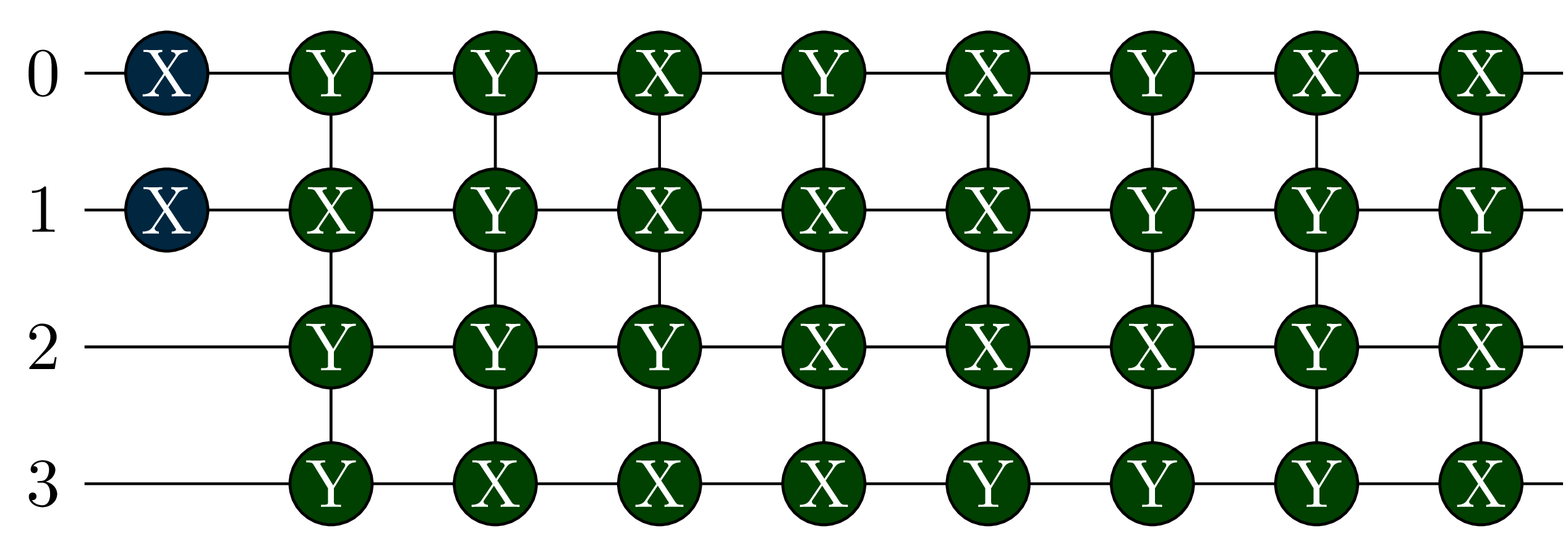}
    \caption{Jordan-Wigner encoded quantum circuit for the unitary coupled cluster operator for the H$_2$ molecule in the minimal basis. }
    \label{fig:ucc_h2}
    \end{subfigure}
\caption{Circuits for the excitation gate and the full unitary coupled cluster operator for the H$_2$ molecule in the minimal basis. The circuits shown uses a compressed representation, where each gate in the circuit represents the circuit for the exponential map of the Pauli term as discussed in Figure~\ref{fig:exp_map}. For example the ``$\qpx \qpy \qpy \qpy$'' gate corresponds to $e^{-\imag \thetahalf \pauli{x}(0)\pauli{y}(1)\pauli{y}(2)\pauli{y}(3)}$ }
\label{fig:h2}
\end{figure}

We next show how to construct the circuit corresponding to an arbitrary double excitation operator and the full UCC ansatz. The resulting circuits are shown in Figure~\ref{fig:ucc_d_exc} and Figure~\ref{fig:ucc_h2} respectively.
\begin{lstlisting}[language=Python]
# construct the unitary for the double excitation fermionic operator
idx = [(0,2),(1,3)]
excitation_gate = mol.make_excitation_gate(idx, angle="d")
tq.draw(excitation_gate)
\end{lstlisting}
\begin{lstlisting}[language=Python]
# create the full UCC circuit
ucc_circuit = mol.make_uccsd_ansatz(trotter_steps = 1)
tq.draw(ucc_circuit)
\end{lstlisting}

We also list all the other software packages in Table~\ref{tab:code} where UCC and UCC-based methods have been implemented.
\begin{table*}[]
    \centering
    \begin{tabular}{ccc}
        \toprule
        Method & Package & Implementation \\
        \midrule
        \multirow{7}{*}{UCCSD} 
        &\textsc{openfermion} & \href{https://github.com/quantumlib/OpenFermion}{github.com/quantumlib/OpenFermion}~\cite{McClean2020openfermion}\\
        &\textsc{tequila}& \href{https://github.com/aspuru-guzik-group/tequila}{github.com/aspuru-guzik-group/tequila}~\cite{kottmann2021tequila}\\
        &\textsc{pennylane} &
        \href{https://github.com/PennyLaneAI/pennylane}{github.com/PennyLaneAI/pennylane}~\cite{bergholm2020pennylane, arrazola2021universal}\\
        &\textsc{qdk} & \href{https://github.com/microsoft/quantum}{github.com/microsoft/quantum}\\
        &\textsc{qforte }& \href{https://github.com/evangelistalab/qforte}{github.com/evangelistalab/qforte}~\cite{stair2021qforte} \\
        &\textsc{qiskit} & \href{https://github.com/QISKit/}{github.com/qiskit/}~\cite{Qiskit} \\
        &\textsc{xacc} & \href{https://github.com/eclipse/xacc}{github.com/eclipse/xacc}~\cite{xacc2020} \\
        \midrule
        \multirow{6}{*}{ADAPT}
        & \textsc{adapt-vqe} & \href{https://github.com/mayhallgroup/adapt-vqe}{github.com/mayhallgroup/adapt-vqe}~\cite{grimsley2019adaptive} \\
        & \textsc{tequila} & \href{https://github.com/aspuru-guzik-group/tequila}{github.com/aspuru-guzik-group/tequila}~\cite{kottmann2021feasible, kottmann2021tequila}\\
        &\textsc{pennylane} &
        \href{https://github.com/PennyLaneAI/pennylane}{PennyLaneAI/pennylane}~\cite{bergholm2020pennylane, delgado2021variational}\\
        &\textsc{qebab} & \href{https://github.com/hanschanhs/QEBAB}{github.com/hanschanhs/qebab}~\cite{chan2021molecular} \\
        &\textsc{qforte }& \href{https://github.com/evangelistalab/qforte}{github.com/evangelistalab/qforte}~\cite{stair2021simulating, stair2021qforte} \\
        &\textsc{qiskit} & \href{https://github.com/QISKit/}{github.com/qiskit/}~\cite{Qiskit} \\
        \midrule
        $k$-UpCCGSD & \textsc{tequila}& \href{https://github.com/aspuru-guzik-group/tequila}{github.com/aspuru-guzik-group/tequila}~\cite{kottmann2021tequila}\\
        \midrule
        PNO-UpCCGSD &\textsc{tequila}& \href{https://github.com/aspuru-guzik-group/tequila}{github.com/aspuru-guzik-group/tequila}~\cite{kottmann2021reducing, kottmann2021tequila}\\
        SPA &\textsc{tequila}& \href{https://github.com/aspuru-guzik-group/tequila}{github.com/aspuru-guzik-group/tequila}~\cite{kottmann2021optimized,kottmann2021tequila}\\
        \midrule
        \multirow{2}{*}{QPE}
        & \textsc{qdk} & \href{https://github.com/microsoft/quantum}{github.com/microsoft/quantum}\\
        &\textsc{qforte }& \href{https://github.com/evangelistalab/qforte}{github.com/evangelistalab/qforte}~\cite{stair2020multireference, stair2021qforte} \\
        \bottomrule
    \end{tabular}
    \caption{Available implementations of individual approaches discussed in this work. See~\cite{bharti2021noisy} for a list of general quantum simulation packages.}
    \label{tab:code}
\end{table*}


\section{UCC-based ans\"atze on quantum computers}\label{sec:ucc-flavors}
In this section, we will review parametrisations building upon UCC that have been introduced until now with the purpose of generating parametrised quantum circuits for quantum algorithms, in particular for the near-term. 
We will consider different UCC-based parametrisations, including adaptive ansatz generation, ans\"atze based on pair approximation, downfolded ans\"atze, and other approximations and ans\"atze related to UCC.
In addition to that, we briefly mention non-UCC based parametrised quantum circuits that can be used for quantum chemistry on quantum computers. 
 
\subsection{Adaptive generation of a UCC-type ansatz}
This section is designated to the adaptive generation of UCC-based parametrised quantum circuits, which is carried out by successively augmenting an initial ansatz with operators from a given pool (e.g., singles and doubles excitations) based on a specific criterion. This way, one builds a factorised ansatz without the need to resort to a Trotter-Suzuki approximation. The adaptive nature typically is able to somewhat circumvent the order problem as well, as the factorised form of individual exponentiated excitations is not assembled in some arbitrary, fixed order but the next operation is added ``by importance'', following a certain metric such as the norm of the gradient of the energy with respect to the excitations. As showcased on a few examples in Ref.~\citen{grimsley2019adaptive}, their adaptive approach outperforms certain fixed ordering schemes. We present the details of most common adaptive techniques in what follows. 
\rereviewed{}{At this point we point out that one could also resort to classical quantum chemistry calculations to obtain an educated guess for the ordering that may be faster obtained than by an adaptive procedure, e.g. throughout a previous MP2 or CC(SD) computation, such as also considered in Refs.~\citen{mcclean2016theory,romero2018strategies,grimsley2019adaptive}.} 

\subsubsection{Qubit Coupled Cluster}\label{subsubsec:qcc}
The Qubit Coupled Cluster (QCC) method~\cite{Ryabinkin.JCTC.2018,Ryabinkin.JCTC.2019} was proposed as a cost-effective way to implement the coupled cluster ansatz directly in the qubit space. 
It uses unitary transformations with generators based on Pauli strings to add correlation effects into a mean-field wavefunction and has been shown to yield accurate energies with shallow quantum circuits when compared to UCCSD~\cite{Ryabinkin.JCTC.2018,Ryabinkin.JCTC.2019}.

\newcommand{\mfwfn}{\ensuremath{\ket{\Omega}}}
\newcommand{\mfwfnInd}[1]{\ensuremath{\ket{\Omega_{#1}}}}
The mean-field wavefunction $\mfwfn$ is written as a product of single-qubit coherent states as 
\begin{align}
    \mfwfn = \prod_{i=1}^N \mfwfnInd{i},
\end{align}
where $N$ is the number of qubits and $\mfwfnInd{i} = \cos \left( \tau_i /2 \right) \ket{0} + \exp (\imag \phi_i) \sin \left( \tau_i /2 \right) \ket{1}$ is the single qubit coherent state, with $\phi_i$ and $\tau_i$ as the azimuthal and polar angles on the Bloch sphere for qubit $i$, respectively, and $\ket{0}$ and $\ket{1}$ are the eigenstates of the $\operator{s}_z(i) = \pauli{z}^{(i)}/2$ operator.
The mean-field can be set to a Hartree-Fock determinant when $\tau_i \in \{0,\pi\}$ and $\phi_i = 0$. 

The correlation effects are added to the mean-field wavefunction using Pauli strings as multi-qubit rotations~\cite{Ryabinkin.JCTC.2018,Ryabinkin.JCTC.2019,Ryabinkin.2020.JCTC} of the form
\begin{align}
\uunitary(\theta) = \prod_{i=1}^{N_{ent}} \exp \left( -\frac{\imag \theta_i \operator{P}_i}{2} \right) ,
\end{align}
with $N_{ent}$ the number of entangling gates, $\theta = (\theta_1, \theta_2, \ldots, \theta_{N_{ent}})^T$ a vector of real parameters and $\operator{P}_i$ Pauli strings. The resulting QCC wavefunction $\ket{\Psi(\theta,\Omega)}$, after the application of $\uunitary(\theta)$ to the mean-field, has the form
\begin{align}
    \ket{\Psi(\theta,\Omega)} = \uunitary(\theta) \mfwfn.
\end{align}
The expectation value $ E = \braket{\Omega | \uunitary(\theta)^\dagger \ham \uunitary(\theta) | \Omega} $ can then be minimised to obtain the optimal parameters $\theta^\star$ and $\Omega^\star$, which provide the approximate ground state and energy.

The total number of possible Pauli strings to choose from to construct the QCC ansatz is $4^N - 3N -1$, with $N$ the number of qubits. 
The exponential growth of the number of Pauli strings poses a problem to the scalability of the QCC approach, so the authors proposed a screening procedure that can be executed efficiently on a classical computer.
The selection of Pauli strings is based on the energy gradients with respect to the corresponding amplitudes~\cite{Ryabinkin.JCTC.2018},
\begin{align} \label{Eq.QCC.DE1}
    \frac{dE}{d\theta_k}\bigg\vert_{\theta=0} = \imag \braket{ \Omega | [\ham,\operator{P}_k] |\Omega},
\end{align}
which can be evaluated on a classical computer. 

In addition, when $\mfwfn$ is a Hartree-Fock state and the Jordan-Wigner transformation is used, the set of Pauli strings with non-zero gradients is readily known.~\cite{Ryabinkin.JCTC.2018,Ryabinkin.JCTC.2019} Construction of these Pauli strings requires at least two $\pauli{x}$ operators in occupied orbitals and in virtual orbitals, then an odd number of $\pauli{x}$ are to be switched to $\pauli{y}$ and the remaining qubits (orbitals) can either experience an identity or $\pauli{z}$-operation.
For example, in the case of two occupied and two
virtual orbitals, eight contributing Pauli strings are obtained, as shown in in Figure~\ref{fig:h2}.
These Pauli strings have the same absolute energy gradient and can be obtained from the JW transformation of the UCC doubles excitation operator. 
Therefore, initial gradient screening of a Hartree-Fock state scaling is $\order{N_\text{occ}^2 N_\text{virt}^2}$,
where $N_\text{occ}$ is the number of occupied orbitals and $N_\text{virt}$ is the number of virtual orbitals due to double excitations. 
For a Hartree-Fock state only double
excitations have non-zero gradients; single excitations have zero gradient due to the Brillouin theorem and higher level excitations due to the Slater-Condon rules~\cite{szabo1982modern}.
However, inclusion of Pauli strings pertaining to zero-gradient-excitations will improve the energy when combined with the entanglers obtained from the initial screening.
This approach can be more efficient 
than UCC by using one entangler per excitation, but can break particle number and total spin symmetries when dealing with highly correlated systems.

The QCC method has been successfully applied to calculate energies of different molecules of interest~\cite{Ryabinkin.JCTC.2018,Ryabinkin.JCTC.2019}. For these molecules, high levels of accuracy have been obtained with the inclusion of a small number of entanglers.
For the H$_2$ molecule in a minimal basis where two qubits where tapered off (compare to \cite{bravyi2017tapering}), QCC with a single entangler can
reproduce the Full CI energy~\cite{Ryabinkin.JCTC.2018}. Furthermore, more complicated
systems like the symmetric stretching of the water molecule (6-31G Pople basis set and
tapering off two qubits) can be simulated with
eight entanglers. Considering the amount of two-qubit gates in the ansatz, it was found that QCC is more efficient than UCCSD at slightly lower accuracy.

Next, we discuss extensions to the QCC method that have been proposed to further overcome some of the challenges while making the procedure more efficient.

\paragraph{Iterative Qubit Coupled Cluster Method}
The main challenge for the QCC method is finding a compact set of generators that provide fast and systematic convergence to chemical
accuracy. This is because of the fact that fixed-rank operators pertaining to fermionic excitations span a polynomial-large space ($O(N^4)$ for doubles) while random sampling might get the algorithm trapped in a barren plateau. 
To overcome this challenge, the iterative qubit coupled cluster method (iQCC)\cite{Ryabinkin.2020.JCTC} was proposed, where the concept of the direct interaction set (DIS) was introduced. It is a set of all possible Pauli strings that guarantee lowering the energy, and is constructed by calculating first order derivatives using Eq. (\ref{Eq.QCC.DE1}).

Then, instead of a single-step construction of $\uunitary(\theta)$, a smaller $\uunitary^{(j)}(\theta^j)$ is constructed using the current DIS to include the operators with larger gradients. Next, its parameters are optimized and the generated $\uunitary^{(j)}(\theta^j)$ is used to transform the current Hamiltonian $\ham^{(j)}$ at iteration $j$:
\begin{align}
    \ham^{(j+1)} = \left( \uunitary^{(j)}(\theta^j) \right)^{\dagger} \ham^{(j)} \uunitary^{(j)}(\theta^j).
\end{align}
It has been demonstrated that using such an iterative approach one can converge to more accurate ground-state energies using shallow quantum circuits at the expense of additional quantum measurements of the intermediate $\hat{H}^{(j)}$. However, the iQCC method does have an exponential growth of intermediate $\hat{H}^{(j)}$ upon dressing. For this reason, the number of iterations and problem size is limited by current classical computers size and capabilities. 
To overcome this exponential growth, one can use techniques described in Ref.~\citen{lang2021unitary} the unitary transformations dressing the Hamiltonian are constructed from involutory linear combinations, meaning their square equals the identity. This way, the number of terms in the Hamiltonian grows only polynomially.

\paragraph{Qubit Coupled Cluster with Singles and Doubles}

An alternative to constructing the ansatz using unitary operations with Pauli strings as generators, is to use particle preserving exchange gates~\cite{xia2020qubit}.
Previously, these exchange gates have been called qubit excitations~\cite{yordanov2020efficient}, particle-exchange gates~\cite{izmaylov2020order}, ``A'' gates~\cite{gard2020efficient} or orbital-rotations~\cite{anselmetti2021local}.
For example, a single particle conservation exchange gates (or Qubit excitation) is defined as:
\begin{align}
\uunitary_{ex,i}^{a}(\theta) &= e^{\theta\left(\pauli{+}^{(a)}\pauli{-}^{(i)} - \pauli{+}^{(i)}\pauli{-}^{(i)}\right)} \\ \nonumber
&= e^{-\imag \theta\left(\pauli{x}^{(a)} \pauli{y}^{(i)} - \pauli{y}^{(a)} \pauli{x}^{(i)}\right)} \\ \nonumber
&= \left[ \begin{matrix} 
1 & 0 & 0 & 0 \\
0 & \cos \theta &-\sin \theta & 0 \\
0 & \sin \theta & \cos \theta & 0 \\
0 & 0 & 0 & 1 \\
\end{matrix} \right],
\end{align}
where $\theta$ is the rotation parameter and the qubit $i$ is partially excited to the qubit $a$. 
Additionally, this is equivalent to a UCC single excitation/de-excitation in the Jordan-Wigner encoding without the inclusion of the parity operator and can be extended to higher order excitations by replacing fermionic creation and annihilation operators with $\pauli{\pm} = \frac{1}{2}\left(\pauli{x} \pm \imag \pauli{y}\right)$ in the same manner.
For the double qubit-excitations, another approximation can be found through a particle conserving combination of four Pauli strings.~\cite{izmaylov2020order}

The gate operations can be reduced from an upper bound of $O(N^5)$ of UCCSD to $O(N^4)$ (for a JW transformation)~\cite{xia2020qubit} by using these particle preserving exchange gates. 
It was shown numerically in Ref.~\citen{xia2020qubit} that using this method, one can achieve similar accuracy as UCCSD for different systems (BeH$_2$, H$_2$O, N$_2$, H$_4$ and H$_6$), indicating that removal of parity terms has little effect on energy calculations for small molecular systems.

\subsubsection{Adaptive Derivative-Assembled Pseudo-Trotter VQE (ADAPT-VQE)}\label{subsubsec:adapt}
The ADAPT-VQE proposed in Ref.~\citen{grimsley2019adaptive} comprises an algorithm based on VQE that adaptively grows a UCC-type ansatz from a predefined set of cluster operators. 
This procedure is also interesting when looking at the order problem of Trotterised UCC, since it does not split up the exponential of a sum of operators but resembles a form of disentangled UCC as in Eq.~\eqref{eq:disucc} \cite{Evangelista2019,izmaylov2020order}, adaptively adding new operators. 
Reference~\cite{Grimsley2020ordering} numerically investigates the robustness against operator ordering for adaptive ansatz construction schemes and notes that use of the ``sequential gradient ordering'' scheme, which is a variant of ADAPT-VQE that restricts the operator pool to single instances of each operator (i.e. sampling without replacement), consistently led to low energies. 
In contrast to that, random operator ordering may lead to chemically inaccurate energies, even after ansatz optimisation. The same dependency of the ordering of excitation operators was found in Ref.~\citen{grimsley2019adaptive}.

Motivated by the fact, that a FCI-wavefunction can be written as a potentially infinitely long product of only one- and two-body terms \cite{Evangelista2019}
\begin{equation}\label{eq:fcibyinfiniteoneandtwo}
 	\ket{\psi_\mathrm{FCI}} = \prod_{k=1}^\infty \prod_{pq}\left( e^{\theta_{pq}(k)\exgen_{p,q}}\prod_{rs} e^{\theta_{pqrs}(k)\exgen_{pq,rs}}\right) \ket{\psi_\mathrm{HF}},  
\end{equation}  
the algorithm grows an ansatz from iteration to iteration by operators taken from a certain operator pool $\{\exgen_m\}$. The operator pool typically consists of one- and two-body generalised excitation generators, $\{\exgen_m\}  = \{\exgen_{p,q}, \exgen_{pq,rs}\}$.
In Ref.~\citen{Evangelista2019}, it was proposed to use the non-generalised forms $\{\exgen_m\}=\{\exgen_{i,a}, \exgen_{ij,ab}\}$ only, given that a generalised cluster operators do not yield an exact parametrisation. 
\rereviewed{Additionally, it is possible to augment the pool by higher excitations, such as three-, four-, and higher- body terms, which is expected to be beneficial in particular by strongly correlated systems.}{Even though Eq.~\eqref{eq:fcibyinfiniteoneandtwo} is formally exact, it can be challenging to apply in practice due to potentially infinite circuit depths or slow convergence in parameter optimization. This is since the number of parameters decomposition in Eq.~\eqref{eq:fcibyinfiniteoneandtwo} grows infinitely large for $k\to\infty$, different than a Trotter decomposition which leaves the number of parameters constant for increasing Trotter number or order of decomposition. In particular for strongly correlated systems, augmentation by higher-order excitations is thus expected to be beneficial}. Associated to each operator, one defines a set of parameters $\{\theta_m\}$, $\theta_m\in\mathbb{R}$. 
The adaptive parametrisation is  assembled as follows: Given a reference $\ket{\psi^{(0)}}$, which should inherit the correct particle number (e.g. the Hartree-Fock state), the ansatz is extended by one operator from the pool per iteration. 
The current parametrisation in the $k$-th iteration is then
\begin{align}\label{eq:adapt-wfn}
	\ket{\psi^{(k)}} = e^{\theta_k\exgen_k} \ket{\psi^{(k-1)}} 
	=  e^{\theta_{k}\exgen_{k}} e^{\theta_{k-1}\exgen_{k-1}} \cdots e^{\theta_1\exgen_1}    \ket{\psi^{(0)}}, \\ \exgen_1,\ldots,\exgen_k \in \big\{\exgen_m\big\}. \nonumber
\end{align}
The operator to be added in iteration $k$ needs to be chosen by a certain criterion; Ref.~\citen{grimsley2019adaptive} uses the absolute value of the gradient of the energy with respect to the parameter associated to a certain operator from the pool at iteration $k$, i.e. $\max_{\theta_m\in\{\theta_m\}} \lvert \pdv{E^{(k)}}{\theta_m} \rvert $. Derivatives with respect to the parameters are obtained $ \pdv{E^{(k)}}{\theta_m}  \braket{\psi^{(k)} | [\ham,\exgen_m] | \psi^{(k)}     }$. Other screening criteria may be considered as well, and one may or may not allow re-utilisation of operators.
As a result of $K$ iterations of this procedure, one obtains an \emph{ordered} set of excitations $\mathcal{A}_K= \{\mu^{(1)}, \mu^{(2)}, \ldots \mu^{(K)}  \} $ with associated optimal parameters.

To summarise, ADAPT-VQE constitutes a UCC-type parametrisation, which in contrast to naive UCCSD aims to adaptively build a parametrisation that (albeit with potentially infinitely many operators) is able to make up a FCI wavefunction. Given that the ansatz is constructed in a way that the ``next important'' operator from a given pool is used, the order problem coming from the usual Trotterisation of a UCC ansatz can be bypassed or at least attenuated. In practice numerical evidence was given that the ADAPT ansatz is able to provide higher accuracy than naive UCCSD at shallower circuits. 

To reduce the number of gates and circuit depth required for ADAPT-VQE, Ref.~\citen{tang2019qubitadapt} introduces qubit-ADAPT-VQE. While this is not to be mistaken for QCC introduced earlier, the basic idea is similar, namely to primarily use easy-to-implement qubit operators (here the Pauli matrices) for the operator pool instead of fermionic operators. While fermionic operators can be seen as ``parameter-efficient'', because for every excitation there is exactly one parameter, their implementation requires a large amount of CNOT gates, cf. section~\ref{sec:qubit_perspective}. On the other hand, Ref.~\citen{tang2019qubitadapt} suggest to use an operator pool, such that the individual operators are comprised of individual Pauli strings after encoding the fermionic operators, while keeping only Pauli strings with an odd number of $Y$-operators (ensuring real, antisymmetric wavefunctions). Additionally, it was found that chains of $\quoperator{Z}$-operations in the JW encoding can be dropped. This way, qubit-ADAPT-VQE allows for significantly lower CNOT-counts as fermionic operator pools in ADAPT-VQE (``gate-efficient'' approach). However, using a qubit operator pool also enlarges the size of the operator pool and thus the number of parameters; to improve upon this, Ref.~\citen{tang2019qubitadapt} discusses strategies to reduce the size of the operator pool.

\subsubsection{Projective Quantum Eigensolver (PQE)}\label{subsubsec:pqe}

So far, the adaptive UCC-type parametrization have been set out in a way to be deployed within VQE, in accordance to the variational procedure in form of the minimization of the Rayleigh-Ritz functional, cf. section~\ref{sec:variational_approaches}.
In contrast to that, and in close resemblance to common projective implementations of classical coupled cluster as described in subsection~\ref{subsec:projectiveclassical}, Ref.~\citen{stair2021simulating} propose a projective eigensolver. They have further developed a selected projective quantum eigensolver (SPQE), that can be seen as a projective equivalent to the variational ADAPT-VQE~\cite{grimsley2019adaptive}. 

The formal procedure goes as follows: One seeks a similarity transformed Hamiltonian $\ham'(\theta) = \uunitary^\dagger(\theta) \ham \uunitary(\theta) $ such that an approximation to the ground-state energy is given by $E^\text{PQE} (\theta) = \braket{\Phi_\text{ref} | \ham'(\theta)| \Phi_\text{ref}}$ for some reference $\ket{\Phi_\text{ref}}$. To this end, one needs to construct a unitary transformation $\uunitary(\theta) = e^{\theta_1 \exgen_1} \cdots e^{\theta_n \exgen_n}$ given an operator pool $\{\exgen_m\}$. This unitary circuit gives rise to an \emph{ansatz} $\mathcal{A}$, consisting of an ordered set of excitations.

Then, an approximation to the ground-state is achieved by solving the residual equations $r_m \to 0$  where 
\begin{equation}\label{eq:pqe-residuals}
    r_m = \braket{\Phi_m | \uunitary(\theta^{(n)}) ^{\dagger} \ham \uunitary(\theta^{(n)}) | \Phi_\text{ref} } \quad \forall m\in\mathcal{A} 
\end{equation} 
This can be done by a simple fixed-point iteration (for details see \cite{stair2021simulating}). 

Reference~\cite{stair2021simulating} proposes an efficient method to measure the residuals $r_m$, avoiding the Hadamard-test, that comes at a cost of two energy expectation values, surpassing most analytic gradient evaluation strategies in a VQE context except exploiting the shift-rule~\cite{kottmann2021feasible}. 

Within one (macro-)iteration, one solves the residual equations within a set of micro-iterations to obtain $E^{\text{PQE}}$.
If the procedure is to be carried out non-adaptively, there is only one macro-iteration and a fixed ansatz.
For an adaptive / selective method, one starts with a certain ansatz (e.g. the identity operation) and selects the ``next most-important'' excitation from $\{\exgen_m\}$ in each macro-iteration to be added to the ansatz, while in the micro-iterations one evaluates the energy and thus determines the optimal parameters by solving~\eqref{eq:pqe-residuals}.
Operators to be added in the subsequent iterations are chosen based on the magnitude of the associated residual $r_m$.
To do this efficiently, \cite{stair2021simulating} suggest a measurement scheme using approximate residual states attempting to resemble \eqref{eq:pqe-residuals} by replacing the Hamiltonian with a truncated time evolution on a quantum computer. 
Note that the formulation in \cite{stair2021simulating} is restricted to excitations that yield orthogonal states built from $\ket{\phiref}$.

As previously mentioned, instead of building a parametrized wavefunction at macro-iteration $n$, the (S)PQE aims to grow a parametrized, similarity transformed Hamiltonian
\begin{equation}
    \ham' \gets e^{-\theta_k \exgen_k} \cdots e^{-\theta_2 \exgen_2} e^{-\theta_1 \exgen_1} \ham  e^{\theta_1 \exgen_1} \cdots e^{\theta_2 \exgen_2} \cdots e^{\theta_k \exgen_k}    
\end{equation}
Formally, this is equivalent to  a parametrised wavefunction with fixed $\ham$ such as in ADAPT-VQE and in fact, this is implemented in form of a unitary circuit using standard, electronic Hamiltonian $H$ to obtain expectation values for the energy. Yet as compared to ADAPT-VQE as in~\eqref{eq:adapt-wfn},  the SPQE-ansatz is built in reverse order. This choice was found to be the superior choice in the projective approach. 

Simulation results in~\cite{stair2021simulating} indicate that while SPQE yields results of similar accuracy as ADAPT-VQE, the measurement cost of SPQE in comparison to VQE-based approaches is significantly less. Although the number of residual vs. gradient evaluations are comparable, the cost to obtain a residual can be significantly less than a the cost of a gradient measurement. Despite being non-variational, SPQE may take up a similar place in the landscape of near-term quantum algorithms as projective CC in classical quantum chemistry.

\subsection{Pair-UCC based ans\"atze}
In this section, we look at ans\"atze based on pair-approximations, which allow to portray more compact wavefunctions. This is helpful especially in the near term, where the number of available qubits is not abundant. 
\subsubsection{$k$-UpCCGSD}\label{subsubsec:k-up}
Although a non-adaptive, fixed parametrisation, the $k$-UpCCGSD ansatz by Ref.~\citen{Lee2019} describes rather compact wavefunctions leading to lower gate counts and circuit depth than naive UCCSD. 
A pair-UCCGD ansatz is given by excitation operators 
$ \exgen^{\text{pair}}_{{p}{q}} = \cre{p_\uparrow}\an{q_\uparrow}\cre{p_\downarrow}\an{q_\downarrow}  - \mathrm{h.c.}$ between spin-orbitals $p,q$, or an equivalent set of excitations $\mathcal{E}_{\mathrm{GD}}^\mathrm{pair}=\big\{({p}_\uparrow {p}_\downarrow,{q}_\uparrow {q}_\downarrow) : 1 \le p < q \le M   \big\} $.
While not aiming to provide an exact parametrisation of an electronic wavefunction, the underlying idea is to heuristically extend the restricted expressibility of a pair-UCCD  by the inclusion of singles and generalised excitations as well as by $k$-fold subsequent application of the same, but independently parametrised unitary circuits. 
This way, the parametrisation is hoped to overcome at least some of the disadvantages of pair-CCD, such as incapability of breaking multiple bonds, loss of invariance with respect to unitary transformations within the occupied-occupied and virtual-virtual subspaces and inability to fully recover the dynamical correlation of CCD. 

Formally, the so-called $k$-UpCCGSD wavefunction reads
\begin{equation}\label{eq:k-up-wfn}
   \ket{k\text{-}\mathrm{UpCCGSD}} = \left(\prod_{\alpha=1}^k e^{ \operator{T}(\alpha) -\operator{T}^\dagger(\alpha) }\right) \ket{\Phi_0},
\end{equation}
with $\alpha=1,\ldots, k$ independently parametrised cluster operators
\begin{align}
    \operator{T}(\alpha) &= \operator{T}_1(\alpha) + \operator{T}_{2,\mathrm{pair}}(\alpha) =
    \sum_{(p, q) \in \mathcal{E}_\mathrm{GS}\cup \mathcal{E}_{\mathrm{G D}}^\mathrm{pair} } \theta_{pq}(\alpha) \exgen_{pq}.
\end{align}

Evidently, the $k$-fold independent application of the UpCCGSD-parametrisation leads \reviewed{}{to} a linear top-up in terms of gate count. However, restricting the doubles excitations to pairs only leads to considerable savings; thanks to a much sparser set of doubles excitations, the gate count depends only quadratic instead of approximately quartic on the number of spin-orbitals $M$, amounting to $\order{kM^2}$. The complexity in terms of circuit depth can even be reduced to $\order{kM}$ as compared to cubic for standard UCC(G)SD because when exploiting the pair-structure, one can divide the operations such that they only operator on different set of spin-orbitals (pairs) enabling execution in parallel. \cite{Lee2019}
\reviewed{}{In order to implement the ansatz in Eq.~\eqref{eq:k-up-wfn}, one usually invokes an additional approximation using Trotterisation.}

Reference~\cite{Lee2019} provides results for the dissociation of water and the nitrogen molecule, among others. In addition to usually provided ground-state, they also showcase simulation results for the computation of excited states. Their results coincide with outlined expectations: The $k$-UpCCGSD ansatz, except for cases when UCCSD is exact, yields better energies than a naive UCCSD parametrisation, especially in case of excited state calculations. Although circuit depth is reduced, this comes at the cost of additional amplitudes/parameters given the introduction of general excitations and $k$-fold independent repetition of the circuit.
Beyond the introducing paper itself, the ansatz has e.g. been applied in Refs.~\citen{Huggins_2020,greene2021K-UPCC-APPLICATION}. While the former rather focuses on benefits of the therein proposed non-orthogonal VQE, Ref.~\citen{greene2021K-UPCC-APPLICATION} applies the $k$-UpCCGSD for a thorough study of computing excited and ionised states. Their results are promising and findings about the performance of the $k$-UpCCGSD parametrisation itself are in alignment with what is described above. 

\subsubsection{Separable Pair Approaches}
Next, we introduce a descendant of the previously described $k\text{-UpCCGSD}$; here, excitations are not only restricted to pair-excitations but rather a pair-natural-orbital (PNO) \cite{edmiston1968pno1,meyer1971pno2,meyer1973pno3,ahlrichs1975pno4} form of the orbital basis is assumed. Reference~\cite{gonthier2020identifying} uses a similar basis in form of frozen natural orbitals (FNOs)~\cite{barr1970fno,taube2005pno}, which are a re-canonicalised form of MP2-PNOs. However, this recanonicalisation effectuates a loss of the pair-structure, i.e., benefits beyond reduction of the number of qubits are cannot be exploited here. In the case of PNOs, one can restrict excitations to take place only between PNOs and their reference orbitals.
\cite{kottmann2021reducing} 

In order to make use of the PNO-based ansatz, one needs to generate a basis consisting of  HF reference orbitals $\{i\}$, while for each $(ij)$-pair there are associated \textit{orthonormalised} PNOs  $\tilde{\mathcal{S}}_{ij} = \bigcup_{\tilde{a}_{ij}=1}^{r_{ij}} \{ \ket{\tilde{a}_{ij}} \}$. 
Assuming such a problem-specific basis, it is possible to restrict to pair-excitations going from reference orbitals to their PNOs via $\uunitary_{\tilde{\mathrm{ D}}}$. Additionally one can think of generalised excitations within each PNO $\uunitary_{\mathrm{G}\tilde{\mathrm{D}}}$ and general single excitations in the spirit of Ref.~\citen{Lee2019}. Then, one obtains the following parametrised state:
\begin{equation}
    \ket{\text{PNO-UpCC(GS)D}} = \uunitary_{\mathrm{S}} \uunitary_{\mathrm G \tilde{\mathrm D}} \uunitary_{\tilde{\mathrm D}} \uunitary_{\mathrm{HF}} \ket{0}
\end{equation}
with the excitation gates
\begin{align}
    \uunitary_{\tilde{\mathrm D}}  = \prod_{i=1}^{N_{\mathrm{el}/2}} \prod_{a\in \tilde{\mathcal{S}}_{ii}} 
    \reviewed{\exp \left\{ \frac{\theta}{2} \exgen^\text{pair}_{aa,ii} \right\}}{\exp \left\{ {\theta} \exgen^\text{pair}_{aa,ii} \right\}}, \\
    \uunitary_{\mathrm G \tilde{\mathrm D}}  = \prod_{i=1}^{N_{\mathrm{el}/2}} \prod_{a,b\in \tilde{\mathcal{S}}_{ii}} 
    \reviewed{\exp\left\{ \frac{\theta}{2} \exgen^\text{pair}_{aa,bb} \right\}}{\exp\left\{ {\theta} \exgen^\text{pair}_{aa,bb} \right\}}
    , \\
    \uunitary_{\mathrm{S}} = \sum_{p,q} \reviewed{\exp{\frac{\theta}{2} \exgen_{p,q}}}{\exp{{\theta} \exgen_{p,q}}.}.
\end{align}
The pair-excitation generator $\exgen^\text{pair}_{aa,ii}$ is defined as
\begin{equation}
    \exgen^\text{pair}_{aa,ii} = 
    \cre{a_\uparrow}\an{i_\uparrow}\cre{a_\downarrow}\an{i_\downarrow} - \mathrm{h.c.}
\end{equation}
We further have the usual the singles excitation generator $\exgen_{p,q} = \cre{p}\an{q} - \cre{q}\an{p}$ and is able to excite between arbitrary orbitals. 

Using such a parametrisation, it was showcased in Ref.~\citen{kottmann2021reducing} that a PNO-UpCCGSD--type wavefunction can drastically reduce the number of parameters and CNOT-gates in comparison to the already compact UpCCGSD \cite{Lee2019}. Of course this is only possible, if the PNO approximation delivers a good approximation of the system in question, and requires generation of the PNOs (either using multiresolution analysis as in Refs.~\citen{kottmann2020direct,kottmann2021reducing} or using an analogue implementation for Gaussian orbitals or other numerical representations). Alternatively the orbitals could also be constructed from a generalised valence bond approach, similar as in Ref.~\citen{larsson2020minimal}. 

The so called separable pair ansatz~\cite{kottmann2021optimized} was introduced as a generalisation of PNO-UpCC(G)SD
\begin{equation}
    \ket{\psi_{\mathrm{SP}}} = \prod_{k=1}^{N_e/2} \ket{\psi_k} = \prod_{k=1}^{N_e/2} \uunitary_k(\theta_k) \uunitary_\mathrm{RHF} \ket{00\cdots 0}
\end{equation}
such that  the individual pair-functions $\ket{\psi_k}$ can be factorised as $\ket{\psi_k} = \sum_{m,n} c_{mn}^k \ket{\phi^k_m}\otimes\ket{\phi^k_n}$ with an orthonormal basis $\{\phi^k_m\}$ and $\uunitary_k$ correspond to one- and two-electron excitations.
As pointed out in Ref.~\citen{kottmann2021optimized}, this ansatz (as well as the PNO-UpCC(G)SD) leads to a classically efficiently solvable model that does not suffer from the curse of dimensionality. Compared to similar classically solvable models (e.g. UpCCD ~\cite{elfving2021hcb}) the corresponding circuits are significantly lower in depth. Hence, while a quantum advantage in the long term is not to be expected, this ansatz can prove valuable for state preparation.  

\subsection{Double Unitary CC}
The Double Unitary Coupled Cluster (DUCC) formalism was introduced in Refs.~\citen{kowalski2018properties, kowalski2021dimensionality, bauman2019downfolding} as an alternate representation of the CC equations, where one can construct active space representation of the Hamiltonian by removing higher fermionic degrees of freedom while reproducing the exact energy of the quantum system. 
The UCC ansatz which can then be decomposed into a product of two unitary transformations according to a system embedding sub-algebra (SES), as follows:
\begin{align}
   \ket{\psi_{\mathrm{UCC}}} = e^{\operator{T} - \operator{T}^{\dagger}} \ket{\Phi_{\mathrm{HF}}} = e^{\operator{\sigma}_{ext}} e^{\operator{\sigma}_{int}} \ket{\Phi_{\mathrm{HF}}}
\end{align}
where, ${\operator{\sigma}_{ext}}$ and ${\operator{\sigma}_{int}}$ are \reviewed{anti-Hermitian}{antihermitian} cluster operators, $\operator{\sigma}_{ext}^\dagger = - \operator{\sigma}_{ext}$ and $\operator{\sigma}_{int}^\dagger = - \operator{\sigma}_{int}$, defined as:
\begin{align}
    {\operator{\sigma}_{ext}} = \operator{T}_{ext} - \operator{T}^{\dagger}_{ext} \\
    {\operator{\sigma}_{int}} = \operator{T}_{int} - \operator{T}^{\dagger}_{int}
\end{align} 
The cluster operator $\operator{T}_{int}$ represents operators within the active space only (i.e., all spin indices within the active space), whereas $\operator{T}_{ext}$ represents operator acting outside the active space (i.e., at least one spin index outside the active space).  

The corresponding energy of the quantum system can be then written as,
\begin{equation}
    E =\bra{\Phi_{\mathrm{HF}}} e^{- \operator{\sigma}_{int}} e^{-\operator{\sigma}_{ext}} H  e^{\operator{\sigma}_{ext}}  e^{\operator{\sigma}_{int}} \ket{\phi_{\mathrm{HF}}}
\end{equation}
and can be obtained by diagonalising the effective Hamiltonian, $\ham^{eff(DUCC)}_{ext}$, in the space of the projection operator $\operator{P} + \operator{Q}_{int}$, where $\operator{P} = \ket{\Phi_{\mathrm{HF}}}\bra{\Phi_{\mathrm{HF}}}$ is the projection operator onto the reference wavefunction, and $\operator{Q}_{int}$ is the projection operator spanned by all the excited configuration generated with acting $\operator{T}_{int}$ on the reference wavefunction. 
The effective Hamiltonian $\ham^{eff(DUCC)}_{ext}$ can be expressed as: 
\begin{equation}
    \ham^{eff(DUCC)}_{ext} = (\operator{P} + \operator{Q}_{int}) \ham^{DUCC}_{ext} (\operator{P} + \operator{Q}_{int})
\end{equation}
where, $\ham^{DUCC}_{ext} = e^{-\operator{\sigma}_{ext}} H  e^{\operator{\sigma}_{ext}}$ is the downfolded Hamiltonian. 
This technique has been shown to be resource efficient in Ref.~\citen{metcalf2020resource}, as by carefully choosing the active space and thus the effective Hamiltonian, one can capture the effect of the full orbital space using a small active space.
It has been further developed for the VQE algorithm in the generalised UCC formalism in Ref.~\citen{bauman2020variational}.

\subsection{Further UCC-related approaches }
In this section, we will take a look at other approximations and ans\"atze related to UCC, not grouped into common themes, like active space, orbital optimisations, parameter initialisation, among others.

\subsubsection{UCC within active spaces}
The cost of implementing UCC for solving larger quantum chemistry problems require significant quantum resources, so using approximations to lower this cost can be a viable option.
An approximation designed to reduce the quantum resource is the complete active space (CAS) approach~\cite{roos1980complete}. 
This involves dividing the full set of orbitals (within the basis set) into sets of inactive and active orbitals, dictated by the nature of the orbitals and the overlap between them. During a computation, the occupation of the inactive orbitals remains unchanged.  The quality of the computation however is highly sensitive to the choice of this active space and in general it is not a priori clear which orbitals to select; an attempt to automatize this is given in Ref.~\citen{stein2016automated}.

The CAS approximations can be applied to UCC to reduce the computational cost for different quantum chemistry problems in a number of ways, including treating static correlation effects, generating better reference wavefunctions, and generating initial guess for larger active space calculations, among others~\cite{romero2018strategies}. 
These can lead to reduced circuit sizes  and UCCSD parameters, and can increase rate of convergence for the full calculation. 
A lot of different studies have successfully used this approach to carry out different UCC calculations on quantum computers~\cite{romero2018strategies}.

\subsubsection{Orbital Optimised UCC}\label{subsubsec:oo-ucc}
The Orbital Optimised UCC (OO-UCC)~\cite{sokolov2020quantum, mizukami2020orbital, Yalouz_2021} is another variant of the traditional UCCSD, where the doubles part is optimised variationally on a quantum computer and singles contributions are optimised on a classical computer using the one-particle (1-RDMs) and two-particle reduced density matrices (2-RDMs) from the runs on the quantum device. The OO-UCCD approach was also extended for singlet and pair UCCD approaches in Ref.~\citen{sokolov2020quantum}. The authors show how using the orbital optimisation techniques allows for restoration of significant fraction of the correct energy value. An additional advantage is access to geometry derivatives thanks to the fully variational nature of both the reference and the amplitudes.

Within UCC, orbital optimised approaches implicitly assume a specific form of the ansatz where the singles operators act last on the total unitary operation
\begin{equation}
    \ket{\Psi} = e^{\exgen_{\text{S}}}  e^{\exgen_{\text{D}}} \ket{\Phi_0}.
\end{equation}
Now the terminal singles block $e^{\exgen_{\text{S}}}$ is formally absorbed into the Hamiltonian
\begin{align}
    \tilde{H}\left({\theta}_{\text{S}}\right) = e^{\exgen_{{\text{S}}}^\dagger}\ham e^{\exgen_{\text{S}}},\label{eq:transformed_H} 
\end{align}
where we denoted the dependency of the transformed Hamiltonian on the angles of the singles rotations $\exgen_{\text{S}}$.\\
Usually at this stage, the singles block is assumed to consist of spin-adapted singles rotations
\begin{align}
    \exgen_{\text{S}}  = \sum_{pq} \theta^q_p \left(\operator{E}^q_p - \operator{E}^p_q\right), \quad \operator{E}_{p}^{q} = \cre{p\uparrow}\an{q\uparrow} + \cre{p\downarrow}\an{q\downarrow}, 
\end{align}
which corresponds to the restriction of the original angles $\theta_{p\uparrow}^{q\uparrow} = \theta_{p\downarrow}^{q\downarrow} $ and $\theta_{p\sigma}^{q\sigma'} \propto \delta_{\sigma\sigma'}$ (\textit{i.e.} all angles corresponding to spin-flips vanish). In this form, the transformation of the second-quantised Hamiltonian corresponds to a rotation of the molecular orbitals. Instead of simultaneously optimising
\begin{align}
    E = \min_{{\theta}_{\text{S}},{\theta}_{\text{D}}}\expval{\tilde{H}\left({\theta}_{\text{S}}\right)}_{\uunitary\left({\theta}_{\text{D}}\right)},
\end{align}
the two sets of parameters are optimised sequentially in an iterative loop. The ${\theta}_{\text{D}}$-optimisation remains the standard VQE procedure with the rotated Hamiltonian $\tilde{H}$, which for this part is constructed by rotating the orbitals and re-constructing the qubit Hamiltonian (see for example the appendix of Ref.~\citen{Yalouz_2021}). The ${\theta}_{\text{S}}$ usually adds further approximations where Eq.~\eqref{eq:transformed_H} is expanded and truncated after second order
\begin{align}
    e^{-\exgen_{\text{S}}}\ham e^{\exgen_{\text{S}}} = \ham + \comm{\ham}{\exgen_{\text{S}}} + \frac{1}{2}\comm{\comm{\ham}{\exgen_{\text{S}}}}{\exgen_{\text{S}}} + \dots
\end{align}
Taking the gradient of the corresponding expectation value with the optimised UCCD wavefunction of the current step with respect to the angles $\boldsymbol{\theta}_\text{S}$, leads to the Newton-Raphson equations
\begin{align}
    -\boldsymbol{H}\boldsymbol{\theta}_\text{S} = \boldsymbol{g}
\end{align}
that can then be solved classically to approximate the optimal angles for the orbital rotations. Here, the matrices $\boldsymbol{H}$ and $\boldsymbol{g}$ are the electronic gradient and Hessian
\begin{align}
    g_p^q = \frac{\partial }{\partial \theta_p^q} e^{-\exgen_{\text{S}}} \ham e^{\exgen_{\text{S}}},\quad h_{pr}^{qs} = \frac{\partial^2}{\partial \theta_p^q \partial \theta_r^s} e^{-\exgen_{\text{S}}}\ham e^{\exgen_{\text{S}}},
\end{align}
which require the evaluation of expectation values of commutators and double commutators of the Hamiltonian with the spin-adapted singles generators $\operator{E}_p^q$. 

\subsubsection{Multicomponent UCC}
The multicomponent UCC (mcUCC) parametrisation  \cite{Pavosevic2021multicomponent} generalises the electronic UCC ansatz to models that also explicitly include positrons, where the Hamiltonian of the system consequently includes positronic kinetic energy, positron-proton, positron-positron and positron-electron terms. Starting with an uncorrelated reference $\ket{\Phi_\text{ref}^-\Phi_\text{ref}^+} = \ket{\Phi_\text{ref}^-}\otimes\ket{\Phi_\text{ref}^+}$, one obtains the parametrised wavefunction
\begin{equation}
    \ket{\psi_\mathrm{mcUCC}} = e^{\operator{T}(\theta)-\operator{T}^\dagger(\theta)}\ket{\Phi_\text{ref}^-\Phi_\text{ref}^+},
\end{equation}
while the cluster operators $\operator{T}(\theta)$ is defined as before~\eqref{eq:cluster-op} and additionally includes positron excitations. From the doubles level, this also includes operations that perform simultaneous excitations in the electron and positron space.  

A similar approach, yet with the other component except for the electronic state being a photon number state, has been introduced in Ref.~\citen{pavosevic2021polaritonic} in order to tackle the field of polaritonic chemistry. 

\subsubsection{Unitary Cluster-Jastrow ansatz}

Jastrow factors are a tool  to account for the correlation cusp originating from the Coulomb potential in electronic wavefunctions \cite{kato1957eigenfunctions,kong2012explicitly} best known from Quantum Monte Carlo (QMC) techniques \cite{foulkes2001qmc}. They aim to provide efficient trial wave functions to accelerate convergence of QMC methods.  Building on a formulation of coupled cluster inspired by \reviewed{the}{these} Jastrow factors coined Cluster-Jastrow \cite{neuscamman2013clusterjastrow},  
Ref.~\citen{matsuzawa2020jastrow} proposed a unitary form named Unitary Cluster-Jastrow (UCJ) to allow an implementation on a quantum computer.  \reviewed{In particular, they applied their ansatz, which requires $\order{N^2}$ parameters, in form of a VQE procedure.}{In particular, applied to the VQE procedure, their ansatz requires $\order{N^2}$ parameters, in form of a VQE procedure.} 

In the Cluster-Jastrow ansatz, the doubles excitation operator $T_2$ is decomposed as follows:
\begin{equation}\label{eq:cluster-jastrow}
    \operator{T}_2 \approxeq \operator{T}^{CJ}  = \sum_{pqrs, \sigma\tau} \left(\sum_{jl} U_{pj} U_{qj} J_{jl}^{(\sigma\tau)} U_{rl} U_{sl}\right) \cre{p_\sigma} \an{q_\sigma} \cre{r_\tau} \an{s_\tau}
\end{equation}
where $U_{pj}$ are unitary single-particle rotations and $J_{jl}^{(\sigma\tau)}$ matrix elements filled with Jastrow factors. Then, one can write 
\begin{equation}
    e^{\operator{T}^{CJ}} = e^{-\operator{K}} e^{\operator{J}} e^{\operator{K}}
\end{equation}
with 
\begin{equation}
    \operator{J} = \sum_{jl, \sigma\tau} J^{(\sigma\tau)}_{jl} \cre{j_\sigma} \an{j_\sigma} \cre{l_\tau} \an{l_\tau}, \quad \operator{K} = \sum_{p<q,\sigma} K_{pq}\big(\cre{p_\sigma}\an{q_\sigma} - \cre{q_\sigma}\an{p_\sigma}\big).
\end{equation}
\reviewed{It is obvious that}{For real $J_{jl}^{\sigma\tau}$}, $e^{\operator{J}}$ cannot be a unitary operation \reviewed{for real $J$}, which is why Ref.~\citen{matsuzawa2020jastrow} enforce \reviewed{the}{its} matrix elements to be  purely imaginary, \reviewed{}{re-named to $\mathcal{J}^{(\sigma\tau)}_{jl} $}. It further proves convenient to choose $\mathcal{K}_{pq}$ such that $\operator{K}$ is \reviewed{anti-Hermitian}{antihermitian}. This leads to 
\begin{equation}
    \operator{J} = \sum_{jl, \sigma\tau} \mathcal{J}^{(\sigma\tau)}_{jl} \cre{j_\sigma} \an{j_\sigma} \cre{l_\tau} \an{l_\tau}, \quad \operator{K} = \sum_{pq,\sigma} \mathcal{K}_{pq}\cre{p_\sigma}\an{q_\sigma}.
\end{equation}
where we note that $\operator{J},\operator{K}$ depend only on fermionic number operators. Hence, using the Jordan-Wigner encoding, this ansatz can be implemented without the Trotter-Suzuki approximation as it leads to Pauli strings comprised of only identity- and $\quoperator{Z}$-operations.

Motivated by a similarity between the CJ-cluster amplitudes in Eq.~\eqref{eq:cluster-jastrow} and a low-rank approximation throughout a tensor-hypercontraction scheme of generalized CC doubles amplitudes \cite{hohenstein2012thc-cc}, the following procedure is suggested in Ref.~\citen{matsuzawa2020jastrow}: \reviewed{Taking $k$-fold, independent repetitions of the UCJ ansatz obtaining amplitudes $\sum_{x=1}^k \left(\sum_{jl} U_{pj}^x U_{qj}^x \mathcal{J}_{jl}^{(\sigma\tau),x} U_{rl}^x U_{sl}^x\right)$ that are able to reproduce UCCGSD for increasing $k$, while the tensor $\mathcal{J}^{\sigma\tau)}_{jl}$ can be  substituted by a low-rank approximation.}{Taking $k$-fold, independent repetitions of the UCJ ansatz yields amplitudes $\sum_{x=1}^k \left(\sum_{jl} U_{pj}^x U_{qj}^x \mathcal{J}_{jl}^{(\sigma\tau),x} U_{rl}^x U_{sl}^x\right)$ that, for increasing $k$, are able to reproduce UCCGSD. Within the procedure, the tensor $\mathcal{J}^{(\sigma\tau)}_{jl}$ can be  substituted by a low-rank approximation.} 
Restrictions on the range of indices $p,q,r,s$ in Eq.~\eqref{eq:cluster-jastrow} allow to retrieve analogues to e.g. pair-restricted parametrizations such as UpCCGSD.
Similarly to the $k$-fold product form of $k$-UpCCGSD in Ref.~\citen{Lee2019}, one can also think of a product form $\prod_{x=1}^k e^{-\operator{K}_x} e^{\operator{J}_x} e^{\operator{K}_x}$. This is not the same as the exponential of a sum of independent UCJ copies because the individual layers do not commute, but is expected to be more expressive than the single-product form and rather retrieve the FCI-limit. 

In their numerical studies, Ref.~\citen{matsuzawa2020jastrow} compared the $k$-UCJ ansatz with a low-rank approximation of the doubles excitations using singular value decomposition (SVD) and the $k$-UpCCGSD which comes at a comparable cost in terms of parameters and gate count. They found that UCJ substantially outperforms the \reviewed{their}{}  low-rank approximation by SVD\reviewed{and}{; further,}  $k$-UCJ yields accurate results, namely chemical accuracy for {N$_2$}) for rather low $k=3$ and approaching FCI for $k\gtrapprox8$.    To overcome local minima during variational optimisation, they suggest to follow Ref.~\citen{neuscamman2013clusterjastrow} and use an initial guess based on an anti-symmetric geminal power (AGP) wavefunction.

\subsubsection{Spin and orbital symmetries for unitary coupled cluster with singles and doubles}
One can make electronic calculations of high symmetry molecules more efficient by
exploitation of point group symmetries. This can reduce the number of floating point operations necessary, when applied to classical 
quantum chemistry calculations, by a factor of $h^2$, with $h$ the order of the molecular 
point group~\cite{Stanton.JCP.1991}.
In this subsection, spin and orbital symmetries  are used to significantly reduce the number of excitations in the UCCSD ansatz for highly 
symmetric molecules like H$_2$, LiH, H$_2$O and N$_2$. 
When solving for the amplitudes, the amplitudes pertaining to the 
dropped excitations will be zero and do not contribute to the energy. Therefore, finding these amplitudes 
before compilation and execution of, e.g. a variational optimisation of the parameters, is crucial to save resources. Previously, UCCSD excitations were  
dropped when having zero (or below a certain tolerance) MP2 amplitudes and an important reduction to 
the number of two-qubit gates was obtained.~\cite{romero2018strategies,Kuhn.JCTC.2019} However, the MP2 amplitudes
can only be used to drop double excitations. Additionally, this involves a previous classical calculation. The following procedure is useful for the identification of
negligible single and double excitations and can be applied as well to higher order excitations.

Implementation of spin symmetry for neutral closed-shell molecules is straightforward. In these molecules, 
there is an equal number of $\uparrow$ and $\downarrow$ electrons (or $\alpha,\beta$). Then, excitations that keep the balance between 
$\uparrow$ and $\downarrow$ electrons are the only ones that should be implemented in order to satisfy spin symmetry.
We use $\operator{\sigma}_1$ to denote single antihermitian (de-)excitations,
\begin{align}\label{E.S.Singles}
\operator{\sigma}_1({\theta}) &= \sum_{i,a \in \{\uparrow\}} \theta_i^a (
\cre{a} \an{i} - \cre{i} \an{a}) \nonumber \\
 &+ \sum_{\bar i,\bar a \in \{\downarrow\}} \theta_{\bar i}^{\bar a}(
\cre{\bar a} \an{\bar i} - \cre{\bar i} \an{\bar a}).
\end{align}
Further, $\hat{\sigma}_2$ applies for double antihermitian (de-)excitations,
\begin{align}\label{E.S.Doubles}
\operator{\sigma}_2({\theta}) &= \sum_{i<j,a<b \in \{\uparrow\}} \theta_{i,j}^{a,b}(
 \cre{a} \cre{b} \an{j} \an{i}
-\cre{i} \cre{j} \an{b} \an{a} )\nonumber \\
 &+ \sum_{\bar i<\bar j,\bar a<\bar b \in \{\downarrow\}} \theta_{\bar i,\bar j}^{\bar a,\bar b}(
 \cre{\bar a} \cre{\bar b} \an{\bar j} \an{\bar i}
-\cre{\bar i} \cre{\bar j} \an{\bar b} \an{\bar a} )\nonumber \\
 &+ \sum_{\substack{i,\bar j,a,\bar b \\ i,a \in \{\uparrow\};\bar j,\bar b \in \{\downarrow\} }} \theta_{i,\bar j}^{a,\bar b}(
 \cre{a} \cre{\bar b} \an{\bar j} \an{i}
-\cre{i} \cre{\bar j} \an{\bar b} \an{a}),
\end{align}
where ${\theta}$ is a set of real parameters $\{\theta_i^a,\theta_{i,j}^{a,b}\}$.

Orbital symmetry is more subtle to implement and we will expect some basic understanding of group theory of the reader to follow this, as found for example in Refs.~\citen{dresselhaus2008group,tinkham1964group}. First, each molecule must be
sorted into a point group. For example, H$_2$ and N$_2$ are $D_{\infty h}$,
LiH is $C_{\infty v}$ and H$_2$O is $C_{2v}$. Furthermore, 
the largest concise point group is used, which is $D_{2 h}$ for H$_2$ and N$_2$
and $C_{2v}$ for LiH and H$_2$O. The $D_{2h}$ group has eight different
irreducible representations while the $C_{2v}$ has four, see
Table \ref{T.OSymm} for a list of irreducible representations.
The multiplication tables for these point groups are \reviewed{also needed,see Tables \ref{T.d2h} and \ref{T.c2v}}{given in the Appendix}. 

\begin{table}
\begin{center}
\caption{Irreducible representations for the $D_{2h}$ and $C_{2v}$ point
groups.\label{T.OSymm}}
\begin{tabular}{cc}
\hline
\hline
$D_{2h}$ & $C_{2v}$ \\
\hline
$A_{g}$ & $A_1$ \\
$A_{u}$ & $A_2$ \\
$B_{1g}$ & $B_1$ \\
$B_{2g}$ & $B_2$ \\
$B_{3g}$ & \\
$B_{1u}$ & \\
$B_{2u}$ & \\
$B_{3u}$ & \\
\hline
\hline
\end{tabular}
\end{center}
\end{table}



In addition, we have an orbital symmetry
operator $\operator{s}_e$, which maps one orbital (or two orbitals) to
the irreducible representation of that orbital (or the product of the
two irreducible representations of the two orbitals). For example, if $i,j$ irreducible
representations are $B_1,B_2$, respectively, then we have
$\operator{s}_e(i) = B_1$, $\operator{s}_e(j) = B_2$ and $\operator{s}_e(i,j) = A_2$.

Single excitations can only occur between orbitals belonging to the same
irreducible representation. Double excitations occur when the (tensor) product of the 
 irreducible representations of the occupied orbitals is equal to the
product of the irreducible representations of the virtual orbitals. Additionally, the products of irreducible representations that partake in the excitation should contain the fully symmetric irreducible representation.  In combination
with the spin symmetry, we obtain the $\tilde{\sigma}_1$ operator for singles,
\begin{align}\label{E.S2.Singles}
\tilde{\sigma}_1({\theta}) &=
\sum_{\substack{i,a \in \{\uparrow\} \\ \operator{s}_e(i) = \operator{s}_e(a)}} \theta_i^a(
\cre{a} \an{i} - \cre{i} \an{a} )\nonumber \\
 &+ \sum_{\substack{\bar i,\bar a \in \{\downarrow\} \\ \operator{s}_e(\bar i) = \operator{s}_e(\bar a)}} \theta_{\bar i}^{\bar a}(
\cre{\bar a} \an{\bar i} - \cre{\bar i} \an{\bar a}),
\end{align}
and the $\tilde{\sigma}_2$ operator for doubles,
\begin{align}\label{E.S2.Doubles}
\tilde{\sigma}_2({\theta})
 &= \sum_{\substack{i<j,a<b \in \{\uparrow\} \\ \operator{s}_e(i,j) = \operator{s}_e(a,b)}} \theta_{i,j}^{a,b}(
 \cre{a} \cre{b} \an{j} \an{i}
-\cre{i} \cre{j} \an{b} \an{a}) \nonumber \\
 &+ \sum_{\substack{\bar i<\bar j,\bar a<\bar b \in \{\downarrow\} \\ \operator{s}_e(\bar i,\bar j) = \operator{s}_e(\bar a,\bar b)}} \theta_{\bar i,\bar j}^{\bar a,\bar b}(
 \cre{\bar a} \cre{\bar b} \an{\bar j}  \an{\bar i}
-\cre{\bar i} \cre{\bar j} \an{\bar b}  \an{\bar a} )\nonumber \\
 &+ \sum_{\substack{i,\bar j,a,\bar b \\ i,a \in \{\uparrow\};\bar j,\bar b \in \{\downarrow\} \\ \operator{s}_e(i,\bar j) = \operator{s}_e(a,\bar b) }} \theta_{i,\bar j}^{a,\bar b}(
 \cre{a} \cre{\bar b} \an{\bar j} \an{i}
-\cre{i} \cre{\bar j} \an{\bar b} \an{a}).
\end{align}
These operators yield the same numerical results and greatly reduce
the number of excitations needed to encode in a quantum computer, reducing the two-qubit gate count too as well as the number of parameters in a subsequent optimisation procedure. 

To display the power of using spin and orbital symmetries, we have calculated the number of single and double 
excitations needed for H$_2$, LiH, H$_2$O and N$_2$ using correlation--consistent double, triple and 
quadruple $\zeta$ basis sets,\cite{dunning1989a,prascher2011a}
(cc-pVDZ, cc-pVTZ and cc-pVQZ).
When using spin symmetry, the number of excitations is reduced by 50\% for single and by 47-61\% for double
excitations. In addition, when using both spin and orbital symmetry, the number of excitations is reduced by 
76-92\% for single and by 82-95\% in the case of double excitations. The best reductions are obtained for the 
N$_2$ molecule due to its high symmetry (highest concise abelian group is $D_{2h}$). 

\subsection{A general overview of other ans\"atze}
UCC ans\"atze correspond to physically motivated circuits equipped with methods for setting informative initial guesses for the variational parameters.
However, these circuits often cannot be realised on near-term quantum devices due to their prohibitive depths even for describing small molecules.
\reviewed{}{Some recent advances have been made to reduce the depth of these circuits by using correlated energy functionals~\cite{peng2021variational} based on the Peeters,
Devreese, and Soldatov (PDS) formalism. However, it can suffer from large number of measurements as we scale to larger systems.}
 So, alternatives to UCC ans\"atze have since been explored and proposed in literature, several of which drastically reduce the circuit depth requirements for small problem sizes, though at the cost of leading to a more challenging parameter optimisation task.
We provide brief descriptions of several alternative ans\"atze here.

The ``Hardware-Efficient Ansatz'' (HEA) corresponds to a multi-layered parameterized quantum circuit comprised of gates and qubit connectivity native to a particular quantum device \cite{kandala2017hardware}. These circuits have been shown numerically and experimentally to be effective for small problem instances in VQE. However, HEA circuits are not tailored to the problem-of-interest, e.g. do not account for symmetries, and thus can become more difficult to optimise for larger problems or have low overlap with the ground state. The HEA corresponds to the following form:
\begin{align}\label{eq:hea}
    \ket{\psi_\text{HEA}(\theta)} = \prod_{L} \uunitary_{\text{ent}}(\theta_L^{(\text{ent})}) \prod_{n=1}^{N} \uunitary_{\text{rot}}(\theta_{n, L}^{(\text{rot})}) \ket{\Phi_{\text{ref}}}
\end{align}
where $L$ indexes the circuit layers and $n$ indexes the qubits. The circuit generally consists of $L$ repetitions of a layer of local rotation gates $\uunitary_{\text{rot}}$ followed by a layer of entangling operations $\uunitary_{\text{ent}}$ (which may or may not be parameterized).
Inspired by HEA and ADAPT-VQE, authors of Ref.~\citen{tang2019qubitadapt} developed ``qubit-ADAPT-VQE,'' in which the operator pool comprises low-depth gate operations native to the device. Compared to ADAPT-VQE for estimating ground state energies of small molecular systems, qubit-ADAPT-VQE generated circuits with significantly fewer CNOTs though they required more parameters. 

Another class of ans\"atze that are physically motivated, like UCC, are the Variational Hamiltonian Ans\"atze (VHA) \cite{wecker2015progress}.
The VHA is a parameterized version of the Trotterised evolution under a problem Hamiltonian:
\begin{align}\label{eq:vha}
    \ket{\psi_\text{VHA}(\theta)} = \prod_{L} \prod_{j} e^{-\imag \theta_{j, L} \ham_j} \ket{\Phi_{\text{ref}}}
\end{align}
where $\sum_j H_j$ is the problem Hamiltonian and $L$ indexes the number of circuit layers. 
Circuits realizing VHA can be very deep due to the number of terms in the Hamiltonian but can be efficient in the number of parameters if the terms in the Hamiltonian are efficiently grouped. 
In addition, a recent study showed that VHAs are minimally impacted by barren plateaus in optimisation landscapes \cite{wiersema2020exploring}.

The Low-Depth Circuit Ansatz (LDCA) was developed as a low-depth alternative to UCC ans\"atze \cite{DallaireDemers2019}.
\reviewed{}{That is, similar to qubit-ADAPT-VQE \cite{tang2019qubitadapt} and Qubit Coupled-Cluster (QCC) \cite{Ryabinkin.JCTC.2018}, LDCA is constructed directly from Pauli words, resulting in lower-depth circuits.}
The LDCA circuit has a highly nested gate structure (see Ref.~\citen{DallaireDemers2019} for circuit diagram), which was inspired by a circuit implementing fermionic Gaussian states.
Specifically, $\quoperator{Z}\quoperator{Z}$-type entanglers (where both $\quoperator{Z}$ act on different qubits) were inserted to extend the circuit to prepare non-Gaussian fermionic states.
The resulting circuit is also hardware-efficient for superconducting circuit architectures with linear qubit connectivity and tunable couplers.
Numerically, LDCA circuits were shown to be effective for preparing ground states of strongly correlated fermionic systems.
The resulting LDCA circuit corresponds to circuit depth which scales linearly with system size.
The parameter count scales quadratically with system size, which has limited applications of LDCA to small systems.
However, a recent work introduced an optimiser that extended the use of LDCA to larger systems by finding an effective parameter subset that minimises the energy \cite{sim2020adaptive}.

Inspired by quantum optimal control theory and the VHA, the Quantum Optimal Control-Inspired Ansatz (QOCA) considers a parameterized form of the Hamiltonian:
\begin{align}\label{eq:qoca}
    \ham_{\text{QOCA}}(t) = \ham_{\text{prob}} + \sum_k c_k(t) \ham_k,
\end{align}
where $\ham_{\text{prob}}$ is the problem Hamiltonian and $\{ \ham_k \}$ is a set of drive terms that do not commute with the problem Hamiltonian \cite{choquette2020quantum}. 
The drive terms are parameterized by $c_k(t)$.
The QOCA then corresponds to the following form:
\begin{align}\label{eq:qoca}
    \ket{\psi_\text{QOCA}(\theta)} = \prod_{L} \bigg( \prod_{j} e^{-\imag \theta_{j, L}^{(\text{prob})} \ham_j} \prod_{k} e^{-\imag \theta_{k, L}^{(\text{drive})} \ham_k} \bigg) \ket{\Phi_{\text{ref}}}
\end{align}
where $\ham_{\text{prob}} = \sum_j \ham_j$, and there are $L$ repetitions of alternatively applying $\uunitary_\text{prob} = \prod_{j} e^{-i \theta_{j, L}^{(\text{prob})} H_j}$ and $\uunitary_{\text{drive}} = \prod_{k} e^{-i \theta_{k, L}^{(\text{drive})} H_k}$.
Unlike the VHA, the addition of drive terms allows the ansatz to temporarily leave the symmetry sector to potentially accelerate optimisation. 
The QOCA was tested for preparing the ground states of instances of the Fermi-Hubbard model, in which it outperformed both the HEA and VHA by preparing ground states with higher fidelity. 

\reviewed{}{
\subsection{Progress on approximating energies with different UCC ans\"atze}
Since the proposal of using UCC ansatz within the VQE framework in Ref.~\citen{peruzzo2014variational}, it has successfully demonstrated experimentally~\cite{o2016scalable,shen2017quantum,hempel2018quantum} for calculating ground state energies of different small molecules such as H$_2$, LiH, and HeH$^+$.
The improvements in the implementation of the UCC ansatz and different adaptations~\cite{kottmann2021optimized, Lee2019, kottmann2021reducing, romero2018strategies} have allowed simulations of larger molecules, such as BeH$_2$ (12 qubits), H$_2$O (12 qubits), H$_4$ (16 qubits), N$_2$ (16 qubits), BH (22 qubits), and C$_2$H$_6$ (28 qubits).
While, we have made huge improvements from the first demonstrations, a lot of effort still needs to be put in before one can implement these ans\"aste on a physical device for complex molecules.
}

\section{A unified perspective on Unitary Coupled Cluster}
\begin{figure*}
    \centering
    \includegraphics[height=.95\textheight]{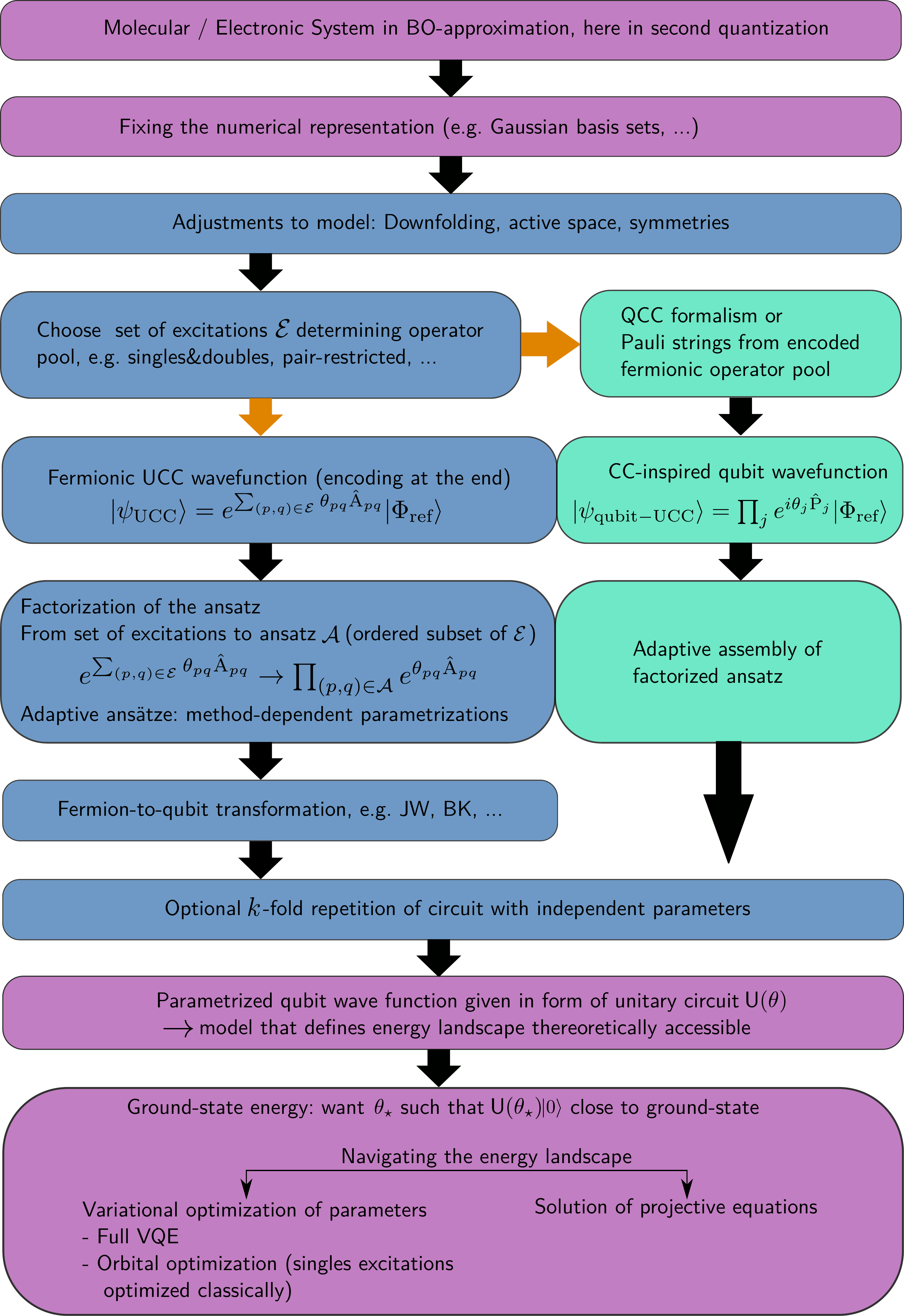}
    \caption{Overview of Unitary Coupled-Cluster in different flavors.}
    \label{fig:ucc-unif-overview}
\end{figure*}
As elucidated in Sec.~\ref{sec:introduction}, (unitary) coupled cluster makes up a specific form to describe an often purely electronic wavefunction. In that sense, (U)CC are a family of approximate models, which range from untruncated excitation degree corresponding to full configuration interaction until singles excitations, allowing only for orbital rotations. Such a model can then be used to look for, e.g., the ground-state wavefunction. This is enabled by degrees of freedom that are built into the model. In the case of (U)CC, the cluster amplitudes $t_a^i, t_{ab}^{ij}$ as in Eq.~\eqref{eq:cluster-op} serve for this purpose. 
While the term ``CC method'', or recently in the context of quantum computing also ``UCC method'', frequently appears in the literature, we consider (U)CC as a \textit{model} in form of a parametrised wavefunction, which is except for the case of adaptive procedures (ADAPT-VQE in section~\ref{subsubsec:adapt}, SPQE in section~\ref{subsubsec:pqe}) not directly dependent on the method that is used to compute the desired outcome.

In order to make a UCC parametrisation feasible on a quantum computer, there are a few additional steps necessary, which have been described in section~\ref{subsec:ucc-ansatz}. Moreover, one can invoke additional tweaks and approximations, that lead to the different flavors described in section~\ref{sec:ucc-flavors} and combinations / generalisations thereof. 

At this point, we would like to consolidate this review with a unifying perspective on the different approaches from section~\ref{sec:ucc-flavors}.  As indicated in the above paragraph, UCC aims to express the wavefunction of a quantum system. Yet to represent such a wavefunction on quantum hardware, one needs to find a unitary transformation to do so. This transformation should, especially during the NISQ era of quantum computing, consist of as few operations and require as few qubits as possible. 
And while UCC can be considered an ansatz that is efficient in the number of parameters, its vanilla version leads to rather deep circuits when using the construction described in section~\ref{subsec:ucc-ansatz}. Therefore, different ans\"atze based on UCC have been developed to make the procedure more efficient and hardware-friendly. 
Having a parametrised wavefunction available, one might e.g. desire an approximation to the system's ground-state energy. To achieve this, one needs to determine the set of parameters $\theta_\star$ such that the computational goal is fulfilled as good as possible. 
Similar strategies can be thought of for other interesting properties such as excited~\cite{Lee2019, kottmann2021feasible} or vibrational~\cite{ollitraut2020hardware} states. 

Figure~\ref{fig:ucc-unif-overview} visualises the hierarchy of steps to go from a generic UCC wavefunction to a qubit wavefunction, that is directly implementable, potentially up to compilation of unitary operators into one- and two-qubit gates.

The main ingredient necessary is a specific system represented by its abstract Hamiltonian operator. As a first step towards a computational solution, one needs to fix a numerical representation for the abstract Hamiltonian. In second quantisation, a popular choice is a basis projection onto Gaussian basis sets, where molecular orbitals are obtained via linear combination of atomic orbitals based on a mean-field procedure. Alternatively, one can think of a directly determined, system-wide pre-optimised molecular basis in form of pair-natural orbitals \cite{kottmann2020direct}, frozen natural orbitals \cite{gonthier2020identifying} or intrinsic atomic orbitals \cite{barison2020quantum}. Beyond that, however not as common in second-quantised methods, one can also employ plane-waves, Slater-type orbitals, and others.

Next, further adjustments to the discretised Hamiltonian can be chosen to reduce model complexity. While the exploitation of symmetries does not lead to an approximation, the choice of an active space or a downfolding procedure in general do. Then, one can choose the comprehensiveness of the operator pool, effectively defining the accessible part of the energy landscape. For naive UCCSD, this would be e.g. $\mathcal{E}_{\mathrm{SD}}$ as in Eq.~\eqref{eq:excitation-set-sd}. 

At this point, one can decide to stay in the fermionic picture for longer, or directly transfer the operator pool into a set of qubit operators (either in the QCC-picture described in section~\ref{subsubsec:qcc}  or in form of Pauli strings obtained by encoding a fermionic excitation pool). 
Using specific compression techniques -- such as generic techniques described in Ref.~\citen{rubin2021compressing} or the unitary Cluster Jastrow ansatz~\cite{matsuzawa2020jastrow} -- one can reduce the number of tensor factors to represent a given operator pool and thus improve efficiency of adaptive, pool-based methods and ease trial state preparations.
The choice of operator pool either defines a fermionic UCC wavefunction or a qubit wavefunction, where the primitive excitations of Eq.~\eqref{eq:basic_fermionic_unitary} are, if one cares to think of the typical CC parametrisation, still ``hidden'' as an exponential of a sum and corresponding unitary is not directly implementable. Since individual excitations, both fermionic or qubit, in general do not commute, one typically raises a Trotter-Suzuki decomposition by writing the exponential of a sum as a product of exponentials. 
As discussed in section~\ref{subsubsec:order}, the ordering of operators can highly influence the quality of the model, which is why adaptive assembly procedures have been developed. 
If still in the fermionic picture, exponential of fermionic excitations need to be encoded and compiled to a qubit operation, using one of the typical encodings (Jordan-Wigner, Bravyi-Kitaev, etc.) and techniques described in section~\ref{subsec:ucc-ansatz}. Whereas the implementation of such exponentials leads to comparably deep circuits, exponentials of Pauli operators, as available when moving the qubit picture early on, can be implemented much easier and have thus been introduced as a more hardware-friendly alternative. 
As a last step, one can, in the spirit of classical neural networks, ``repeat layers'' $k$ times, as proposed in Ref.~\citen{Lee2019}. 

By these means, one obtains an (up to compilation) implementable, parametrised qubit wavefunction in form of a unitary circuit, which defines the accessible energy landscape. One may then choose different ways to move along the landscape in the search for the desired state, e.g. the one corresponding to the minimal energy. This can be carried out in the spirit of variational optimisation (section~\ref{sec:variational_approaches}) of the parameters. Here, one may choose to de-couple the singles optimisation to obtain on-the-fly orbital-optimisation, which has been discussed in section~\ref{subsubsec:oo-ucc}. An alternative to the variational procedure is given by the solution of projective equations (section~\ref{subsec:projectiveclassical}).

The final wavefunction is then influenced by \emph{both} the set of approximations to the parametrised model wavefunction (UCC) and the method to determine the optimal parameters. In the case of adaptive schemes such as ADAPT-VQE (section~\ref{subsubsec:adapt}) or SQPE (section~\ref{subsubsec:pqe}), model and method go hand in hand. 

\section{Outlook}
This review is an attempt to bridge the knowledge gap of both scientists working on quantum computation and theoretical chemistry.
It contains some basic explanation of coupled cluster theory for classical simulations.
Furthermore, it describes how unitary coupled cluster theory can help advance chemistry calculations using quantum devices. 
It discusses how different adaptions of UCC can help resolve some of the issues associated with calculating eigenenergies of molecular systems on quantum computers -- that is, e.g. 
as a parametrised state in variational algorithms and a state preparation procedure for quantum phase estimation. 
However, for it to reach its potential there are many caveats to overcome; some of which are pointed out in the following.

A challenge in near-term implementations of quantum algorithms is the exponential growth of error with increasing number of circuit gates. This growth is related to a finite fidelity for each gate and absence of error correction. To partially address this problem in the near-term, error mitigation techniques have been used. To facilitate use of these techniques even further, one can think of developing generator pools for UCC-based procedures that are more amenable to error mitigation.

Beyond that, to find the optimal parameters of UCC-based methods, one needs to take into account calculations of energy gradients whose computational cost depends on the eigenspectra of generators. Some work has already been done to reduce this cost,\cite{kottmann2021feasible,izmaylov2021analytic,wierichs2021general} but one can  further reduce the cost by developing new protocols involving modified generator pools.

Another shortcoming of UCC-based methods with fixed excitation rank of fermionic generators (e.g. UCCSD) is that even though the number of generators in the  first-order trotterised form grows only polynomially with the system size, the necessary chemical accuracy cannot always be achieved in general. This has to be overcome before one can use such methods to reliably solve strongly correlated systems, such as stretched nitrogen or chromium dimers.

Other exciting directions are in further exploration of  multi-reference states using hybrid UCC and non-UCC ans\"atze and  combining different adaptions of UCC-based methods, such as projective quantum eigensolver with qubit operator pools.

Finally, we hope that this review is a medium to bring together scientists from the classical and quantum community to take inspiration from each others experiences and research to further advance both quantum computing as well as theoretical chemistry, similarly as done in Refs.~\citen{chen2021QuantumInspired} and~\citen{filip2020stochastic} which adapted techniques from quantum into a classical algorithm.

\section*{Conflicts of interest}
There are no conflicts to declare.

\section*{Acknowledgements}
We thank Chong Sun for providing valuable feedback and discussions on the manuscript. 
A.A.-G. acknowledges the generous support from Google, Inc.  in the form of a Google Focused Award. A.A.-G. also acknowledges support from the Canada Industrial Research Chairs  Program and the Canada 150 Research Chairs Program. A.F.I. acknowledges financial support from the Google Quantum Research Program, Zapata Computing Inc., the Early Researcher Award from the Ontario Ministry of Research and Innovation, and the Natural Sciences and Engineering Research Council of Canada. P.S. acknowledges support by a fellowship within the IFI program of the German Academic Exchange Service (DAAD).  We thank the generous support of Anders G. Fr\o{}seth. Resources used in preparing this research were provided, in part, by the Province of Ontario, the Government of Canada through CIFAR, and companies sponsoring the Vector Institute <www.vectorinstitute.ai/partners>.

\bibliography{ucc.bib}
\bibliographystyle{apsrev4-1}

\section*{Appendix}
\section*{Multiplication tables for $C_{2v}$ and $D_{2h}$}
For the sake of completeness, we state the multiplication tables for the point groups $C_{2v}$ and $D_{2h}$ that the main text refers to here:

\begin{table}[h]
\begin{center}
\caption{Multiplication table for the $C_{2v}$ point group.}\label{T.c2v}
\begin{tabular}{c | cc cc}
& $A_1$ & $A_2$ & $B_1$ & $B_2$ \\
\hline
$A_1$ & $A_1$ & $A_2$ & $B_1$ & $B_2$ \\
$A_2$ & $A_2$ & $A_1$ & $B_2$ & $B_1$ \\
$B_1$ & $B_1$ & $B_2$ & $A_1$ & $A_2$ \\
$B_2$ & $B_2$ & $B_1$ & $A_2$ & $A_1$ \\
\end{tabular}
\normalcolor
\end{center}
\end{table}

\begin{table}[h]
\begin{center}
\caption{Multiplication table for the $D_{2h}$ point group.}\label{T.d2h}
\begin{tabular}{c | cc ccc ccc}
& $A_{g}$ & $A_{u}$ & $B_{1g}$ & $B_{2g}$ & $B_{3g}$ & $B_{1u}$ & $B_{2u}$ & $B_{3u}$ \\
\hline
$A_{g}$ & $A_{g}$ & $A_{u}$ & $B_{1g}$ & $B_{2g}$ & $B_{3g}$ & $B_{1u}$ & $B_{2u}$ & $B_{3u}$ \\
$A_{u}$ & $A_{u}$ & $A_{g}$ & $B_{1u}$ & $B_{2u}$ & $B_{3u}$ & $B_{1g}$ & $B_{2g}$ & $B_{3g}$ \\
$B_{1g}$ & $B_{1g}$ & $B_{1u}$ & $A_{g}$ & $B_{3g}$ & $B_{2g}$ & $A_{u}$ & $B_{3u}$ & $B_{2u}$ \\
$B_{2g}$ & $B_{2g}$ & $B_{2u}$ & $B_{3g}$ & $A_{g}$ & $B_{1g}$ & $B_{3u}$ & $A_{u}$ & $B_{1u}$ \\
$B_{3g}$ & $B_{3g}$ & $B_{3u}$ & $B_{2g}$ & $B_{1g}$ & $A_{g}$ & $B_{2u}$ & $B_{1u}$ & $A_{u}$ \\
$B_{1u}$ & $B_{1u}$ & $B_{1g}$ & $A_{u}$ & $B_{3u}$ & $B_{2u}$ & $A_{g}$ & $B_{3g}$ & $B_{2g}$ \\
$B_{2u}$ & $B_{2u}$ & $B_{2g}$ & $B_{3u}$ & $A_{u}$ & $B_{1u}$ & $B_{3g}$ & $A_{g}$ & $B_{1g}$ \\
$B_{3u}$ & $B_{3u}$ & $B_{3g}$ & $B_{2u}$ & $B_{1u}$ & $A_{u}$ & $B_{2g}$ & $B_{1g}$ & $A_{g}$ \\
\end{tabular}
\normalcolor
\end{center}
\end{table}
\normalcolor

\section*{Reduction of CNOT gate count for UCCSD by using the tableau representation and Jordan-Wigner transformation}\label{A.Tableu}

The structure of single and double excitations obtained using the JW transformation
is easy to compile using the tableau representation and its algebra, which will be explained in more detail in the subsequent section. Reductions of at least 50\% in the number of CNOT gates can be obtained for both singles and doubles when comparing with the typical exponential map of Pauli strings that has been described in section~\ref{sec:qubit_perspective}. Here, a matrix consisting of 
mutually commuting Pauli strings is first diagonalized and then exponentiated instead of exponentiating each 
Pauli string individually.

When the Jordan-Wigner transformation is used to encode UCC excitations, a non-local parity operator 
is needed to introduce the required phase change by a string of $\qpz$ operators between the excitation 
qubits (for single excitations) or between the occupied and between the virtual qubits (for double excitations) \cite{seeley2012bravyikitaev}.
For this reason, the scaling of the CNOT count increases by the number of qubits ($N$) in addition with the $N_\text{occ}^2N_\text{virt}^2$ bottleneck of
double excitations for UCCSD, yielding an upper-bound of $O(N^5)$ number of CNOT gates. However, due to its structure, a 
simple compilation of UCCSD operators can be performed based on the tableau representation, which reduces the CNOT gate count significantly.

The compilation of UCCSD excitations is separated in two steps. First, the parity operator is factorised for a set of four
commuting Pauli strings. Second, the remaining qubit excitations are diagonalized with Clifford gates. 

\subsection*{Tableau representation and its algebra}
We use a binary array to represent matrices consisting of Pauli strings.
This array has three blocks ($[X,Z,s]$), where each row is a Pauli string. The $j$-th
components of the $i$-th Pauli string are obtained from ($X_{i,j},Z_{i,j}$)
and defined as
\begin{align}
(X_{i,j},Z_{i,j}) &=
\left\{ \begin{matrix}
(0,0)\ j \textrm{-th qubit has } \identity^{(j)} \\
(0,1)\ j \textrm{-th qubit has } \pauli{z}^{(j)} \\
(1,0)\ j \textrm{-th qubit has } \pauli{x}^{(j)} \\
(1,1)\ j \textrm{-th qubit has } \pauli{y}^{(j)} \\
\end{matrix} \right. .
\end{align}
In addition, the $s$ block is a single column containing the sign
of the Pauli string. For example, a matrix describing the addition of
two Pauli strings ($+\pauli{y}^{(0)} \pauli{z}^{(1)} \pauli{x}^{(2)}-\pauli{x}^{(0)}\pauli{z}^{(1)}\pauli{y}^{(2)}$ )
can be represented with the following tableau:
\begin{align}\label{E.T.1}
\left[ \begin{matrix}
+\pauli{y}^{(0)} \pauli{z}^{(1)} \pauli{z}^{(2)} \pauli{x}^{(3)} \\
-\pauli{x}^{(0)} \pauli{z}^{(1)} \pauli{z}^{(2)} \pauli{y}^{(3)} \\
\end{matrix} \right]
=
\left[ \begin{matrix}
1 & 0 & 0 & 1 \\
1 & 0 & 0 & 1 \\
\end{matrix} \left|
\begin{matrix}
1 & 1 & 1 & 0 \\
0 & 1 & 1 & 1 \\
\end{matrix} \right|
\begin{matrix}
0 \\
1 \\
\end{matrix} \right] .
\end{align}

Representation of a linear combination of Pauli strings in the tableau
representation ($\bm{P}$) is useful to display the correlation information
between these operators in a more visual way. Then, it becomes easier
to apply unitary operators from the Clifford group ($\bm{C}$) to simplify
$\bm{P}$ into either all $\pauli{z}$ or diagonal operators ($\bm{Z}$). The Clifford
group in $\mathbb{C}^2$ is generated by the Hadamard, phase and conditional-NOT gates ($\hada$, $\quoperator{S}$ and
CNOT, respectively). Both $\hada$ and $\quoperator{S}$ are one-qubit gates. For example,
$\hada(a)$ applies $\hada$ to qubit $a$. The CNOT-gate (also $\quoperator{CX}$)
is a two qubit gate; it applies $\pauli{x}$ to qubit $b$ to the $\ket{1}$-part of  qubit $a$.
Furthermore, the $\quoperator{CZ}(a,b)$-gate applies $\pauli{z}$ to qubit
$b$ conditioned on qubit $a$ being in $\ket{1}$; it is equal to $\hada(b)\quoperator{CX}(a,b)\hada(b)$.

We can define $\bm{C}$ as a product of several Clifford gates
working in an initial  linear combination of Pauli strings ($\bm{P}_1$) such that
\begin{align}\label{E.tableau-transformation}
\bm{P}_1 = \bm{C} \bm{P}_2 \bm{C}^{\dagger} ,
\end{align}
where $\bm{P}_2$ is the updated tableau pertaining the $\bm{C}$ unitary transformation,
see Table \ref{T.tableau} for a summary of tableau algebra when using Clifford gates.

\begin{table}
\begin{center}
\caption{Tableau sign and block update when applying Clifford gates.\label{T.tableau}}
\begin{tabular}{ccc}
\hline
\hline
Gate & Sign update & Block update \\
\hline
$\hada(a)$  & $s_i = s_i \oplus X_{i,a}\otimes Z_{i,a}$ & swap($X_{i,a}$,$Z_{i,a}$) \\
$\quoperator{S}(a)$  & $s_i = s_i \oplus X_{i,a}\otimes (Z_{i,a}\oplus 1)$ & $Z_{i,a} = X_{i,a}\oplus Z_{i,a}$ \\
$\quoperator{CX}(a,b)$  & $s_i = s_i \oplus X_{i,a}\otimes Z_{i,b}\otimes$ & $Z_{i,a} = Z_{i,a}\oplus Z_{i,b}$\\
&$(X_{i,b}\oplus Z_{i,a}\oplus 1)$& $X_{i,b} = X_{i,a}\oplus X_{i,b}$\\
$\quoperator{CZ}(a,b)$  & $s_i = s_i \oplus X_{i,a}\otimes X_{i,b}\otimes$ & $Z_{i,a} = X_{i,b}\oplus Z_{i,a}$\\
&$(Z_{i,a}\oplus Z_{i,b})$& $Z_{i,b} = X_{i,a}\oplus Z_{i,b}$\\
\hline
\hline
\end{tabular}
\end{center}
\end{table}

For example, when $\bm{P}_1$ is the tableau from Eq. (\ref{E.T.1}) ($+\pauli{y}^{(0)} \pauli{z}^{(1)} \pauli{z}^{(2)}
\pauli{x}^{(3)}-\pauli{x}^{(0)} \pauli{z}^{(1)} \pauli{z}^{(2)} \pauli{y}^{(3)}$ )
and $\bm{C}$ equals to $\quoperator{CX}(1,2)$, the following is obtained:
\begin{align}\label{E.T.example}
\left[ \begin{matrix}
1 & 0 & 0 & 1 \\
1 & 0 & 0 & 1 \\
\end{matrix} \left|
\begin{matrix}
1 & 1 & 1 & 0 \\
0 & 1 & 1 & 1 \\
\end{matrix} \right|
\begin{matrix}
0 \\
1 \\
\end{matrix} \right]
=
\bm{C}
\left[ \begin{matrix}
1 & 0 & 0 & 1 \\
1 & 0 & 0 & 1 \\
\end{matrix} \left|
\begin{matrix}
1 & 0 & 1 & 0 \\
1 & 0 & 1 & 1 \\
\end{matrix} \right|
\begin{matrix}
0 \\
1 \\
\end{matrix} \right]
\bm{C}^{\dagger},
\end{align}
or,
\begin{align}\label{E.T.example2}
\left[ \begin{matrix}
+\pauli{y}^{(0)} \pauli{z}^{(1)} \pauli{z}^{(2)} \pauli{x}^{(3)} \\
-\pauli{x}^{(0)} \pauli{z}^{(1)} \pauli{z}^{(3)} \pauli{y}^{(3)} \\
\end{matrix} \right]
=
\quoperator{CX}(1,2)
\left[ \begin{matrix}
+\pauli{y}^{(0)} \pauli{z}^{(2)} \pauli{x}^{(3)} \\
-\pauli{x}^{(0)} \pauli{z}^{(3)} \pauli{y}^{(3)} \\
\end{matrix} \right]
\quoperator{CX}(1,2) ,
\end{align}
where $\quoperator{CX}(a,b)^{\dagger}=\quoperator{CX}(a,b)$ is used (also, in the future $\hada(a)^{\dagger}=\hada(a)$ and $\quoperator{CZ}(a,b)=\quoperator{CZ}(a,b)^{\dagger}$).
Following up with the $\quoperator{CZ}(2,3)$ gate will remove the $\pauli{z}$ operators completely,
\begin{align}\label{E.T.example2}
\left[ \begin{matrix}
+\pauli{y}^{(0)} \pauli{z}^{(2)} \pauli{x}^{(3)} \\
-\pauli{x}^{(0)} \pauli{z}^{(3)} \pauli{y}^{(3)} \\
\end{matrix} \right]
=
\quoperator{CZ}(2,3)
\left[ \begin{matrix}
+\pauli{y}^{(0)} \pauli{x}^{(3)} \\
-\pauli{x}^{(0)} \pauli{y}^{(3)} \\
\end{matrix} \right]
\quoperator{CZ}(2,3) ,
\end{align}
which yields the same results obtained previously for the removal of the parity operator (a continuous non-local 
string of $\pauli{z}$ operators obtained with the JW transformation).
This is useful because $\bm{C}\bm{C}^{\dagger}=\identity$ and several Pauli strings have the same parity operator 
(or similar) and large reductions of two-qubit gate counts are obtained using this compilation technique.

\subsection*{Tableau compilation of single excitations/de-excitations}
Grouping four mutually commutative Pauli strings within the same spatial-orbitals yields a better reduction for single 
excitations/de-excitations than compilation of one single excitation at a time. 
In this subsection the spin-orbital notation is different, 
the occupied $\alpha$ spin-orbitals is $i$ ($a$) and the occupied (virtual) $\beta$ spin-orbital
is $\bar{i}$ ($\bar{a}$).
Then, the first group to compile contains excitations/de-excitations within the same spin,
$\hat{a}_{a}^{\dagger} \hat{a}_{i} - \hat{a}_{i}^{\dagger} \hat{a}_{a}
+\hat{a}_{\bar{a}}^{\dagger} \hat{a}_{\bar{i}} - \hat{a}_{\bar{i}}^{\dagger} \hat{a}_{\bar{a}}$,
and the Jordan-Wigner transformation yields,
\begin{align}
\text{JW}(\cre{a} \an{i} +\cre{\bar{a}} \an{\bar{i}} - h.c.) 
&= \imag\frac{1}{2} \bigotimes_{k=\bar{i}+1}^{a-1} \pauli{z}^{(k)} 
(+\pauli{y}^{(i)}\pauli{z}^{(\bar{i})}\pauli{x}^{(a)} \nonumber \\
&-\pauli{x}^{(i)}\pauli{z}^{(\bar{i})}\pauli{y}^{(a)}
 +\pauli{y}^{(\bar{i})}\pauli{z}^{(a)}\pauli{x}^{(\bar{a})}
 -\pauli{x}^{(\bar{i})}\pauli{z}^{(a)}\pauli{y}^{\bar{a}} ) ,
\end{align}
where the spatial-orbital parity operator ($\bigotimes_{k=\bar{i}+1}^{a-1} \pauli{z}^{(k)}$) can be factorised from
a tableau of pseudo-qubit-excitations.
First, the parity operator is factorised using $\uunitary_{s,ss,p}$,
\begin{align}
\bigotimes_{k=\bar{i}+1}^{a-1} \pauli{z}^{(k)} 
\left( \begin{matrix}
+\pauli{y}^{(i)}\pauli{z}^{(\bar{i})}\pauli{x}^{(a)}  \\
+\pauli{y}^{(\bar{i})}\pauli{z}^{(a)}\pauli{x}^{(\bar{a})} \\
-\pauli{x}^{(i)}\pauli{z}^{(\bar{i})}\pauli{y}^{(a)}  \\
-\pauli{x}^{(\bar{i})}\pauli{z}^{(a)}\pauli{y}^{\bar{a}} \\
\end{matrix} \right)
 =
\uunitary_{s,ss,p}
\left[ \begin{matrix}
 +\pauli{y}^{(i)}\pauli{z}^{(\bar{i})}\pauli{x}^{(a)}  \\
 +\pauli{y}^{(\bar{i})}\pauli{z}^{(a)}\pauli{x}^{(\bar{a})} \\
 -\pauli{x}^{(i)}\pauli{z}^{(\bar{i})}\pauli{y}^{(a)}  \\
 -\pauli{x}^{(\bar{i})}\pauli{z}^{(a)}\pauli{y}^{\bar{a}} \\
\end{matrix} \right] 
\uunitary_{s,ss,p}^{\dagger},
\end{align}
where $\uunitary_{s,ss,p}$ is a unitary transformation that contains $(L+1)$ two-qubit gates ($L$ is the span of the $k$-index), 
\begin{align}
\uunitary_{s,ss,p} = \left(\bigotimes_{k = i + 1}^{\bar{a} -2} \quoperator{CX}(k,k+1)\right) \quoperator{CZ}(a-1,a) \quoperator{CZ}(a-1,\bar{a}) .
\end{align}
Second, the remaining tableau (pertaining pseudo-single-qubit-excitations) is diagonalised with $\uunitary_{s,ss}$
\begin{align}
\left[ \begin{matrix}
 +\pauli{y}^{(i)}\pauli{z}^{(\bar{i})}\pauli{x}^{(a)}  \\
 +\pauli{y}^{(\bar{i})}\pauli{z}^{(a)}\pauli{x}^{(\bar{a})} \\
 -\pauli{x}^{(i)}\pauli{z}^{(\bar{i})}\pauli{y}^{(a)}  \\
 -\pauli{x}^{(\bar{i})}\pauli{z}^{(a)}\pauli{y}^{\bar{a}} \\
\end{matrix} \right] 
=
\uunitary_{s,ss} \left[ \begin{matrix}
-\pauli{z}^{(i)}\\
-\pauli{z}^{(\bar{i})}\\
+\pauli{z}^{(a)}\\
+\pauli{z}^{(\bar{a})}\\
\end{matrix} \right] \uunitary_{s,ss}^{\dagger} ,
\end{align}
where $\uunitary_{s,ss}$ is a unitary transformation that contains three two-qubit gates,
\begin{align}
\uunitary_{s,ss}=\quoperator{CZ}(\bar{i},a) \hada(*) \quoperator{CZ}(i,a) \quoperator{CZ}(\bar{i},\bar{a}) \quoperator{S}(*) \hada(*).
\end{align}
We used $(*)$ to denote that the corresponding one-qubit gate is applied to all the four pertaining qubits.

The second group contains single excitations/de-excitations with different spin,
$\hat{a}_{\bar{a}}^{\dagger} \hat{a}_{i} - \hat{a}_{i}^{\dagger} \hat{a}_{\bar{a}}
+\hat{a}_{a}^{\dagger} \hat{a}_{\bar{i}} - \hat{a}_{\bar{i}}^{\dagger} \hat{a}_{a}$,
and the Jordan-Wigner transformation yields,
\begin{align}
\text{JW}(\cre{\bar{a}} \an{i} +\cre{a} \an{\bar{i}} - h.c.) 
&= i\frac{1}{2} \bigotimes_{k=\bar{i}+1}^{a-1} \pauli{z}^{(k)} 
(+\pauli{y}^{(i)}\pauli{z}^{(\bar{i})}\pauli{z}^{(a)}\pauli{x}^{(\bar{a})} \\
&-\pauli{x}^{(i)}\pauli{z}^{(\bar{i})}\pauli{z}^{(a)}\pauli{y}^{\bar{a}}
 +\pauli{y}^{(\bar{i})}\pauli{x}^{(a)} 
 -\pauli{x}^{(\bar{i})}\pauli{y}^{(a)} ) ,
\end{align}
where the same spatial-orbital parity operator ($\bigotimes_{k=\bar{i}+1}^{a-1} \pauli{z}^{(k)}$) 
and a second tableau of opposite spin pseudo qubit excitations are obtained.
The parity operator can be factorised with the same unitary transformation ($\uunitary_{s,ss,p}$) and because 
$\uunitary_{s,ss,p}^{\dagger}\uunitary_{s,ss,p}=\identity$ further reduction of two-qubit gates is obtained.
Then, the second tableau is diagonalized with the $\uunitary_{s,os}$,
\begin{align}
\left[ \begin{matrix}
 +\pauli{y}^{(i)}\pauli{z}^{(\bar{i})}\pauli{z}^{(a)}\pauli{x}^{(\bar{a})} \\
 +\pauli{y}^{(\bar{i})}\pauli{x}^{(a)}  \\
 -\pauli{x}^{(\bar{i})}\pauli{y}^{(a)}  \\
 -\pauli{x}^{(i)}\pauli{z}^{(\bar{i})}\pauli{z}^{(a)}\pauli{y}^{\bar{a}} \\
\end{matrix} \right] =
\uunitary_{s,os} \left[ \begin{matrix}
-\pauli{z}^{(i)}\\
+\pauli{z}^{(\bar{i})}\\
-\pauli{z}^{(a)}\\
+\pauli{z}^{(\bar{a})}\\
\end{matrix} \right] \uunitary_{s,os}^{\dagger} ,
\end{align}
where $\uunitary_{s,os}$ is an unitary transformation that contains four two-qubit gates,
\begin{align}
\uunitary_{s,os}=\quoperator{CZ}(\bar{i},\bar{a}) \quoperator{CZ}(a,\bar{a}) \hada(*) \quoperator{CZ}(i,\bar{a}) \quoperator{CZ}(\bar{i},a) \quoperator{S}(*) \hada(*) .
\end{align}

The total number of two-qubit gates must be multiplied by two to complete the unitary transformation, 
application of $\uunitary$ followed by $\uunitary^{\dagger}$. Then the total number of two-qubit gates is $2((L+1)+7)=16+2L$, which greatly improves over the basic exponential map of individual Pauli strings
that requires $32+16L$ two-qubit gates.

\subsection*{Tableau compilation of double excitations/de-excitations}
The JW-transformation of a double excitation/de-excitaion reads,~\cite{romero2018strategies}
\begin{align}
\text{JW}(\cre{a}\cre{b}\an{j}\an{i} - h.c.) &= i\frac{1}{8} \bigotimes_{k=i+1}^{j-1}
\pauli{z}^{(k)}\bigotimes_{c=a+1}^{b-1}\pauli{z}^{(c)} 
(+\pauli{y}^{(i)}\pauli{x}^{(j)}\pauli{x}^{(a)}\pauli{x}^{(b)} 
 +\pauli{x}^{(i)}\pauli{y}^{(j)}\pauli{x}^{(a)}\pauli{x}^{(b)} \nonumber \\
&-\pauli{x}^{(i)}\pauli{x}^{(j)}\pauli{y}^{(a)}\pauli{x}^{(b)}
 -\pauli{x}^{(i)}\pauli{x}^{(j)}\pauli{x}^{(a)}\pauli{y}^{(b)}
 -\pauli{x}^{(i)}\pauli{y}^{(j)}\pauli{y}^{(a)}\pauli{y}^{(b)} \nonumber \\
&-\pauli{y}^{(i)}\pauli{x}^{(j)}\pauli{y}^{(a)}\pauli{y}^{(b)} 
 +\pauli{y}^{(i)}\pauli{y}^{(j)}\pauli{x}^{(a)}\pauli{y}^{(b)}
 +\pauli{y}^{(i)}\pauli{y}^{(j)}\pauli{y}^{(a)}\pauli{x}^{(b)} ) ,
\end{align}
where the parity operator ($\bigotimes_{k=i+1}^{j-1}\pauli{z}^{(k)}\bigotimes_{c=a+1}^{b-1}\pauli{z}^{c}$)
is factorised with $\uunitary_{d,p}$, 
\begin{align}
\bigotimes_{k=i+1}^{j-1}
\pauli{z}^{(k)}\bigotimes_{c=a+1}^{b-1}\pauli{z}^{(c)}
\left( \begin{matrix}
 +\pauli{y}^{(i)}\pauli{x}^{(j)}\pauli{x}^{(a)}\pauli{x}^{(b)} \\
 +\pauli{x}^{(i)}\pauli{y}^{(j)}\pauli{x}^{(a)}\pauli{x}^{(b)} \\
 -\pauli{x}^{(i)}\pauli{x}^{(j)}\pauli{y}^{(a)}\pauli{x}^{(b)} \\
 -\pauli{x}^{(i)}\pauli{x}^{(j)}\pauli{x}^{(a)}\pauli{y}^{(b)} \\
 -\pauli{x}^{(i)}\pauli{y}^{(j)}\pauli{y}^{(a)}\pauli{y}^{(b)} \\
 -\pauli{y}^{(i)}\pauli{x}^{(j)}\pauli{y}^{(a)}\pauli{y}^{(b)} \\
 +\pauli{y}^{(i)}\pauli{y}^{(j)}\pauli{x}^{(a)}\pauli{y}^{(b)} \\
 +\pauli{y}^{(i)}\pauli{y}^{(j)}\pauli{y}^{(a)}\pauli{x}^{(b)} \\
\end{matrix} \right)
=
\uunitary_{d,p}
\left[ \begin{matrix}
 +\pauli{y}^{(i)}\pauli{x}^{(j)}\pauli{x}^{(a)}\pauli{x}^{(b)} \\
 +\pauli{x}^{(i)}\pauli{y}^{(j)}\pauli{x}^{(a)}\pauli{x}^{(b)} \\
 -\pauli{x}^{(i)}\pauli{x}^{(j)}\pauli{y}^{(a)}\pauli{x}^{(b)} \\
 -\pauli{x}^{(i)}\pauli{x}^{(j)}\pauli{x}^{(a)}\pauli{y}^{(b)} \\
 -\pauli{x}^{(i)}\pauli{y}^{(j)}\pauli{y}^{(a)}\pauli{y}^{(b)} \\
 -\pauli{y}^{(i)}\pauli{x}^{(j)}\pauli{y}^{(a)}\pauli{y}^{(b)} \\
 +\pauli{y}^{(i)}\pauli{y}^{(j)}\pauli{x}^{(a)}\pauli{y}^{(b)} \\
 +\pauli{y}^{(i)}\pauli{y}^{(j)}\pauli{y}^{(a)}\pauli{x}^{(b)} \\
\end{matrix} \right]
\uunitary_{d,p}^{\dagger} ,
\end{align}
where $\uunitary_{d,p}$ is an unitary transformation built with $L$ two-qubit gates ($L$ is the addition of $k$ and $c$ spans)
and $\uunitary_{d,p}$ reads,
\begin{align}
\uunitary_{d,p} = \left( \bigotimes_{k = i + 1}^{j-2} \quoperator{CX}(k,k+1) \right) \quoperator{CX}(j-1,a+1) 
\left( \bigotimes_{c = a + 1}^{b-2} \quoperator{CX}(c,c+1) \right) \quoperator{CZ}(b-1,b) .
\end{align}

The remaining Pauli strings is a double qubit excitation and is separated into two tableaus,
\begin{align}
\bm{P}_{d,x} &= \left[ \begin{matrix}
-\pauli{x}^{(i)}\pauli{y}^{(j)}\pauli{y}^{(a)}\pauli{y}^{(b)} \\
-\pauli{y}^{(i)}\pauli{x}^{(j)}\pauli{y}^{(a)}\pauli{y}^{(b)} \\
+\pauli{y}^{(i)}\pauli{y}^{(j)}\pauli{x}^{(a)}\pauli{y}^{(b)} \\
+\pauli{y}^{(i)}\pauli{y}^{(j)}\pauli{y}^{(a)}\pauli{x}^{(b)} \\
\end{matrix}\right]  \\
\bm{P}_{d,y} &= \left[ \begin{matrix}
+\pauli{y}^{(i)}\pauli{x}^{(j)}\pauli{x}^{(a)}\pauli{x}^{(b)} \\
+\pauli{x}^{(i)}\pauli{y}^{(j)}\pauli{x}^{(a)}\pauli{x}^{(b)} \\
-\pauli{x}^{(i)}\pauli{x}^{(j)}\pauli{y}^{(a)}\pauli{x}^{(b)} \\
-\pauli{x}^{(i)}\pauli{x}^{(j)}\pauli{x}^{(a)}\pauli{y}^{(b)} \\
\end{matrix}\right] .
\end{align}
Then, the $\bm{P}_{d,y}$ tableau is diagonalized with $\uunitary_{d,y}$
\begin{align}
\left[ \begin{matrix}
+\pauli{y}^{(i)}\pauli{x}^{(j)}\pauli{x}^{(a)}\pauli{x}^{(b)} \\
+\pauli{x}^{(i)}\pauli{y}^{(j)}\pauli{x}^{(a)}\pauli{x}^{(b)} \\
-\pauli{x}^{(i)}\pauli{x}^{(j)}\pauli{y}^{(a)}\pauli{x}^{(b)} \\
-\pauli{x}^{(i)}\pauli{x}^{(j)}\pauli{x}^{(a)}\pauli{y}^{(b)} \\
\end{matrix}\right]
=
\uunitary_{d,y}
\left[ \begin{matrix}
-\pauli{z}^{(i)}\\
-\pauli{z}^{(j)}\\
+\pauli{z}^{(a)}\\
+\pauli{z}^{(b)}\\
\end{matrix}\right]
\uunitary_{d,y}^{\dagger} ,
\end{align}
where the unitary transformation reads,
\begin{align}
\uunitary_{d,y} = \hada(*) \quoperator{CZ}(i,j) \quoperator{CZ}(i,a) \quoperator{CZ}(i,b) \quoperator{CZ}(j,a) \quoperator{CZ}(j,b) \quoperator{CZ}(a,b) \quoperator{S}(*) \hada(*) .
\end{align}
Furthermore, diagonalisation of the $\bm{P}_{d,y}$ tableau is obtained with 
\begin{align}
\left[ \begin{matrix}
-\pauli{x}^{(i)}\pauli{y}^{(j)}\pauli{y}^{(a)}\pauli{y}^{(b)} \\
-\pauli{y}^{(i)}\pauli{x}^{(j)}\pauli{y}^{(a)}\pauli{y}^{(b)} \\
+\pauli{y}^{(i)}\pauli{y}^{(j)}\pauli{x}^{(a)}\pauli{y}^{(b)} \\
+\pauli{y}^{(i)}\pauli{y}^{(j)}\pauli{y}^{(a)}\pauli{x}^{(b)} \\
\end{matrix}\right]
=
\uunitary_{d,x}
\left[ \begin{matrix}
-\pauli{z}^{(i)}\\
-\pauli{z}^{(j)}\\
+\pauli{z}^{(a)}\\
+\pauli{z}^{(b)}\\
\end{matrix}\right]
\uunitary_{d,x}^{\dagger} ,
\end{align}
where $\uunitary_{d,x}$ is quite similar to $\uunitary_{d,y}$,
\begin{align}
\uunitary_{d,x} = \quoperator{S}(*) \uunitary_{d,y} .
\end{align}

In conclusion, the total number of two-qubit gates needed to double excitations is $24+2L$ ($2(L+6+6)$) which greatly improves 
over the basic compilation requiring $48+16L$ two-qubit gates.
More importantly, the JW complexity due to the parity operator is reduced by 87.5\% for both single and double excitations.

\end{document}